\documentclass[aip,jcp,reprint,numerical,amsmath,amssymb,amsfonts]{revtex4-1}
\usepackage{bm,graphicx,xcolor,hyperref,rotating,lineno,mathtools,multirow,makecell,placeins,physics}
\usepackage[version=4]{mhchem}
\usepackage{hyperref}
\hypersetup{
    colorlinks=true,
    linkcolor=blue,
    filecolor=blue,
    urlcolor=blue,
    citecolor=blue
}

\usepackage[normalem]{ulem}

\newcommand{\br}{\mathbf{r}}
\newcommand{\bp}{\mathbf{p}}

\newcommand{\bx}{\mathbf{x}}
\newcommand{\bX}{\mathbf{X}}
\newcommand{\by}{\mathbf{y}}
\newcommand{\bY}{\mathbf{Y}}
\newcommand{\bH}{\mathbf{H}}
\newcommand{\bg}{\mathbf{g}}
\newcommand{\bU}{\mathbf{U}}
\newcommand{\bA}{\mathbf{A}}
\newcommand{\bI}{\mathbf{I}}

\newcommand{\bZ}{\mathbf{Z}}
\newcommand{{\dno}}{\Delta\text{NO}}
\newcommand{{\dnoof}}{\Delta\text{NO-OF}}
\newcommand{{\dnocs}}{\Delta\text{NO-CS}}
\newcommand{\cl}{\text{cl}}
\newcommand{\open}{\text{op}}
\newcommand{\oc}{\text{oc}}
\newcommand{\na}{n^\alpha}
\newcommand{\nb}{n^\beta}

\makeatletter
\newcommand\footnoteref[1]{\protected@xdef\@thefnmark{\ref{#1}}\@footnotemark}
\makeatother

\newcommand{\UWin}{Department of Chemistry, University of Winnipeg, Winnipeg, Manitoba, R3B 2E9, Canada}
\newcommand{\UMan}{Department of Chemistry, University of Manitoba, Winnipeg, Manitoba, R3T 2N2, Canada}

\begin{document}
%\linenumbers
\preprint{Hn v1}

%	HEAD DATA
\title{${\dno}$ and the complexities of electron correlation in simple hydrogen clusters}
\author{Ismael A. Elayan}
\affiliation{\UWin}
\author{Rishabh Gupta}
\affiliation{\UWin}
\author{Joshua W. Hollett}
\email[Corresponding author: ]{j.hollett@uwinnipeg.ca}
\affiliation{\UWin}
\affiliation{\UMan}
\date{\today}

% ABSTRACT
\begin{abstract}
	The $\Delta$ natural orbital ($\dno$) two-electron density matrix (2-RDM) and energy expression are derived from a multideterminantal wave function.  The approximate $\dno$ 2-RDM is combined with an on-top density functional and a double-counting correction to capture electron correlation.  A trust-region Newton's method optimization algorithm for the simultaneous optimization of $\dno$ orbitals and occupancies is introduced and compared to the previous iterative diagonalization algorithm.  The combination of $\dno$ and two different on-top density functionals, Colle-Salvetti (CS) and Opposite-spin exponential cusp and Fermi-hole correction (OF), is assessed on small hydrogen clusters and compared to density functional, single-reference coupled cluster, and multireference perturbation theory (MRMP2) methods.  The $\dnocs$ and $\dnoof$ methods outperform the single-reference methods, and are comparable to MRMP2.  However, there is a distinct qualitative error in the $\dno$ potential energy surface for \ce{H4} compared to the exact.  This discrepancy is explained through analysis of the $\dno$ orbitals, occupancies and the two-electron density.
\end{abstract}

%	TITLE
\maketitle

%--------------------------
\section{Introduction}
%--------------------------
\label{sec:intro}
In an effort to create efficient electron structure methods, it has become commonplace to combine contrasting models of electron correlation ({\it e.g.} wave function, density functional, density matrix functional) in order to exploit their most effective attributes.\cite{Cizek1969, Jeziorski1981, Andersson1990, Mok1996, Handy2001, Cremer2001, Pulay2011, Becke2013, Crittenden2013, Tsuchimochi2014, Wallace2014, DNO1, RamosCordoba2016, BenavidesRiveros2017, Evangelista2018, ViaNadal2019, Fromager2007, Weimer2008, Toulouse2009, Janesko2009a, Chai2009, Stoyanova2013, Yamanaka2006, Rohr2010, NOFMP2a, NOFMP2b, Perez-Jimenez2007, Manni2014, Carlson2015, Gagliardi2017, Malcolm1998, Kohn2013, Hubert2016, DNO2, Hirao1992, Grimme1999, Grafenstein2000, Angeli2001, Angeli2002, Mostafanejad2020, Furche2008} Current methods that employ such a strategy, vary in computational cost and accuracy, and in most cases their remains room for improvement in both departments.  The most common strategy is to decompose electron correlation into static (strong) and dynamic (weak) correlation, and approach each differently.  The development of these methods involves the design of models that precisely capture each component, and also blend the approaches together in a seamless fashion.  The former aspect can be thoroughly evaluated by applying the method to relatively simple systems, such as small hydrogen clusters. 

Although they are not chemically complex, the electronic structure of \ce{H} atom clusters, particularly linear chains and regular lattices of \ce{H} atoms, can serve as a systematic testing ground for new treatments of electron correlation.\cite{Paldus1993, Kowalski1998PRL, Kowalski1998CPL, Schipper1999, Jankowski1999, VanVoorhis2000, Hachmann2006, AlSaidi2007, Rassolov2007, Sinitskiy2010, Stella2011, Bytautas2011, RamosCordoba2015, Bulik2015, Robinson2012a, Robinson2012b, Robinson2012c, Robinson2012d, Sand2013, Limacher2013, Kats2013, Henderson2014, Motta2017, Motta2020, Giner2020, Marie2021}  Creating a model that accurately describes the surprisingly complex features of \ce{H} cluster potential energy surfaces can prove challenging. There have been several studies of such clusters with an array of methods, including density functional approximations, natural orbital functionals, and truncated post-HF methods.  These systems have been shown to be effective for evaluating the ability of methods to describe both ground and excited state properties. While it is important for these methods to accurately describe the potential energy surface of these ``simple" systems, it is also important to accurately describe the underlying two-electron density. A method that correctly models the two-electron density, as well as the energy and one-electron density, is more likely to be universally successful.

The $\dno$ method combines a cumulant functional approach for the treatment of static correlation with an on-top density functional (or post-HF) to treat dynamic correlation.\cite{DNO1, DNO2}  Amongst other approaches, it most resembles multiconfigurational-pair density funcitonal theory (MC-PDFT).\cite{Manni2014, Carlson2015, Gagliardi2017, Sand2017, Mostafanejad2020}  In which, a multireference one-electron density matrix (1-RDM), and subsequent one-electron density, is used to determine the kinetic, nuclear-potential, and classical Coulomb energy.  The on-top pair density, obtained from the multireference two-electron density matrix (2-RDM), is used to create a modified spin-density\cite{Manni2014} which is inserted into the usual exchange-correlation density functionals ({\it e.g.}~PBE\cite{Perdew1996}) to determine the exchange and correlation energy. The 2-RDM can be obtained from a traditional multireference self-consistent-field \cite{Manni2014, Carlson2015, Gagliardi2017, Sand2017}, a variational 2-RDM-driven complete active-space self-consistent field\cite{Mostafanejad2020}, or another economical alternative such as a pair-coupled cluster doubles calculation.\cite{Garza2015} The $\dno$ method uses an on-top density functional for only the dynamic correlation energy, where the functional depends directly on the on-top density.  The remaining components of the energy are determined using the approximate $\dno$ 2-RDM.

This work begins by introducing a $\dno$ wave function (Subsection \ref{ssec:wf}) from which the $\dno$ two-electron density matrix and energy expression can be derived.  The $\dno$ 2-RDM is then supplemented with an {\it ad hoc} term to correct for missing static correlation between electron pairs (Subsection \ref{ssec:hsc}).  This is followed by a description of the determination of the dynamic correlation energy from the $\dno$ 2-RDM (Subsection \ref{ssec:odf}), and then an approach to reduce correlation double-counting (Subsection \ref{ssec:dc}).  Then a new optimization algorithm for the $\dno$ 2-RDM is described (Subsection \ref{ssec:opt}), which involves the simultaneous optimization of the orbitals and the electron transfer variables.  The performance of the optimization method is then evaluated relative to the previous $\dno$ 2-RDM optimization algorithm. The performance of $\dno$, in describing potential energy surfaces of small \ce{H} clusters, is then compared to established single- and multireference methods (Section \ref{sec:results}).  The two-electron density is then used to analyze the static correlation provided by the $\dno$ model in the rectangular \ce{H4} cluster (Subsection \ref{ssec:rec2sq}).  Finally, the results of the $\dno$ evaluation and analysis are summarized (Section \ref{sec:conc}).

%--------------------------
\section{Theory}
%--------------------------
\label{sec:theory}

%--------------------------
\subsection{The $\dno$ wave function}
%--------------------------
\label{ssec:wf}

The multideterminantal $\dno$ wave function,
\begin{equation}\label{eq:DNOwf}
	\Psi_{\dno} = \sum_I c_I \Psi_I,
\end{equation}
is defined with respect to a reference configuration in which all open-shell (singly occupied) orbitals are considered frozen ({\it i.e.}~singly occupied in all determinants).  Only doubly occupied orbitals are active and all excitations are $2n$-tuples (doubles, quadruples, ...) where no electrons are unpaired. In other words, with respect to the active orbitals, $\Psi_{\dno}$ is seniority-zero.\cite{Bytautas2011, Stein2014, Henderson2014, Poelmans2015, Chen2015, Limacher2015, VanMeer2018, Brzek2019, Vu2020} Also, each active pair of electrons has its own active space, consisting of the doubly occupied orbital in the reference wave function and some number of vacant (virtual) orbitals. Each determinant of the expansion is then defined by $2n$-tuple excitations of the reference configuration,
\begin{equation}
	\Psi_I = \Psi_{i\bar{i}j\bar{j}k\bar{k}\dots}^{a\bar{a}b\bar{b}c\bar{c}\dots}
\end{equation}
where $i,j$ and $k$ refer to reference occupied orbitals, and $a,b$ and $c$ refer to all active orbitals. The absence of a bar, $i$, indicates an $\alpha$-spin orbital while the presence of a bar, $\bar{i}$, indicates a $\beta$-spin orbital.  Considering this, it is useful to define the determinantal expansion coefficients, $c_I$, as a product of coefficients corresponding to specific double excitations of the active electron pairs,
\begin{equation}
	c_I = d_i^a d_j^b d_k^c \dots
\end{equation}

In order to arrive at the $\dno$ formalism, the double excitation coefficients, $d_i^a$, are defined in terms of orbital occupancies, $n_i$, and electron transfer variables, $\Delta_{ia}$,
\begin{equation}
	d_i^a = \delta_{ia} \sqrt{n_i} - \sqrt{\Delta_{ia}},
\end{equation}
where $\delta_{ia}$ is the Kroenecker delta and $\Delta_{ii}=0$.  Orbital occupancies, occupied and vacant (with respect to the reference configuration), are defined as 
\begin{align}
	n_i & = 1 - \sum_a \Delta_{ia}, \\
	n_a & = \sum_i \Delta_{ia}.
\end{align}

The $N$-representable components of the $\dno$ energy expression, $E^{\dno'}$, can then be obtained by expanding the expectation value of the $N$-electron hamiltonian in terms of $n_i$, $\Delta_{ia}$, and one and two-electron integrals over orbitals.  Initially, the energy may be divided into diagonal and off-diagonal contributions,
\begin{align} \label{eq:EDNOp}
	E^{\dno'}	& = \langle \Psi_{\dno} | \hat{H} | \Psi_{\dno} \rangle \nonumber \\
			& = \sum_I c_I^2 H_{II} + \sum_{I\ne J} c_I c_J H_{IJ} ,
\end{align}
where
\begin{equation}
	H_{IJ} = \langle \Psi_I | \hat{H} | \Psi_J \rangle,
\end{equation}
and the electronic hamiltonian (in atomic units) is given by
\begin{equation}
	\hat{H} = \sum_{i=1}^N -\tfrac{1}{2} \nabla^2_i - \sum_{i=1}^N \sum_{A=1}^M \frac{Z_A}{r_{iA}} + \sum_{i=2}^N \sum_{j<i} \frac{1}{r_{ij}},
\end{equation}
where $M$ is the number of nuclei, $Z_A$ is the nuclear charge, $r_{iA}$ is the electron-nucleus distance, and $r_{ij}$ is the electron-electron distance. The diagonal contributions can be expressed as,
\begin{equation}
	\sum_I c_I^2 H_{II} =  E^{\dno}_\text{0-1RDM} + E^{\dno}_\text{pair},
\end{equation}
where $E^{\dno}_\text{0-1RDM}$ is the energy associated with the zeroth-order term in the expansion of the two-electron density matrix (2-RDM) in terms of the one-electron density matrix (1-RDM),\cite{DNO2} which is the basis of cumulant functional theory.\cite{Kutzelnigg1999}  The 0-1RDM energy is expressed as,
\begin{align}
	E^{\dno}_\text{0-1RDM} & = \sum^\cl_p 2n_p h_{pp} + \sum^\open_p n_p h_{pp} \nonumber \\
	& + \sum^\cl_p \sum^\cl_q n_p n_q \left( 2J_{pq} - K_{pq} \right) \nonumber \\
	& + \sum^\cl_p \sum^\open_q n_p n_q \left( 2J_{pq} - K_{pq} \right) \nonumber \\
	& + \sum^\open_p \sum^\open_q \frac{n_p n_q}{2} \left( J_{pq} - K_{pq} \right) ,
\end{align}
where $\cl$ and $\open$ denote the set of closed and open-shell orbitals, respectively. The one-electron integrals over spatial orbitals, $\phi_p$, are defined as,
\begin{equation}
	h_{pp} = \int \phi^*_p(\br) \left[ -\frac{1}{2}\nabla^2 - \sum_{A} \frac{Z_A}{r_A}  \right] \phi^*_p(\br) d\br
\end{equation}
where $\br = (x,y,z)$ and the two-electron Coulomb, $J_{pq}$, and exchange, $K_{pq}$, integrals are defined as,
\begin{equation}
	J_{pq} = \int \frac{\phi^*_p(\br_1)\phi^*_q(\br_2)\phi_p(\br_1)\phi_q(\br_2)}{r_{12}} d\br_1 d\br_2
\end{equation}
and
\begin{equation}
	K_{pq} = \int \frac{\phi^*_p(\br_1)\phi^*_q(\br_2)\phi_q(\br_1)\phi_p(\br_2)}{r_{12}} d\br_1 d\br_2 .
\end{equation}
After expanding the expectation values for the diagonal contributions to $E^{\dno'}$ [first term of Equation \eqref{eq:EDNOp}], simplifying, and removing the contribution of $E^{\dno}_\text{0-1RDM}$, the pair correction energy, $E^{\dno}_\text{pair}$, remains.  The pair correction energy is the first term associated with the cumulant and is given by,
\begin{equation} \label{eq:epair}
	E^{\dno}_\text{pair} = \sum_{pq} \eta_{pq} \left( 2J_{pq} - K_{pq} \right),
\end{equation}
where
\begin{align}
	\eta_{pq} & = \delta_{pq} n_p (1 - n_p) \nonumber \\
	& + \left(1 - \delta_{pq}\right) \Bigg[ O_p V_q \Delta_{pq} \left( n_q - n_p - \Delta_{pq} \right) \nonumber \\
	& \hphantom{\left(1 - \delta_{pq}\right)\Bigg[} + V_p O_q \Delta_{qp} \left( n_p - n_q - \Delta_{qp} \right) \nonumber \\
	& \hphantom{\left(1 - \delta_{pq}\right)\Bigg[} - V_p V_q \sum_r \Delta_{rp} \Delta_{rq} \Bigg] .
\end{align}
The $O_p$ and $V_p$ coefficients are elements of orbital-basis-sized vectors consisting of 1s and 0s,
\begin{equation}
	O_p = \begin{cases} 1\;,\;p \in \mathcal{O} \\ 0\;,\; \text{otherwise} \end{cases}
\end{equation}
\begin{equation}
	V_p = \begin{cases} 1\;,\;p \in \mathcal{V} \\ 0\;,\; \text{otherwise} \end{cases}
\end{equation}
where $\mathcal{O}$ denotes the set of active orbitals that are occupied in the reference wave function, and $\mathcal{V}$ denotes the set of active vacant orbitals.

The off-diagonal terms of the expectation value expansion are responsible for electron correlation.  In the case of $\dno$, excitations are limited to low-lying near-degenerate orbitals which is aimed at capturing the so-called ``static" correlation energy,
\begin{equation}
	\sum_{IJ} c_I c_J H_{IJ} =  E^{\dno}_\text{stat}
\end{equation}
By employing Slater-Condon rules\cite{Szabo1996} for the evaluation of matrix elements between Slater determinants, and considering that the active spaces for the electron pairs are disjoint, the static correlation energy can be simplified to give,
\begin{equation} \label{eq:estat}
	E^{\dno}_\text{stat} = \sum_{pq} \left( \zeta_{pq} - \xi_{pq} \right) L_{pq} ,
\end{equation}
where
\begin{equation}
	\zeta_{pq} = V_p V_q \sum_r \sqrt{\Delta_{rp} \Delta_{rq}} ,
\end{equation}
and
\begin{equation}
	\xi_{pq} = O_p V_q \sqrt{n_p \Delta_{pq}} + O_q V_p \sqrt{n_q \Delta_{qp}} .
\end{equation}
The time-inversion exchange integral is defined as
\begin{equation}
	L_{pq} = \int \frac{\phi^*_p(\br_1)\phi^*_p(\br_2)\phi_q(\br_1)\phi_q(\br_2)}{r_{12}} d\br_1 d\br_2 .
\end{equation}

The $\dno$ wave function contains only $2n$-tuple excitations, in which the electrons remained paired, and each electron pair is excited to its own (restricted) active space.  Therefore, $E^{\dno'}$ is an upper bound to the more general doubly-occupied configuration interaction (DOCI) energy.\cite{Bytautas2011,Bytautas2015}

%--------------------------
\subsection{The high-spin correction}
%--------------------------
\label{ssec:hsc}

As defined above, the $\dno$ wave function and the subsequent 2-RDM do not account for the correlation between active electron pairs (interpair correlation).  The high-spin correction is added to $\dno$ to account for this correlation exactly in the strong correlation limit ({\it e.g.}~multiple bond dissociation).  The term is {\it ad hoc}, and therefore does not guarantee $N$-representability below the strong correlation limit.  Nevertheless, it proved effective previously\cite{DNO2} and is added to the $N$-representable $\dno$ energy, derived in Subsection \ref{ssec:wf}, to give the total $\dno$ energy, 
\begin{equation}
	E^{\dno} = E^{\dno'} + E^{\dno}_\text{HSC} .
\end{equation}
The high-spin correction energy is defined as
\begin{equation} \label{eq:ehsc}
	E^{\dno}_\text{HSC} = -\sum_{pq} \kappa_{pq} K_{pq},
\end{equation}
where
\begin{multline} \label{eq:kappa}
	\kappa_{pq} = (1-\delta_{pq}) \Bigg[ \sum_{r \ne s} \xi_{pr} \xi_{ps}\\
	+ W_p \frac{n_p}{2\sqrt{2}} \sum_r \xi_{qr} + W_q \frac{n_q}{2\sqrt{2}} \sum_r \xi_{pr} \Bigg] ,
\end{multline}
where the last two terms have been added here for modelling open-shell systems.  The coefficient $W_p$ indicates if the orbital is singly-occupied,
\begin{equation}
	W_p = \begin{cases} 1\;,\;p \in \mathcal{S} \\ 0\;,\; \text{otherwise} \end{cases}
\end{equation}
where $\mathcal{S}$ is the set of singly-occupied orbitals.  By following the same reasoning as the high-spin correction for the doubly-occupied active orbitals (first term of Equation \ref{eq:kappa}), a factor of $\frac{1}{2}$ for the last two terms would be predicted using the strong-correlation limit.\cite{DNO2}  However, that results in too severe of a correction and therefore, for this first application of $\dno$ to open-shell systems, it has been reduced by a factor of $\sqrt{2}$.

%--------------------------
\subsection{Dynamic correlation}
%--------------------------
\label{ssec:odf}

As mentioned above, the $\dno$ wave function is designed to capture static correlation, which means dynamic correlation must be accounted for by some other means.  In this study, the dynamic correlation energy is captured using an on-top density functional (ODF).  The dynamic correlation energy, $E^{\dno}_\text{dyn}$, is added to the total $\dno$ energy to give the total energy including static and dynamic correlation,
\begin{equation}
	E^{\dno\text{-dyn}} = E^{\dno} + E^{\dno}_\text{dyn} .
\end{equation}
Within the current formalism, it is necessary to remove extra correlation that is erroneously included, or ``double counted".  Therefore, the total dynamic correlation energy is the sum of the dynamic correlation energy supplied by the ODF and a double-counting correction,
\begin{equation}
	E^{\dno}_{\text{dyn}} = E_\text{ODF}[\Gamma^{\dno}(\br,\br)] + E_\text{DC}[\Gamma^{\dno}(\br,\br)] ,
\end{equation}
where both are a function of the two-electron on-top density, $\Gamma(\br,\br)$. The on-top density is the probability that two-electrons may be found at the same coordinate, $\br$.  It may be calculated from the 2-RDM, which is given by
\begin{multline} 
	\Tilde{\Gamma}(\bx_1,\bx_2,\bx_1',\bx_2') = \frac{N(N-1)}{2} \int \Psi^*(\bx_1',\bx_2',\bx_3,\dots,\bx_N) 
	\\ 
	\times \Psi(\bx_1,\bx_2,\bx_3,\dots,\bx_N) d\bx_3 \dots d\bx_N,
\end{multline}
where $\Psi$ is the $N$-electron wave function and $\bx =(\br,\omega)$ is a combination of spatial $\br$ and spin $\omega$ coordinates. The spinless 2-RDM may be obtained through integration over the spin-coordinates,
\begin{equation}
	\Gamma(\br_1,\br_2,\br_1',\br_2')
	= \int \left. \Tilde{\Gamma}(\bx_1,\bx_2,\bx_1',\bx_2') \right|_{\substack{\omega_1' = \omega_1\\ \omega_2' = \omega_2}} d\omega_1 d\omega_2,
\end{equation}
and the aforementioned two-electron density is the diagonal, $\Gamma(\br_1,\br_2) = \Gamma(\br_1,\br_2,\br_1,\br_2)$.  The spinless 2-RDM can then be resolved into different spin-components corresponding to the relative spin of electrons 1 and 2,
\begin{equation} 
\begin{split}
	\Gamma(\br_1,\br_2,\br_1',\br_2') 
	& =
	\Gamma^{\alpha\alpha}(\br_1,\br_2,\br_1',\br_2') + \Gamma^{\beta\beta}(\br_1,\br_2,\br_1',\br_2') 
	\\
	& + \Gamma^{\alpha\beta}(\br_1,\br_2,\br_1',\br_2') + \Gamma^{\beta\alpha}(\br_1,\br_2,\br_1',\br_2'),
\end{split}
\end{equation}
where $\alpha$ and $\beta$ denote spin-up and spin-down electrons, respectively.

The 2-RDM, and its components, can also be expanded in an orbital basis,
\begin{equation} \label{eq:2RDMbasis}
	\Gamma(\br_1,\br_2,\br_1',\br_2') = \sum_{pqrs} \Gamma_{pqrs} \phi^*_p(\br_1')\phi^*_q(\br_2')\phi_r(\br_1)\phi_s(\br_2).
\end{equation}
where, in this case, the $\phi_p$ are the $\dno$ orbitals.  Following a procedure similar to the derivation of $E^{\dno'}$, but using second quantization, one can arrive at an expression for the elements of the spin-resolved $\dno$ 2-RDM over the $\dno$ basis,
\begin{equation}
	\Gamma^{\dno,\sigma\sigma}_{pqrs} = \frac{f_{pq}^{\sigma\sigma} n_p n_q + \eta_{pq}}{2} \delta_{pr}^{qs}
\end{equation}
and
\begin{multline} \label{eq:2RDMos}
	\Gamma^{\dno,\sigma\sigma'}_{pqrs} = \frac{f_{pq}^{\sigma\sigma'} n_p n_q + \eta_{pq}}{2} \delta_{pr}\delta_{qs} \\ - \frac{\kappa_{pq}}{2}\delta_{ps}\delta_{qr} + \frac{\zeta_{pr} - \xi_{pr}}{2}\delta_{pq}\delta_{rs},
\end{multline}
where $\sigma=\alpha$ or $\beta$ (with $\sigma \ne \sigma'$), $\delta_{pr}^{qs} = \delta_{pr}\delta_{qs} - \delta_{ps}\delta_{qr}$ and the $f^{\sigma\sigma'}_{pq}$ ensures there are only interactions with $\alpha$ electrons in singly-occupied orbitals,
\begin{subequations}
\begin{align}
	 f^{\alpha\alpha}_{pq} & = 1, \\
	 f^{\alpha\beta}_{pq} & = 1 - W_q, \\
	 f^{\beta\alpha}_{pq} & = 1 - W_p, \\
	 f^{\beta\beta}_{pq} & = \left( 1 - W_p \right)\left( 1 - W_q \right) .
\end{align}
\end{subequations}

Two ODFs are used in this study. The first is a combination of the Opposite-spin exponential cusp (OSEC) and Fermi-hole correction (FHC) functionals, referred to as the OF functional.\cite{OF2018}  The second is the Colle-Salvetti (CS) functional.\cite{Colle1975,Colle1979} The OSEC functional depends on the total $\alpha\beta$ on-top density, whereas the FHC functional depends on the $\alpha\alpha$ and $\beta\beta$ on-top densities (more specifically, their Laplacians),
\begin{multline}
	E_\text{OF}[\Gamma(\br,\br)] = E_\text{OSEC}[\Gamma^{\alpha\beta}(\br,\br) + \Gamma^{\beta\alpha}(\br,\br)] \\ + E_\text{FHC}[\Gamma^{\alpha\alpha}(\br,\br)] + E_\text{FHC}[\Gamma^{\beta\beta}(\br,\br)]
\end{multline}

The FHC functional effectively widens the Fermi-hole, which exists in any 2-RDM that obeys antisymmetry, using an exponential-cusp ansatz for the second-order cusp.  The functional is given by
\begin{widetext}
\begin{multline}
E_\text{FHC}[\Gamma^{\sigma\sigma}] =  \int \Bigg\{ \pi\left(4\delta\left[ 3 \sqrt{\pi} \left(\delta + 16 \delta^3 \right) - 40 \delta^2 -1 \right]e^{-\frac{1}{16\delta^2}} \right.\\ \left. + \sqrt{\pi}\left[ 3\sqrt{\pi} \left(\delta + 24 \delta^3\right) - 48\left(\delta^2 + 4 \delta^4\right) - 1 \right]\left[ 1 + \text{erf}\left(\frac{1}{4\delta}\right) \right] \right) L^{\sigma\sigma}(\br) \Bigg\} \\
/ \Bigg\{ 3 \delta^4 \left(4\left(\delta + 40 \delta^3\right)e^{-\frac{1}{16\delta^2}} + \sqrt{\pi} \left[ 1 + 48 \left( \delta^2 + 4 \delta^4 \right) \right]\left[ 1 + \text{erf}\left(\frac{1}{4\delta}\right) \right] \right) \Bigg\} d\br,
\end{multline}
\end{widetext}
where $L^{\sigma\sigma}(\br)$ is the Laplacian of the two-electron density with respect to the interelectronic coordinate, $\bm{u} = \br_1 - \br_2$, at the electron-electron coalescence point ($u=0$),
\begin{equation}
L^{\sigma\sigma}(\br) = \nabla_{\bm{u}}^2 \left.\Gamma^{\sigma\sigma}(\br+\bm{u}/2,\br-\bm{u}/2) \right|_{u=0},
\end{equation}
where $\br = \frac{\br_1 + \br_2}{2}$.  The $\delta$ determines the correlation length,
\begin{equation}
	\delta \equiv \delta(\br) = \kappa L^{\sigma\sigma}(\br)^{1/8}
\end{equation}
where $\kappa = 2.30$.\footnote{This value differs from the value determined in reference \citenum{OF2018}, $\kappa=1.73$, because it was found to overestimate the correlation energy between electrons of parallel-spin in these \ce{H} clusters.}

The OSEC functional approximates the input $\alpha\beta$ two-electron density as constant in the neighbourhood of the electron-electron coalescence point, and determines the energy associated with introducing an exponential cusp.  The OSEC energy is given by the functional
\begin{widetext}
\begin{equation}
E_\text{OSEC}[\Gamma^{\alpha\beta}(\br,\br)] = \int \frac{2\pi\left(2\lambda \left(\sqrt{\pi} \lambda - 1 \right)e^{-\frac{1}{4\lambda^2}} + \sqrt{\pi} \left(\sqrt{\pi} \lambda - 2\lambda^2 -1 \right)\left[ 1 + \text{erf}\left(\frac{1}{2\lambda}\right) \right] \right) \Gamma^{\alpha\beta}(\br,\br)}{\lambda^2 \left( 2 \lambda e^{-\frac{1}{4\lambda^2}} + \sqrt{\pi}\left( 1 + 2\lambda^2 \right) \left[ 1 + \text{erf}\left(\frac{1}{2\lambda}\right) \right] \right)} d\br,
\end{equation}
\end{widetext}
where the correlation length is determined by $\lambda$,
\begin{equation}
	\lambda \equiv \lambda(\br) = q_\text{OSEC} \rho(\br)^{1/3}
\end{equation}
with $q_\text{OSEC} = 2.54$ and $\rho(\br)$ is the one-electron density.\cite{OF2018}

The Colle-Salvetti ODF depends on the total (sum of all spin components) on-top density,
\begin{multline}
	E_\text{CS}\left[\Gamma\left(\bm{r},\bm{r}\right)\right] = -4 a \int \frac{\Gamma\left(\bm{r},\bm{r}\right)}{\rho\left(\bm{r}\right)}\\  \left( \frac{ 1 + b \rho\left(\bm{r}\right)^{-8/3}e^{-c\rho\left(\bm{r}\right)^{-1/3}} L(\br) }{1 + d \rho\left(\bm{r}\right)^{-1/3} } \right) d\bm{r}
\end{multline}
where $L(\br)$ is the Laplacian of the total on-top density at $u=0$, $a=0.049$, $b=0.132$, $c=0.2533$ and $d=0.349$.\cite{Colle1975}

%--------------------------
\subsection{Double-counting correction}
%--------------------------
\label{ssec:dc}

The ODFs, OF and CS, are both derived from correcting the two-electron density in the neighbourhood of the electron-electron coalescence point, or in the short-range of interelectronic distance. Multideterminantal treatments of electron correlation, such as $\dno$, capture the full range of electron correlation.  Therefore, there is overlap, or double-counting, that can occur when wave function methods are combined with density functional methods in the present manner.  The main motivation for specifically using ODFs with $\dno$, is that as electrons become strongly correlated this information is revealed to the ODF via $\Gamma^{\dno}$.  However, the ODF only depends on the OD, $\Gamma^{\dno}(\br,\br)$, which does not possess information regarding $\dno$ short-range correlation beyond the electron-electron coalescence point (in the neighbourhood of $u=0$).  Because the ODF is not sensitive to the effects of correlation at small $u\ne0$, a correction must be applied.

The general scheme behind the derivation of the OF functional\cite{OF2018} (similar to that of the CS functional\cite{Colle1975}), is the approximation of the exact two-electron density $\Gamma\left(\br_1,\br_2\right)$ by the uncorrelated two-electron density $\Gamma^0\left(\br_1,\br_2\right)$ with a cusp, $c(u)$, inserted at $u=0$,
\begin{equation}
	\Gamma\left(\br_1,\br_2\right) = \Gamma^0\left(\br_1,\br_2\right)\left( 1 - g(u)\left[ 1 - \Phi(\br) c(u) \right] \right)
\end{equation}
where $\Phi(\br)$ is a function that maintains the normalization of $\Gamma\left(\br_1,\br_2\right)$.  The correction is applied in the short-range ({\it i.e.} small $u$) and the range is controlled by the function $g(u)$,
\begin{equation}
	g(u) = e^{-\lambda(\br)u^2},
\end{equation}
which depends on the correlation length $\lambda(\br)$.  The derivation is completed by Taylor-expanding $\Gamma^0\left(\br_1,\br_2\right)$ about $u=0$.  The expansion is zeroth-order for the OSEC functional,
\begin{equation}
	\Gamma^0\left(\br_1,\br_2\right) \approx \Gamma^0\left(\br,\br\right),
\end{equation}
and second-order for the FHC functional.  The CS functional also uses a second-order expansion, but multiple approximations are made to ensure the integrand is well-behaved.

If the ODF is to exclusively account for the correlation energy in the short-range, then the short-range correlation already present in $\Gamma$ must be removed,
\begin{multline}
	E^{\dno}_\text{dyn} = E^{\dno}_\text{ODF}[\Gamma] \\- \frac{1}{2}\int \frac{\left[ \Gamma(\br_1,\br_2) - \Gamma^0(\br_1,\br_2) \right] g(u)}{u} d\br_1 d\br_2 .
\end{multline}
To arrive at a practical expression for the double-counting correction, $\Gamma(\br_1,\br_2)$ and $\Gamma^0(\br_1,\br_2)$, are treated with a zeroth-order Taylor expansion about $u=0$ ({\it i.e.}~constant), identical to the OSEC functional.  Also, the double-counting correction is not required in the strong correlation limit, when correlation causes electrons to completely avoid each other and $\Gamma(\br,\br)=0$, and must be turned off.  Therefore, the correction is mutliplied by an electron-pair specific factor, $\frac{\Gamma_{pp}}{\Gamma^0_{pp}}$, and the resulting double-counting correction is given as
\begin{multline}
	E_\text{DC}^{\dno}\left[\Gamma\left(\br,\br\right)\right] = 2\pi c_\text{DC} \\
	 \int \sum_p O_p \left( \frac{\Gamma_{pp}}{\Gamma^0_{pp}} \right) \frac{\Gamma_{pp}^0\left(\br,\br\right)
	- \Gamma_{pp}\left(\br,\br\right)}{\lambda(\br)^2} d\br
\end{multline}
where $\Gamma^0$ is the uncorrelated two-electron density, which for $\dno$ is given by
\begin{equation}
	\Gamma^{\dno,0} = \Gamma^{\dno}_\text{0-1RDM} + \Gamma^{\dno}_\text{pair} .
\end{equation}
The two-electron densities are decomposed into electron-pair components, which for $u=0$, are given by
\begin{equation}
	\Gamma_{pp}^0\left(\bm{r},\bm{r}\right) = n_p \left| \phi_p(\br) \right|^4 + \sum_q \Delta_{pq} \left| \phi_q(\br) \right|^4,
\end{equation}
and
\begin{multline}
	\Gamma_{pp}\left(\bm{r},\bm{r}\right) = \Gamma_{pp}^0\left(\bm{r},\bm{r}\right) + \sum_{qs} \Delta_{pq}\Delta_{ps} \left| \phi_q(\br) \right|^2 \left| \phi_s(\br) \right|^2\\ - 2\sum_q \xi_{pq} \left| \phi_p(\br) \right|^2 \left| \phi_q(\br) \right|^2 .
\end{multline}
The above correction is added to the OF functional energy to give the total dynamic correlation energy.  In the case of the CS functional,
\begin{equation}
	\lambda(\br) = q_\text{CS} \rho(\br)^{1/3}
\end{equation}
where $q_\text{CS} = 2.29$ (original value from CS derivation).

%--------------------------
\subsection{2-RDM optimization}
%--------------------------
\label{ssec:opt}

The $\dno$ 2-RDM depends upon the $\dno$ orbitals, $\{ \phi_p \}$, and the electron transfer variables, $\{ \Delta_{ia} \}$.  The optimal $\dno$ 2-RDM is found by minimizing $E^{\dno}$ with respect to both $\{\phi_p\}$ and $\{\Delta_{ia}\}$.  The optimization can be performed for both $\{\phi_p\}$ and $\{\Delta_{ia}\}$ simultaneously, using the trust-region method.

In order to obtain the optimal $\dno$ orbitals, $\{\phi_p\}$, the initial set of orthonormalized orbitals, $\{ \Tilde{\phi}_r\}$, are rotated amongst themselves via a unitary transformation,
\begin{equation}
	\phi_p = \sum_r U_{rp} \Tilde{\phi}_r .
\end{equation}
To determine such a transformation, the unitary matrix is parameterized with the exponential of a skew matrix,\cite{HelgakerBook} $\bY$,  
\begin{equation}
	\bU = e^{\bY} ,
\end{equation}
where
\begin{equation}
	Y_{pq} = y_{pq}, \quad Y_{qp} = -y_{pq} .
\end{equation}

Any energy, $E[\by]$, that is a functional of a set of orbitals, may be expanded about the optimal orbitals, in a Taylor series, in terms of the skew matrix parameters, $\by$, 
\begin{equation}
	E[\by] \approx E[\mathbf{0}] + \by \cdot \bg^\phi + \frac{1}{2} \by^\dag \bH^{\phi\phi} \by ,
\end{equation}
and approximated to second-order.\cite{Bozkaya2011, Bozkaya2013, Bozkaya2014, Bozkaya2016}  The gradient, $\bg^\phi$, and hessian, $\bH^\phi$, consist of the first and second derivatives of the energy with respect to the skew-matrix parameters,
\begin{equation}
	g^\phi_{pq} = \left. \frac{\partial E^{\dno}}{\partial y_{pq}} \right|_{\by=0} , \quad H^{\phi\phi}_{pq,rs} = \left. \frac{\partial^2 E^{\dno}}{\partial y_{pq} \partial y_{rs}} \right|_{\by=0} .
\end{equation}
General expressions for the above derivatives may be found elsewhere,\cite{Bozkaya2011} but for $\dno$ specifically see Appendix \ref{app:derivs}.  In the case of $E^{\dno}$, the energy is invariant ($g^\phi_{pq} =0$) with respect to rotations between orbital pairs where both belong to the inactive (frozen) closed-shell, open-shell, or inactive vacant orbital subspaces. Consequently, optimization must be performed with respect to all other possible rotations.

In the case of $\dno$, the energy is also a function of the electron transfer variables, $\{ \Delta_{ia} \}$. To simultaneously optimize the energy with respect to both $\{ \phi_p \}$ and $\{ \Delta_{ia} \}$, the orbital gradient and hessian are combined with those for $\{ \Delta_{ia} \}$.  However, in order to avoid violation of $N$-representability of the 1-RDM, the $\{ \Delta_{ia} \}$ are first parameterized such that they remain between 0 and 1/2,
\begin{equation} \label{eq:theta}
	\Delta_{ia} = \frac{\cos^2 \theta_{ia}}{2} .
\end{equation}
The total gradient and hessian are then given by
\begin{equation}
	\bg = \begin{bmatrix} \bg^\phi \\ \bg^\theta \end{bmatrix} , \quad \bH = \begin{bmatrix} \bH^{\phi\phi} & \bH^{\phi\theta} \\ \bH^{\phi\theta} & \bH^{\theta\theta} \end{bmatrix},
\end{equation}
where
\begin{equation}
	g^\theta_{pq} = \frac{\partial E^{\dno}}{\partial \theta_{pq}} , \quad H^{\theta\theta}_{pq,rs} = \frac{\partial^2 E^{\dno}}{\partial \theta_{pq} \partial \theta_{rs}},
\end{equation}
and
\begin{equation}
	H^{\phi\theta}_{pq,rs} = \frac{\partial^2 E^{\dno}}{\partial y_{pq} \partial \theta_{rs}} .
\end{equation}

Minimization can be performed using the trust-region Newton method, in which an augmented hessian is prepared and a step is taken within a dynamic trust-region radius.  A standard trust-region approach is described in Nocedal and Wright (Algorithm 4.1).\cite{Nocedal2006} The procedure begins with the calculation of an augmented hessian, $\bA$,
\begin{equation}
	\bA = \bH + \epsilon \bI
\end{equation}
where $\bI$ is the identity matrix and $\epsilon$ is chosen such that $\bA$ is positive definite. Then a Newton step, $\bp$, is calculated by solving
\begin{equation}
	\bA \bp = -\bg .
\end{equation}
If the step is within the trust-region radius, $p = \sqrt{\bp\cdot\bp} < t$, then the step is taken, otherwise a subproblem must be solved (Equation 4.5 in reference \citenum{Nocedal2006}) in which a step within the trust-region radius is found iteratively.  Each step, whether pure Newton-Raphson or a solution to the subproblem, is evaluated by comparing the actually change in the function ($E$) to the change predicted by a second-order Taylor expansion.  If the reduction is less than a minimum threshold ({\it e.g.}~25\% of predicted) then no step is taken and $t$ is reduced.  If the reduction is more than the minimum, a step is taken, and if it is more than a maximum threshold  ({\it e.g.}~75\% of predicted) then $t$ is increased.

In order to avoid optimizations to saddle points in $\{ \phi_p \}$ and $\{ \Delta_{ia} \}$ space, a modification can be made before commencing the trust-region algorithm.  The non-augmented hessian is diagonalized to evaluate the eigenvalues,
\begin{equation}
	\bZ^T \bH \bZ = \bH'
\end{equation} 
where $\bH'$ is a diagonal matrix of hessian eigenvalues and $\bZ$ is the matrix that transforms to normal coordinates.  For any normal coordinate (mode) with a negative eigenvalue, $H_{ii}'<0$, and a gradient component with a magnitude less than a threshold, $\left|g_i'\right| < g_0$, the gradient is amplified to promote escape from the saddle point,
\begin{equation}
	\bg_\text{mod} = \bZ \bg_\text{mod}'
\end{equation}
where
\begin{equation}
	g_{\text{mod},i}' = \begin{cases} \text{sign}(g_i')g_0, &\text{ If } H_{ii}'< 0 \text{ and } \left| g_i' \right| < g_0 \\ g_i', &\text{ otherwise}   \end{cases}
\end{equation}
The trust-region method then continues as usual.

When an optimization step is calculated, the unitary matrix for the orbital rotation is prepared by first approximating $e^{\bY}$ to second order,
\begin{equation}
	\bU \approx \mathbf{I} + \bY + \frac{1}{2} \bY \bY^{\dag} ,
\end{equation}
and then ensuring it is unitary by applying Gram-Schmidt orthonormalization.

%--------------------------
\subsection{Conditions of $N$-representability}
%--------------------------
\label{ssec:Nrep}
Due to the addition of the high-spin correction, the $\dno$ 2-RDM is not strictly $N$-representable.  Therefore, it is useful to have some means of evaluating the extent to which $N$-representability might be violated.  The necessary, but not sufficient, $N$-representability conditions that depend only on the 2-RDM are referred to as the $PQG$ (or $DQG$) conditions.\cite{Garrod1964,Mazziotti2006, Mazziotti2012} They require that the following matrices, in a spin-orbital basis and defined in terms of the second quantization, be positive semi-definite ({\it i.e.}~all non-negative eigenvalues)
\begin{subequations}
\begin{align}
	P_{ijkl} & = \langle \Psi | a_j^\dagger a_i^\dagger a_k a_l  | \Psi \rangle, \\
	Q_{ijkl} & = \langle \Psi | a_j a_i a_k^\dagger a_l^\dagger  | \Psi \rangle, \\
	G_{ijkl} & = \langle \Psi | a_j^\dagger a_i a_k^\dagger a_l  | \Psi \rangle,
\end{align}
\end{subequations}
where $\Psi$ is the $N$-electron wave function, $a_i^\dagger$ and $a_i$ are creation and annihilation operators, respectively.  Sufficient conditions for $N$-representability require the calculation of higher-order matrices and are not considered here.\cite{Mazziotti2012,Li2021}  For a spin-resolved 2-RDM over spatial orbitals [\eqref{eq:2RDMps} and \eqref{eq:2RDMos}], and a diagonal 1-RDM, the elements of $\bm{P}$, $\bm{Q}$, and $\bm{G}$ can be expressed as
\begin{subequations}
\begin{align}
	P^{\sigma\sigma'\sigma\sigma'}_{pqrs} = &\; \Gamma^{\sigma\sigma'}_{pqrs}, \\
	Q^{\sigma\sigma'\sigma\sigma'}_{pqrs} = &\; \Gamma^{\sigma\sigma'}_{pqrs} + \frac{\delta_{pr}\delta_{qs}}{2} \left( 1 - n_p - n_q \right)  \nonumber \\
	& - \delta_{\sigma\sigma'} \frac{\delta_{ps}\delta_{qr}}{2} \left( 1 - n_p - n_q \right), \\
	G^{\sigma\sigma'\sigma\sigma'}_{pqrs}  = &\; \frac{\delta_{pr}\delta_{qs}}{2}n_q - \Gamma^{\sigma\sigma'}_{rqps}, \\
	G^{\sigma\sigma\sigma'\sigma'}_{pqrs}  = &\; \Gamma^{\sigma\sigma'}_{qrps} \quad (\sigma \ne \sigma'),
\end{align}
\end{subequations}
where $\sigma = \alpha$ or $\beta$, and $\sigma$ can be equivalent to $\sigma'$ (except for $G^{\sigma\sigma\sigma'\sigma'}_{pqrs}$). The expression of $\bm{P}$, $\bm{Q}$, and $\bm{G}$ in terms of spatial-orbital indices (subscript) requires specifying the spin-function (superscript) used to form the corresponding spin orbital. The matrices can be decomposed into spin-blocks for construction and analysis, which is evident from their structure,\cite{Mazziotti2002}
\begin{subequations}
\begin{align} 
	\bm{P} & =  \begin{bmatrix} \bm{P}^{\alpha\alpha\alpha\alpha} & 0 & 0 & 0 \\
	0 & \bm{P}^{\beta\beta\beta\beta} & 0 & 0 \\
	0 & 0 & \bm{P}^{\alpha\beta\alpha\beta} & 0 \\
	0 & 0 & 0 & \bm{P}^{\beta\alpha\beta\alpha} \end{bmatrix}, \\
	\bm{Q} & =  \begin{bmatrix} \bm{Q}^{\alpha\alpha\alpha\alpha} & 0 & 0 & 0 \\
	0 & \bm{Q}^{\beta\beta\beta\beta} & 0 & 0 \\
	0 & 0 & \bm{Q}^{\alpha\beta\alpha\beta} & 0 \\
	0 & 0 & 0 & \bm{Q}^{\beta\alpha\beta\alpha} \end{bmatrix},	\\
	\label{eq:Gmat}
	\bm{G} & =  \begin{bmatrix} \bm{G}^{\alpha\alpha\alpha\alpha} & \bm{G}^{\alpha\alpha\beta\beta} & 0 & 0 \\
	\bm{G}^{\beta\beta\alpha\alpha} & \bm{G}^{\beta\beta\beta\beta} & 0 & 0 \\
	0 & 0 & \bm{G}^{\alpha\beta\alpha\beta} & 0 \\
	0 & 0 & 0 & \bm{G}^{\beta\alpha\beta\alpha} \end{bmatrix}.
\end{align}
\end{subequations}
The sum of the negative eigenvalues of these matrices can be used to quantify the extent to which they are not positive semi-definite and subsequently violate the $PQG$-conditions.\cite{RodriguezMayorga2017} These sums are denoted by
\begin{equation} \label{eq:Xneg}
	X_\text{neg} = \sum_{\substack{i = 1\\ x_i < 0}}^{K^2} x_i,
\end{equation}
where $K$ is the size of the orbital basis, $\bm{X} = \bm{P}, \bm{Q}$ or $\bm{G}$ and $x_i$ is an eigenvalue of $\bm{X}$.

%-----------------------------------
\section{Method}
%-----------------------------------
\label{sec:method}

%-----------------------------------
\subsection{Reference energies and properties}
%-----------------------------------

The energies from single-reference methods: completely renormalized coupled-cluster (CR-CCL)\cite{Piecuch2002, Piecuch2005} and spin-unrestricted density functional approximation (B3LYP),\cite{Becke1993, Stephens1994, Hertwig1997} and multireference energies: multireference self-consistent field (MRSCF)\cite{Chaban1997} and multireference M{\"o}ller-Plesset second-order perturbation theory (MRMP2),\cite{Hirao1992a, Hirao1992b, Hirao1992c, Hirao1993} were calculated using GAMESS.\cite{GAMESS}  The active space for the MRSCF and MRMP2 calculations was chosen to be all occupied orbitals of the reference configuration, and an equivalent number of the lowest lying vacant (virtual) orbitals. Full configuration interaction (FCI) energies, and 2-RDMs, were determined using Quantum Package 2.\cite{qp2}  All calculations were performed using the cc-pVTZ basis set.\cite{BSE2019, Dunning1989}

%-----------------------------------
\subsection{$\dno$ energies and properties}
%-----------------------------------

The $\dno$, $\dnoof$, and $\dnocs$ calculations were performed with MUNgauss.\cite{mungauss} The active space for each electron pair consisted of the orbital they occupy in the reference configuration and one vacant (virtual) orbital.  The $\dno$ 2-RDMs were optimized using the the trust-region method outlined in Subsection \ref{ssec:opt}.  The iterative subproblem was solved using the dgqt routine from MINPACK-2,\cite{More1983} translated to Fortran 90. Restricted Hartree-Fock orbitals are used as the intial guess, and the intial trust-region radius was $t = 0.05$ with a maximum radius value of 2.00. The gradient threshold, required for saddle-point escape, was $g_0 = 10^{-4}$.  The convergence criteria was $\bar{g} < 10^{-6}$ where $\bar{g} = \frac{|\bg|}{n}$ is the normalized total gradient length, where $n$ is the total number of orbital optimization parameters, $\{ y_{pq} \}$ and $\{ \Delta_{ia} \}$.

The OF and CS energies are calculated post-$\dno$ using an SG-1 grid\cite{Gill1993} and Becke atomic weights.\cite{Becke1988}

The value of the double-counting correction parameter, $c_\text{DC}$, was chosen to minimize the error in the $\dnoof$, and $\dnocs$, \ce{H2} potential energy curves, relative to FCI.  The values were 0.40 and 0.35 for $\dnoof$ and $\dnocs$, respectively.

%-----------------------------------
\section{Results}
%-----------------------------------
\label{sec:results}

%--------------------------
\subsection{Cluster geometries}
%--------------------------

\begin{figure}
	\begin{center}
	\includegraphics[width=0.42\textwidth]{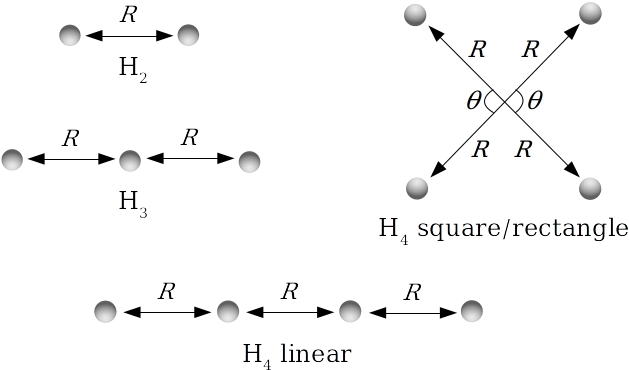}
	\caption{\label{fig:clusters} Geometries of \ce{H} clusters}
	\end{center}
\end{figure}

The geometries of the \ce{H} clusters studied are presented in Figure \ref{fig:clusters}.
The potential energy surfaces were calculated with established single-reference (FCI, CR-CCL and B3LYP) and multireference methods (MRSCF and MRMP2) as well as with $\dno$, $\dnoof$ and $\dnocs$.  The potential energy surface is defined as,
\begin{equation}
	U(\bX) = E(\bX) + V_{nn}(\bX)
\end{equation}
where $E(\bX)$ is the electronic energy provided by the electronic structure method, $V_{nn}(\bX)$, is the nuclear-repulsion energy, and both are a function of the nuclear coordinates, $\bX$.

%--------------------------
\subsection{2-RDM optimization}
%--------------------------

The most time consuming part of the $\dno$ 2-RDM optimization is the transformation of the two-electron integrals from the atomic orbital basis to the $\dno$ orbital basis, which scales as $N_\text{act} K^4$, where $N_\text{act}$ is the number of active orbitals and $K$ is the number of atomic basis functions.  Recently, this cost has been reduced for similar calculations in natural orbital functional methods by introducing the resolution-of-the-identity (DoNOF-RI).\cite{LewYee2021}
\begin{table}
	\caption{\label{tab:2RDMopt} Comparison of the number of integral transformations required to optimize the $\dno$ 2-RDM}
	\begin{ruledtabular}
	\begin{tabular}{lccc}
			& 	& \multicolumn{2}{c}{integral transformations}	\\
				\cline{3-4}
	cluster	& $R$ (\AA) 	& 	ID\footnote{ID = iterative diagonalization of pseudo-Fock matrix}	&	 TR\footnote{TR = trust-region method}	\\
	\hline
	\ce{H2}	&	1.2	&	170	&	10	\\
	\ce{H3}	&	1.2	&	924	&	13	\\
	\ce{H4} linear	&	1.2	&	252	&	47	\\
	\ce{H4} square	&	1.2	&	320	&	20	\\
	\end{tabular}
	\end{ruledtabular}
\end{table}

For the systems studied, the use of the trust-region method, on average, reduced the number of optimization steps (and therefore the number of integral transformations) by an order of magnitude compared to the previous iterative diagonalization algorithm\cite{Piris2009,DNO1,DNO2} (Table \ref{tab:2RDMopt}).  Furthermore, the introduction of an analytical hessian, and a saddle-point escape procedure, enabled the trust-region method to find global minima that the previous iterative-diagonalization algorithm could not locate.

In some cases, in order to determine whether optimized 2-RDMs were indeed global minima, multiple different initial guesses were used.  This includes using orbitals different from RHF, including MRSCF orbitals and FCI natural orbitals, choosing different vacant orbitals for the active space, and also using 2-RDMs (both the orbitals and the electron-transfer variables) from other cluster geometries.  The minima found by the 2-RDM optimization algorithm were invariant with respect to the choice of initial orbitals or which vacant orbitals to include in the active space.  Local minima were only located when an alternative 2-RDM guess was supplied from another cluster geometry ({\it e.g.}~2-RDM from \ce{H4} at $\theta=90^\circ$ for \ce{H4} at $\theta=85^\circ$).

%--------------------------
\subsection{\ce{H2}}
%--------------------------

\begin{figure}
	\begin{center}
	\includegraphics[width=0.48\textwidth]{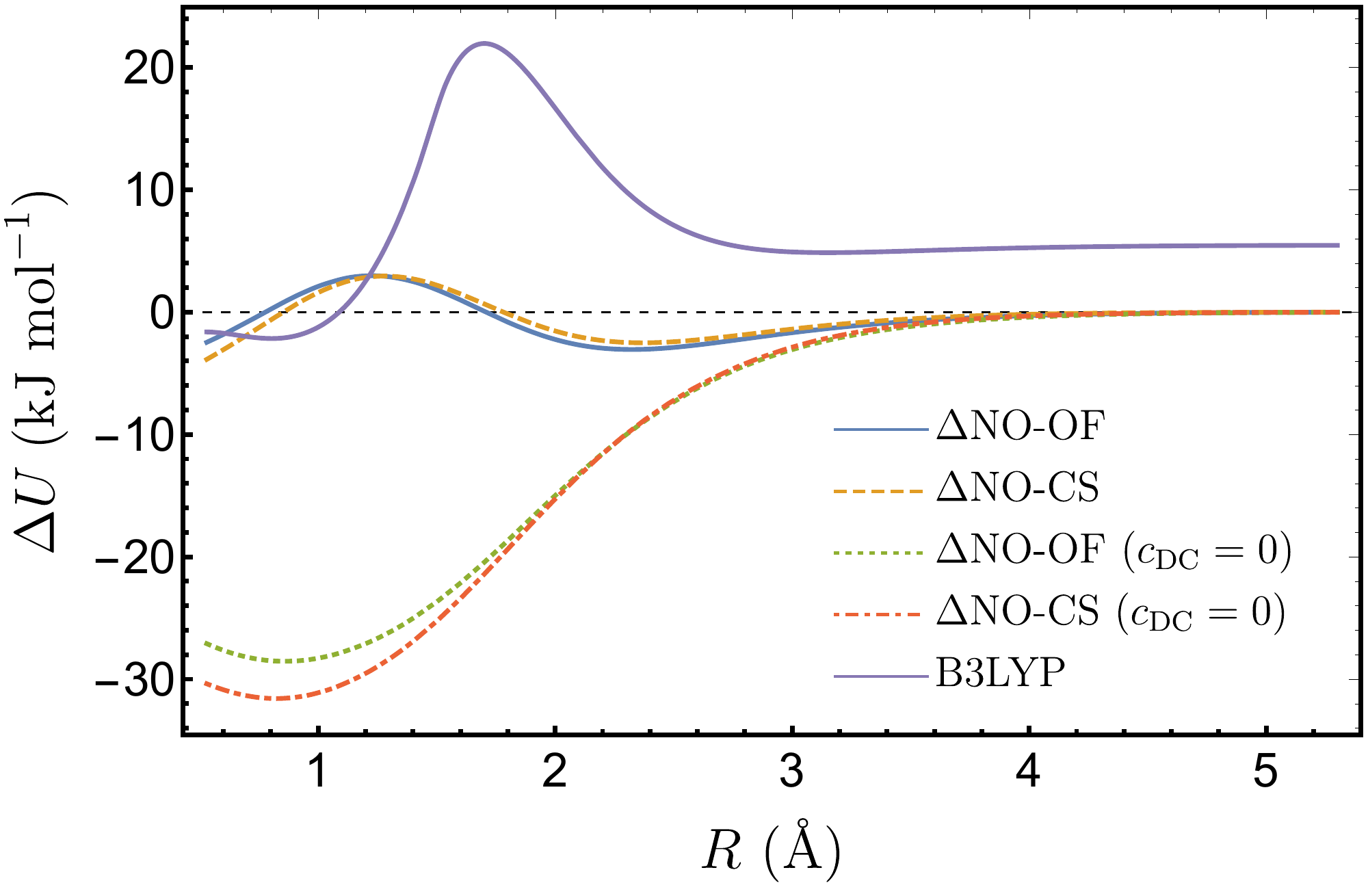}
	\caption{\label{fig:H2error} Error in \ce{H2} potential energy curves compared to FCI ($\Delta U = U - U_\text{FCI}$).}
	\end{center}
\end{figure}

The error in the \ce{H2} potential energy curves relative to the FCI result is presented in Figure \ref{fig:H2error}.
To illustrate the importance of the double-counting correction, the $\dnoof$ and $\dnocs$ result without including the double-counting correction ($c_\text{DC} = 0$) is shown.  Without the double-counting correction, both $\dnoof$ and $\dnocs$ predict $\sim$ 30 kJ mol$^{-1}$ too much correlation energy near the equilibirum bond length ($R = 0.743$ \AA).  However, as the amount a static correlation grows with increasing $R$, the error approaches zero.  The static correlation included in the $\dno$ 2-RDM causes the on-top density to vanish, and therefore there is no superfluous correlation energy at dissociation.  If the double-counting correction is included, the error in $\dnoof$ and $\dnocs$ is dramatically reduced with a maximum error of $\sim$ 3 kJ mol$^{-1}$ (a positive deviation near equilibrium, and negative at stretched bond lengths).  The error in the B3LYP potential energy curve is also included, where the most notable deviation is the energy at the dissociation limit.  The dissociation limit predicted by B3LYP is 5.6 kJ mol$^{-1}$ higher than FCI. 

The equilibrium bond lengths and dissociation energies predicted by each method for the various \ce{H} clusters are presented in Tables \ref{tab:Re} and \ref{tab:DE}.  In the case of \ce{H2}, $\dnoof$ and $\dnocs$ predict equilibrium bond lengths within 0.004 {\AA} of the FCI value, 0.0743 \AA, while the other methods are slightly more accurate.  Also for \ce{H2}, $\dno$ and MRSCF are identical and therefore predict that same bond length (0.756 \AA) and dissociation energy (399 kJ mol$^{-1}$). 
\begin{table}
	\caption{\label{tab:Re} Equilibrium bond lengths (in \AA) of \ce{H} clusters}
	\begin{ruledtabular}
	\begin{tabular}{lcccc}
			& \multicolumn{4}{c}{cluster}	\\
			\cline{2-5}
	method		& \ce{H2}	& \ce{H3}	&\ce{H4} linear&\ce{H4} square \\
	\hline
	$\dno$		&	0.756	&	0.942	&	0.908	&	0.910	\\
	$\dnoof$	&	0.739	&	0.924	&	0.882	&	0.863	\\
	$\dnocs$	&	0.739	&	0.918	&	0.879	&	0.857	\\
	B3LYP		&	0.744	&	0.931	&	0.885	&	0.841	\\
	CR-CCL		&	0.743	&	0.930	&	0.887	&	0.878	\\
	MRSCF		&	0.756	&	0.957	&	0.906	&	0.912	\\
	MRMP2		&	0.745	&	0.934	&	0.889	&	0.868 	\\
	FCI		&	0.743	&	0.930	&	0.887	&	0.864		
	\end{tabular}
	\end{ruledtabular}
\end{table}

Other than $\dno$ and MRSCF, all methods predict fairly accurate bond dissociation energies for \ce{H2}. The high accuracy of $\dnoof$ (0 kJ mol$^{-1}$ error) and $\dnocs$ (1 kJ mol$^{-1}$ error) is due to the calibration of the double-counting correction coefficient, $c_\text{DC}$.  The strategy is to use the bond of \ce{H2} as a prototype for the static and dynamic correlation energy balance between paired electrons.
\begin{table}
	\caption{\label{tab:DE} Dissociation energies (in kJ mol$^{-1}$) of \ce{H} clusters}
	\begin{ruledtabular}
	\begin{tabular}{lcccc}
			& \multicolumn{4}{c}{cluster}	\\
			\cline{2-5}
	method		& \ce{H2}	& \ce{H3}	&\ce{H4} linear&\ce{H4} square \\
	\hline
	$\dno$		&	399	&	317	&	588	&	168	\\
	$\dnoof$	&	454	&	411	&	722	&	288	\\
	$\dnocs$	&	455	&	413	&	718	&	288	\\
	B3LYP		&	461	&	444	&	765	&	349	\\
	CR-CCL		&	454	&	412	&	726	&	292	\\
	MRSCF		&	399	&	327	&	610	&	202	\\
	MRMP2		&	444	&	400	&	713	&	278 	\\
	FCI		&	454	&	412	&	733	&	288		
	\end{tabular}
	\end{ruledtabular}
\end{table}

%--------------------------
\subsection{\ce{H3}}
\begin{figure}
	\begin{center}
	\includegraphics[width=0.48\textwidth]{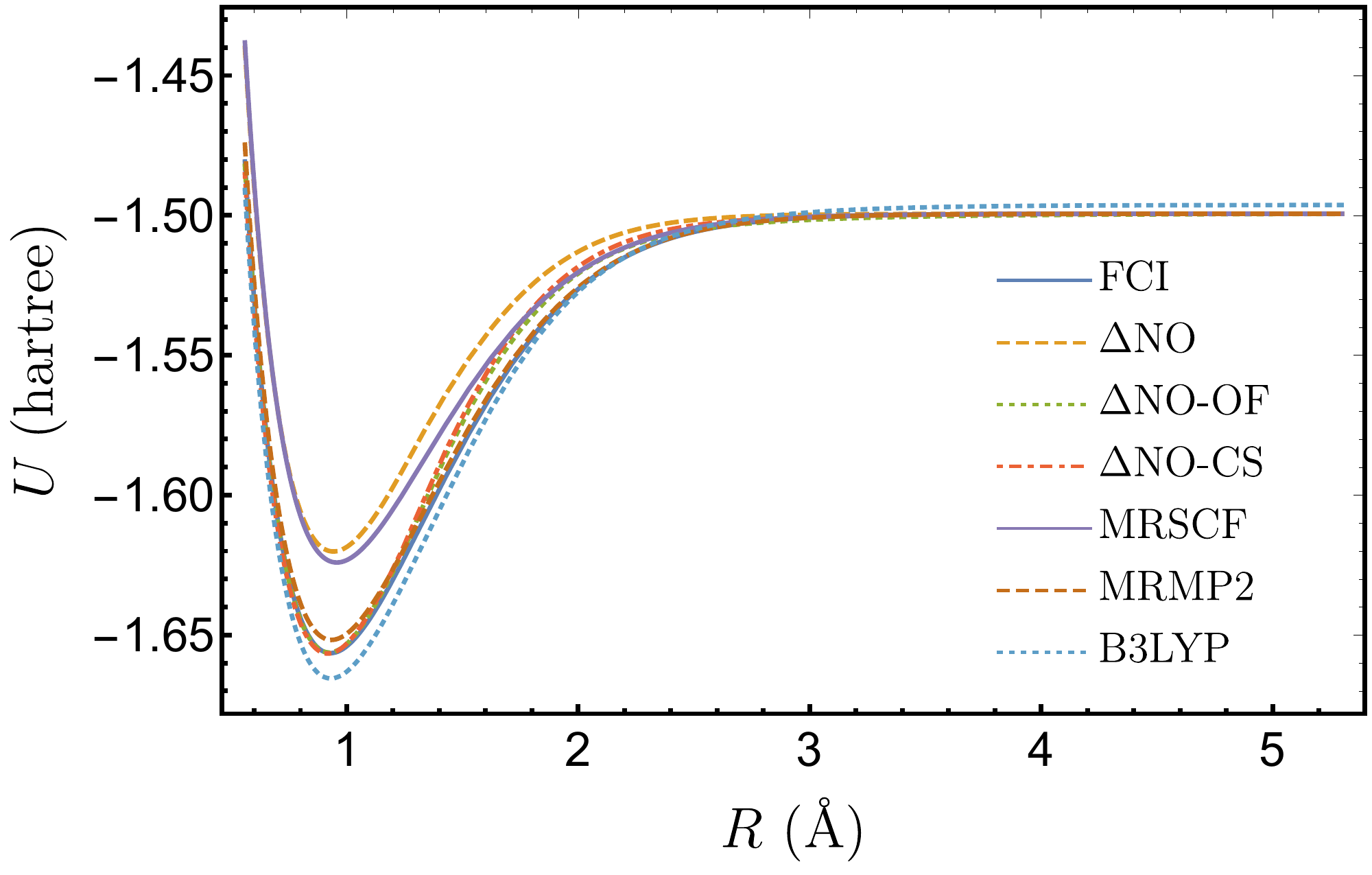}
	\caption{\label{fig:H3} Potential energy of a linear \ce{H3} cluster with neighbouring \ce{H} atoms separated by a distance $R$.}
	\end{center}
\end{figure}

%--------------------------
The potential energy curves (PECs) for \ce{H3}, in a linear arrangement, are presented in Figure \ref{fig:H3}.
The $\dno$ and MRSCF PECs are above the PECs of the other methods designed to also capture dynamic correlation.  Unlike \ce{H2}, the $\dno$ and MRSCF PECs differ and it is apparent that, even with the high-spin correction applied to the unpaired electron (Equation \ref{eq:ehsc}), $\dno$ is missing correlation at moderately stretched bond lengths.  Nevertheless, the correct dissociation limit is reached by $\dno$, $\dnoof$ and $\dnocs$.

\begin{figure}
	\begin{center}
	\includegraphics[width=0.48\textwidth]{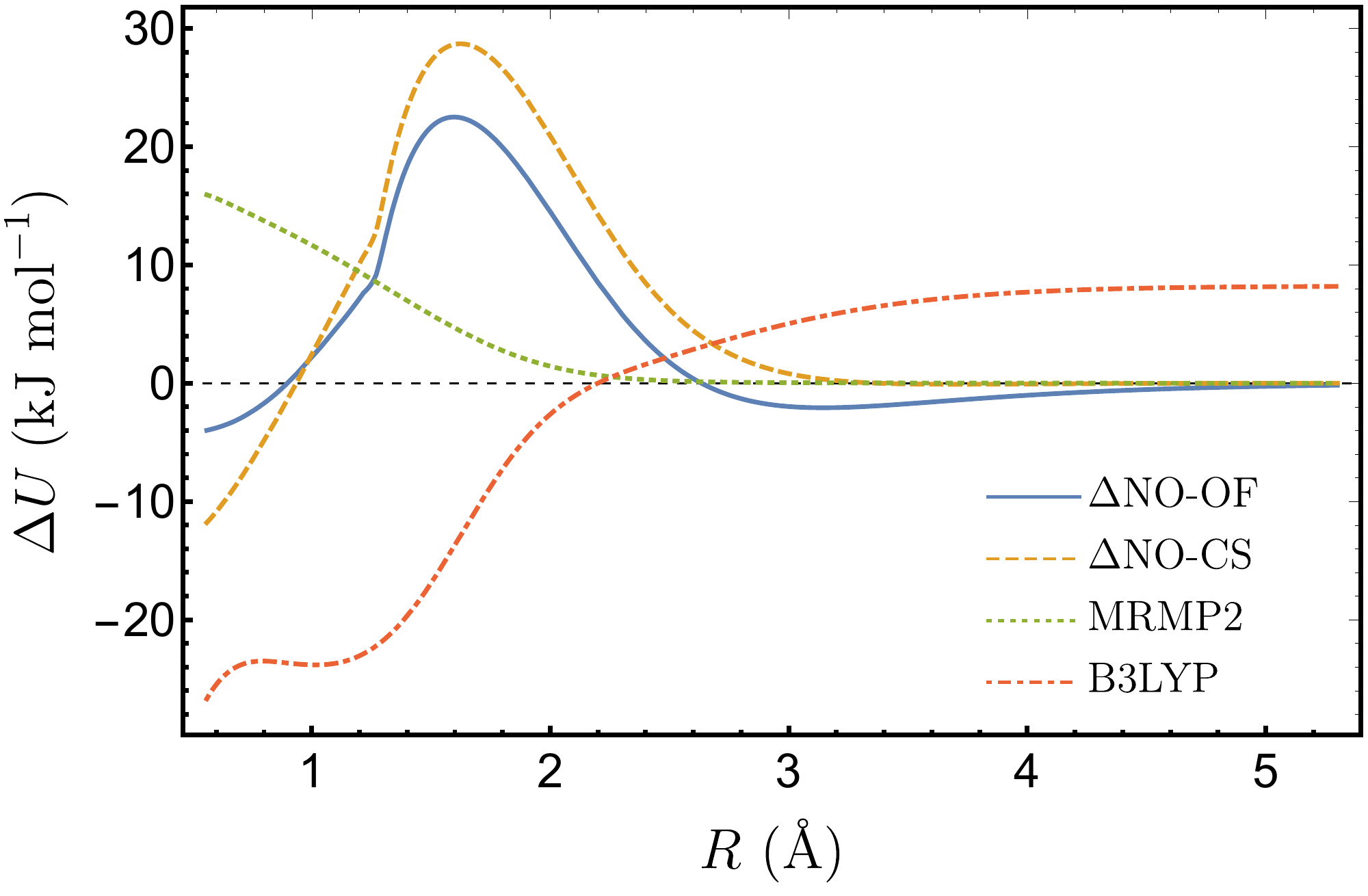}
	\caption{\label{fig:H3error} Error in \ce{H3} potential energy curves compared to FCI ($\Delta U = U - U_\text{FCI}$).}
	\end{center}
\end{figure}

The error in the PECs, for the methods that include both static and dynamic correlation ($\dnoof$, $\dnocs$, MRMP2 and B3LYP), compared to the FCI PEC is shown in Figure \ref{fig:H3error}.
At small $R$, $\dnoof$ and $\dnocs$ overestimate the correlation energy by a maximum of 4 and 11 kJ mol$^{-1}$, respectively.  At moderately stretched bond lengths, both $\dnoof$ and $\dnocs$ underestimate the correlation energy (23 and 29 kJ mol$^{-1}$, respectively), which is due to the missing static correlation in $\dno$ between the electron pair and the unpaired electron.

The largest deviation of the MRMP2 PEC from the FCI PEC is at small $R$ (16 kJ mol$^{-1}$), and the error decreases monotonically as $R$ increases.  In the case of B3LYP, the correlation energy is significantly overestimated around equilibrium $R$ (24 kJ mol$^{-1}$) and like \ce{H2}, the energy at the dissociation limit is incorrect.  These errors are reflected in the predicted equilibrium $R$ values and dissociation energies (Tables \ref{tab:Re} and \ref{tab:DE}).

Similar to \ce{H2}, both $\dno$ and MRSCF overestimate the equilibrium separation.  Whereas, $\dnoof$ and $\dnocs$ slightly underestimate $R_e$, by 0.006 {\AA} and 0.012 \AA, respectively.  The values of $R_e$ from both MRMP2 (0.934 \AA) and B3LYP (0.931 \AA) differ only slightly from the FCI value.  Also, like \ce{H2}, the CR-CCL PEC is equivalent to FCI.

For those methods designed to capture all the correlation energy, it is only the dissociation energy predicted by B3LYP (444 kJ mol$^{-1}$) that differs substantially from the FCI value (412 kJ mol$^{-1}$).

%--------------------------
\subsection{Linear \ce{H4}}
\begin{figure}
	\begin{center}
	\includegraphics[width=0.48\textwidth]{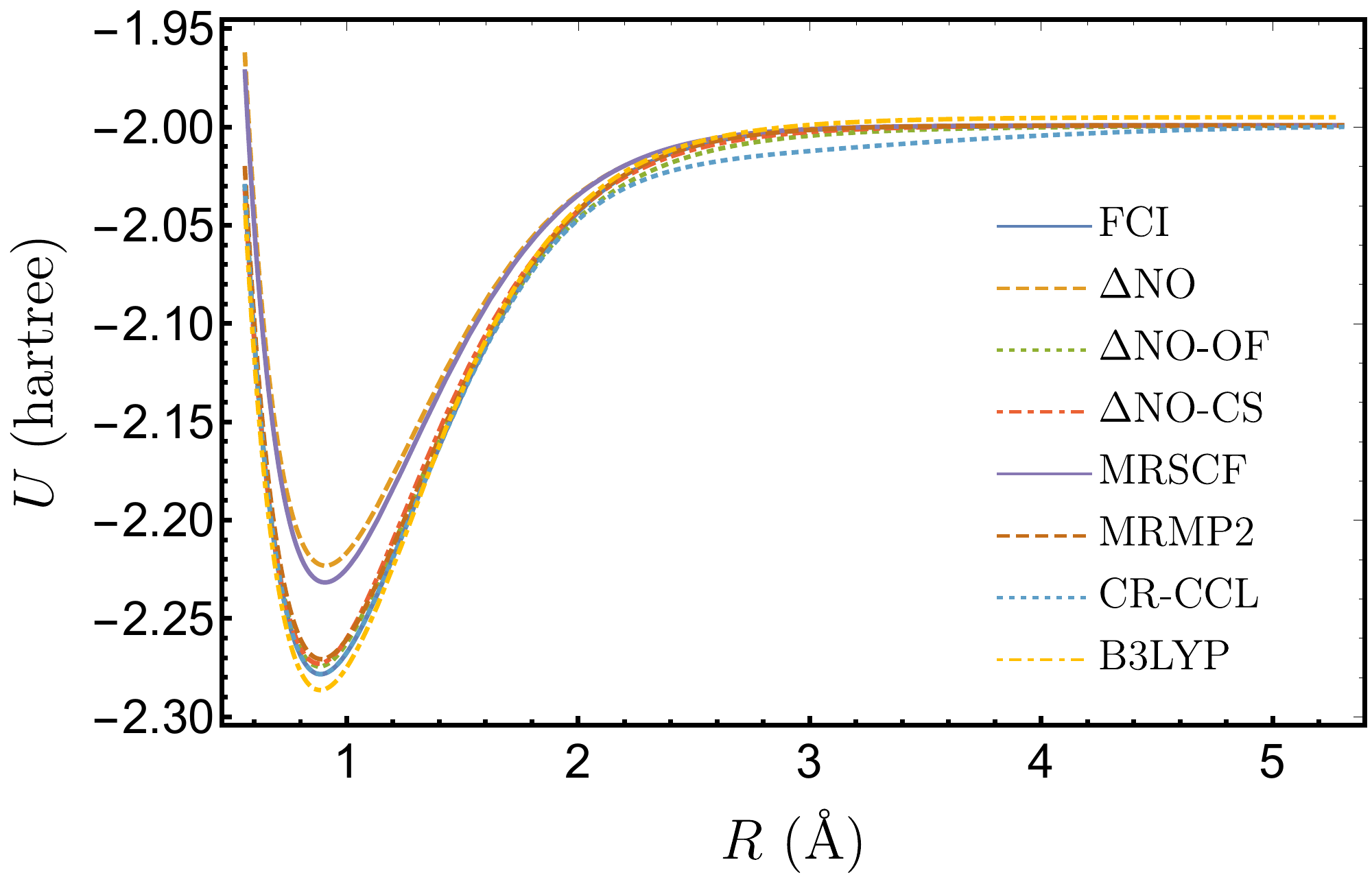}
	\caption{\label{fig:H4lin} Potential energy of a linear \ce{H4} cluster with neighbouring \ce{H} atoms separated by a distance $R$.}
	\end{center}
\end{figure}

%--------------------------
The PECs for the linear \ce{H4} cluster, with interatomic spacing $R$, are presented in Figure \ref{fig:H4lin}. 
Similar to \ce{H3}, the $\dno$ and MRSCF PECs appear above the others (as expected), and the B3LYP PEC lies below the others at equilibrium and above at dissociation.  Interestingly, while the CR-CCL PEC is essentially exact close to $R_e$, and eventually approaches the correct dissociation limit, there is a significant negative deviation from the FCI PEC at large $R$.

\begin{figure}
	\begin{center}
	\includegraphics[width=0.48\textwidth]{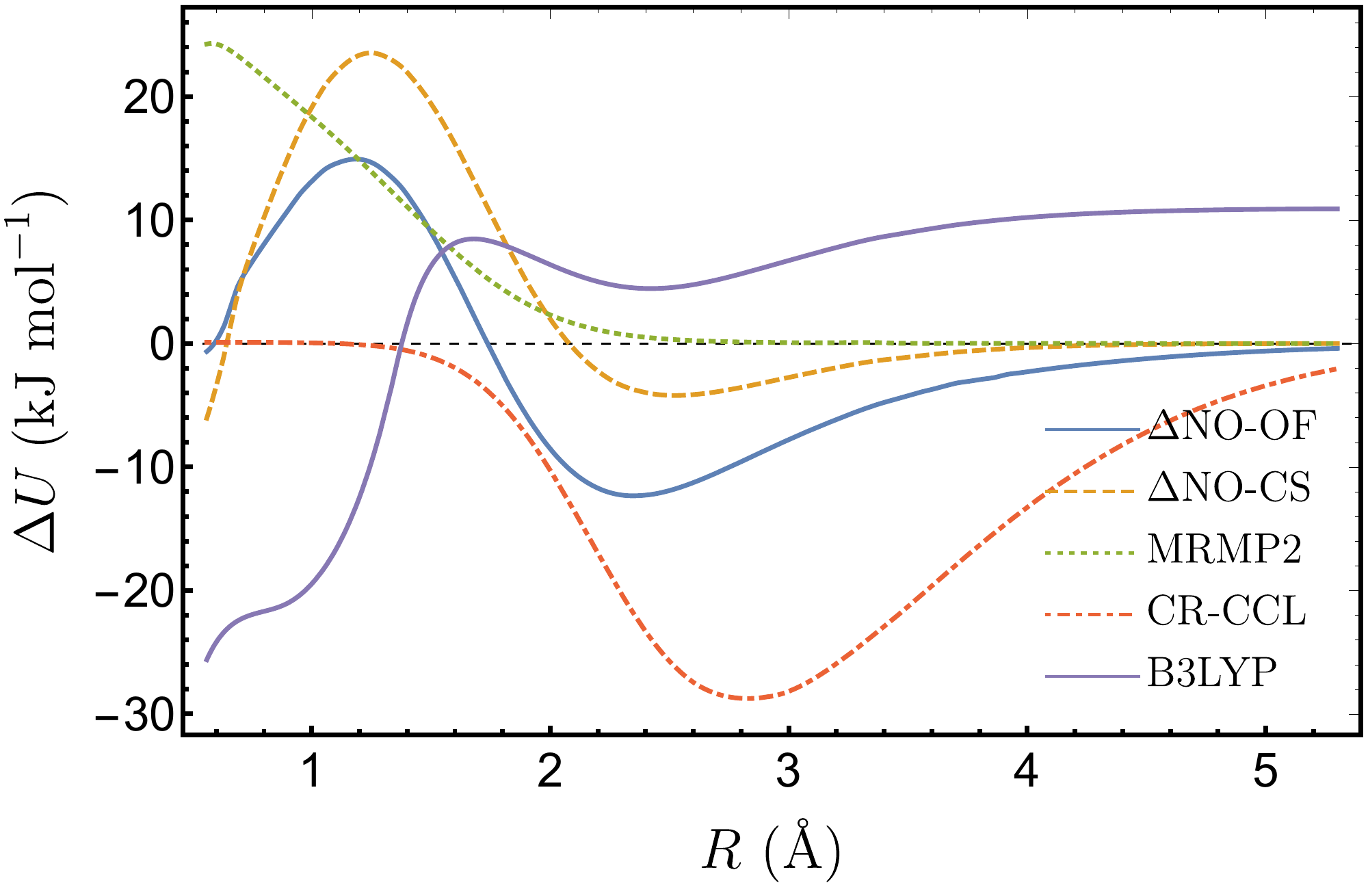}
	\caption{\label{fig:H4linerror} Error in linear \ce{H4} cluster potential energy curves compared to FCI ($\Delta U = U - U_\text{FCI}$).}
	\end{center}
\end{figure}

The error in the linear \ce{H4} PECs, compared to FCI, are explicitly shown in Figure \ref{fig:H4linerror}.
The behaviour of the error in the $\dnoof$ and $\dnocs$ PECs for linear \ce{H4} is the same as that seen for \ce{H2}, although substantially larger.  Both $\dnoof$ and $\dnocs$ underestimate the correlation energy at small $R$, and overestimate at large $R$, while eventually reaching the correct dissociation limit.  Like \ce{H3}, the error in the MRMP2 PEC is largest at small $R$ and decreases as $R$ is increased.  In the case of CR-CCL, the PEC is essentially exact for small $R$ and then begins overestimating the correlation energy at $\sim$ 1.4 \AA.  The error in the B3LYP PEC is similar that seen for both \ce{H2} and \ce{H3}.

With the exception of $\dno$ and MRSCF, all methods provide good estimates of $R_e$ compared to FCI, with $\dnoof$ and $\dnocs$ showing the most deviation (0.005 and 0.008 \AA, respectively).  Both $\dno$ and MRSCF predict larger $R_e$ values, and are within close agreement with each other (differing by 0.002 \AA).

The dissociation energies predicted by each method are more varied.  The dissociation energies predicted by $\dnoof$ and $\dnocs$, 722 kJ mol$^{-1}$ and 718 kJ mol$^{-1}$, respectively, are in decent agreement with the FCI value, 733 kJ mol$^{-1}$.  Only CR-CCL provides a better estimate, 726 kJ mol$^{-1}$.

%--------------------------
\subsection{Square \ce{H4}}
\begin{figure}
	\begin{center}
	\includegraphics[width=0.48\textwidth]{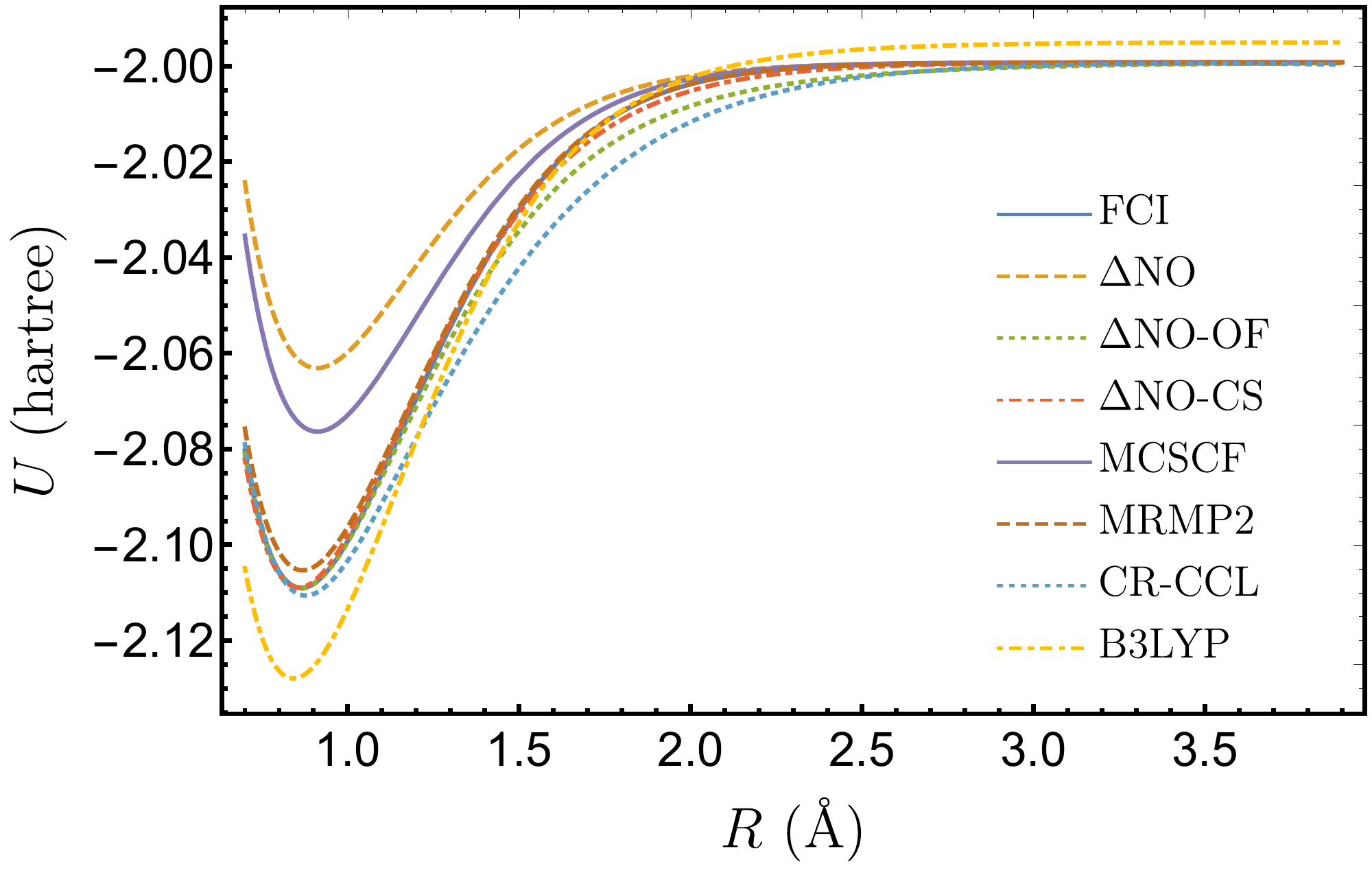}
	\caption{\label{fig:H4sq} Potential energy of a square \ce{H4} cluster where each at is a distance $R$ from the centre (see Figure \ref{fig:clusters}).}
	\end{center}
\end{figure}

\begin{figure}
	\begin{center}
	\includegraphics[width=0.48\textwidth]{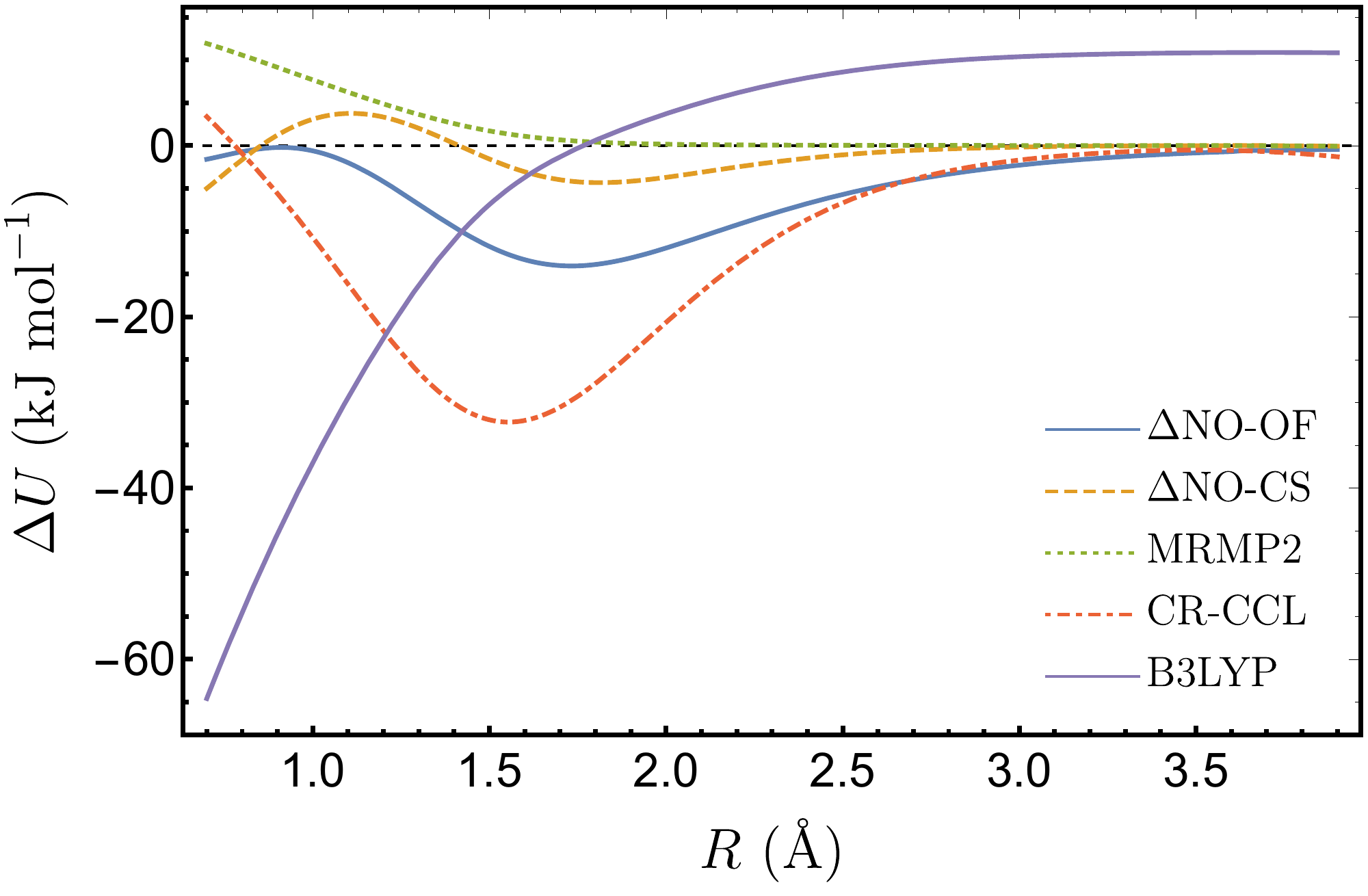}
	\caption{\label{fig:H4sqerror} Error in square \ce{H4} cluster potential energy curves compared to FCI ($\Delta U = U - U_\text{FCI}$).}
	\end{center}
\end{figure}

%--------------------------
The PECs for the dissociation of \ce{H4} in a square arrangement ($\theta = 90^\circ$, see Figure \ref{fig:clusters}) are shown in Figure \ref{fig:H4sq}. 
The square arrangement of \ce{H} atoms results in much more varied performance amongst the methods assessed compared to the linear arrangement.  The B3LYP PEC is far below the others near $R_e$, but above again at the dissociation limit.  There is substantial separation between the $\dno$ and MRSCF PECs near $R_e$.  Furthermore, this system is a known issue for single-reference coupled-cluster methods, and while CR-CCL performs better here than CCSD(T), the error is significant.  The explicit errors are shown in Figure \ref{fig:H4sqerror}.

The error in the $\dnocs$ PEC is relatively small, and provides, arguably, the best agreement with the FCI PEC.  The usual underestimation of the correlation energy at small $R$ by MRMP2 is relatively small for this system, and therefore MRMP2 also provides a good estimate of the FCI PES.  The negative deviation of $\dnoof$ at moderate $R$ values reaches a maximum of $\sim$ 14 kJ mol$^{-1}$ before going to the correct dissociation limit. Whereas, the CR-CCL PES overestimates the correlation energy by more than 30 kJ mol$^{-1}$ before approaching the dissociation limit.

Despite the errors in the PECs, accurate predictions of the FCI $R_e$, 0.864 \AA, are provided by $\dnoof$, $\dnocs$, MRMP2 and CR-CCL, with values of 0.863, 0.857, 0.868 and 0.878, respectively.  As usual, the dissociation energies vary more widely.  Both $\dnoof$ and $\dnocs$ provide very accurate estimates of the dissociation energy, and MRMP2 and CR-CCL differ by only 10 and 4 kJ mol$^{-1}$, respectively.  While B3LYP overestimates the dissociation energy by 61 kJ mol$^{-1}$.   

%--------------------------
\subsection{Rectangle to square \ce{H4}}
%--------------------------
\label{ssec:rec2sq}

\begin{figure*}
	\begin{center}
	\includegraphics[width=0.315\textwidth]{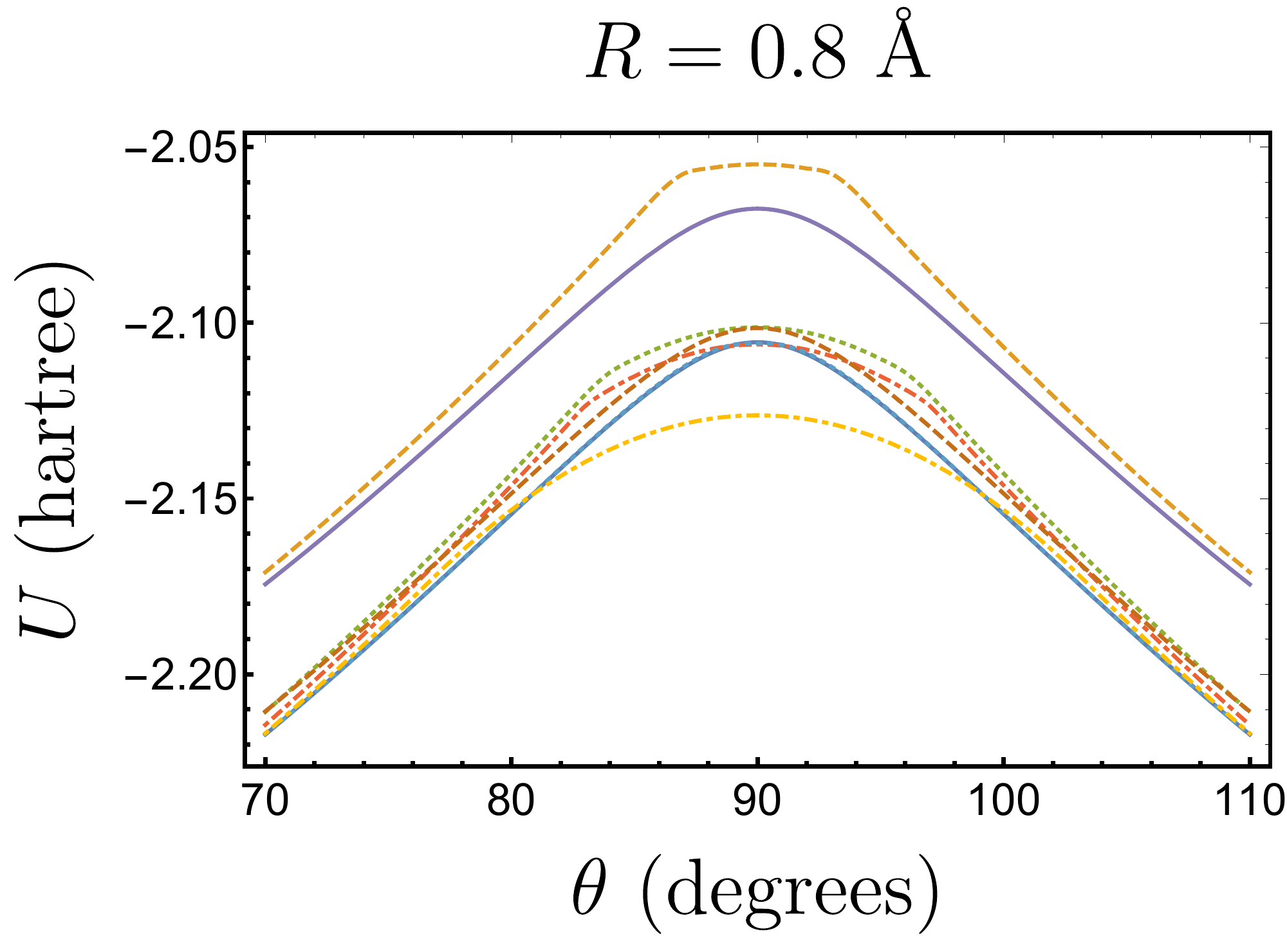}
	\includegraphics[width=0.315\textwidth]{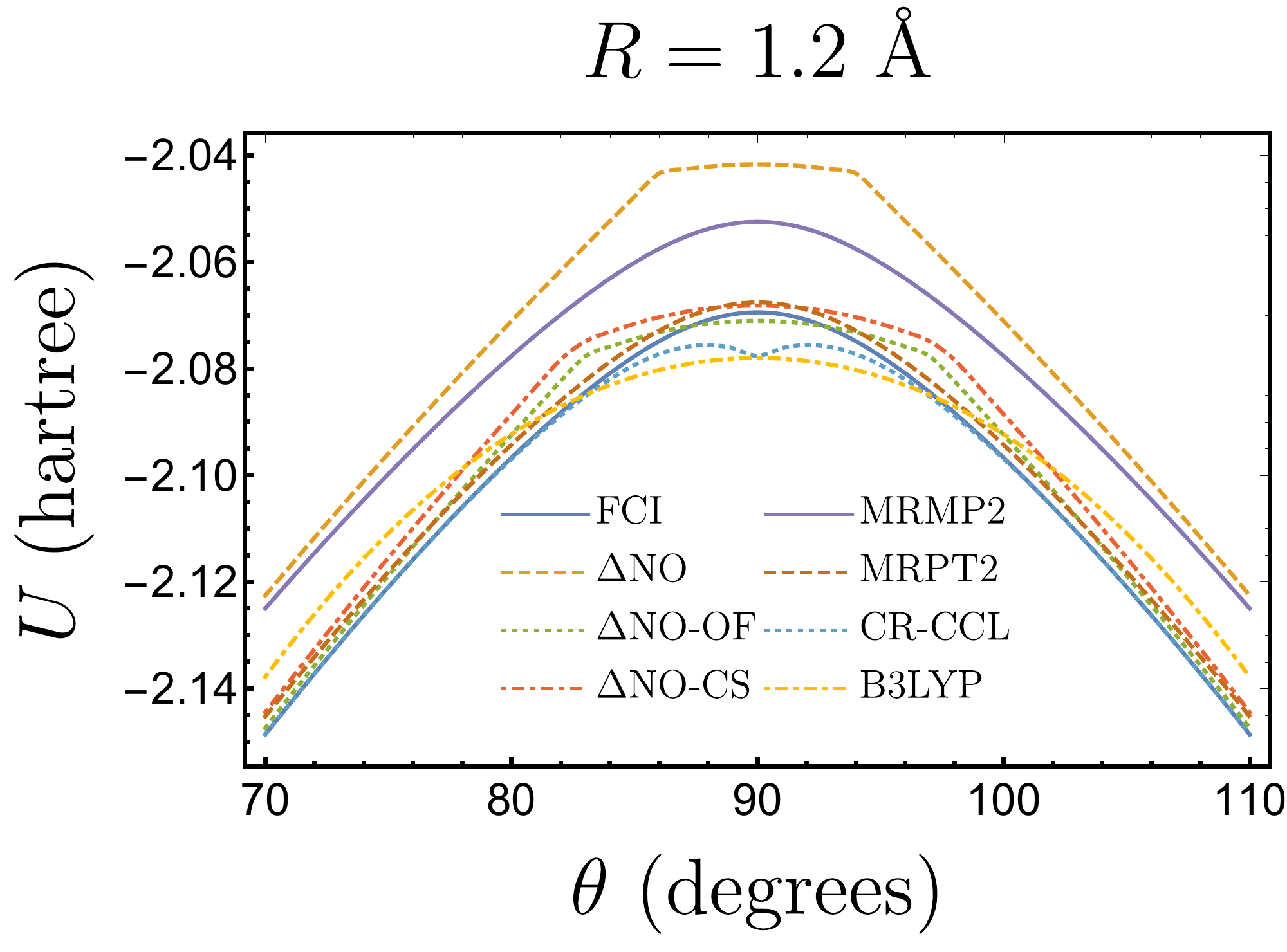}
	\includegraphics[width=0.315\textwidth]{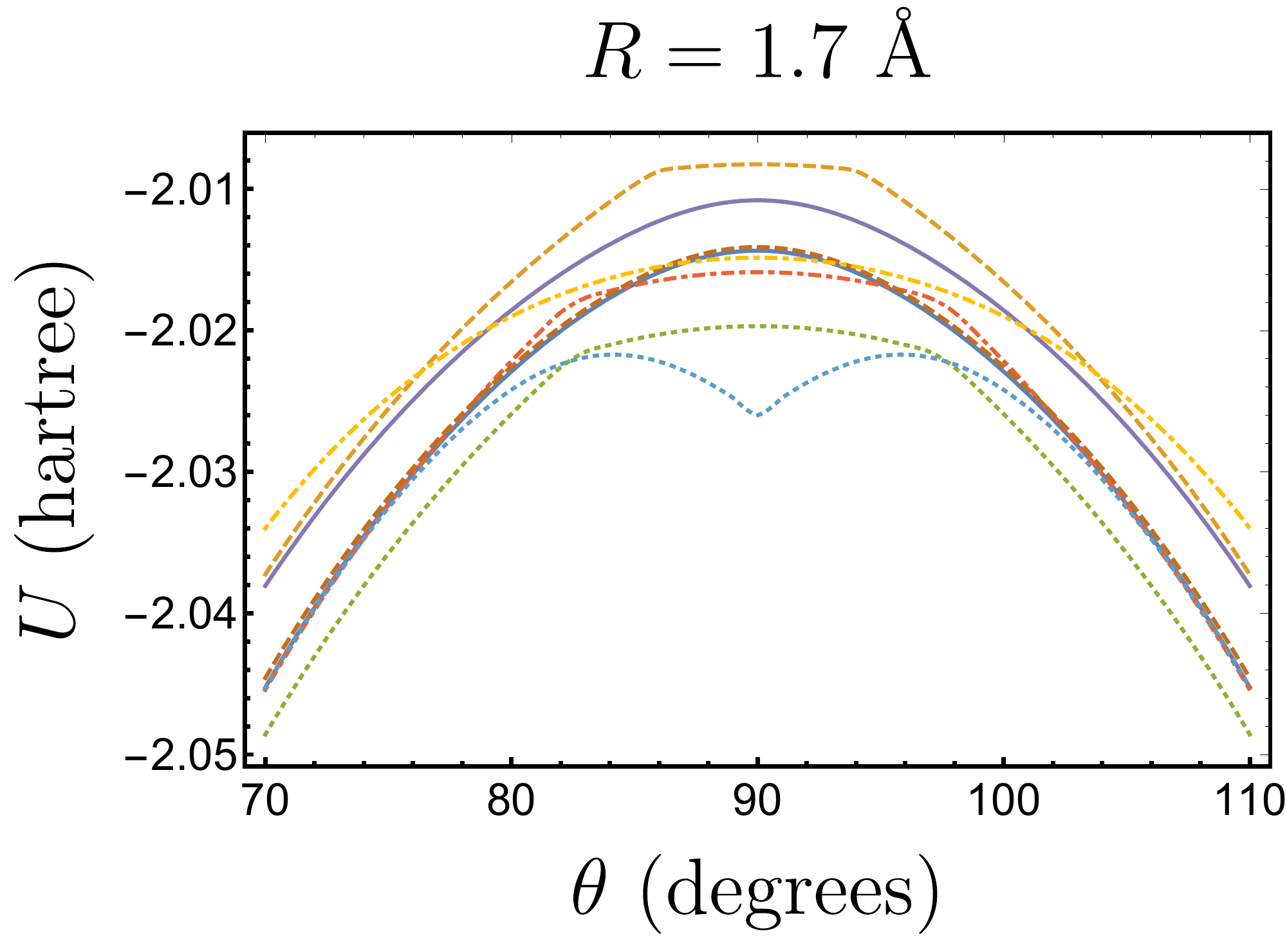}
	\caption{\label{fig:H4theta} Potential energy of \ce{H4} cluster as a function of $\theta$ and $R$  (see Figure \ref{fig:clusters}).}
	\end{center}
\end{figure*}

At fixed $R$, if $\theta$ of the, $D_{4h}$ symmetry, square arrangement of \ce{H4} (Figure \ref{fig:clusters}) is changed from 90$^\circ$, the result is a rectangular, $D_{2h}$ symmetry, arrangement.  The PECs for $\theta=70^\circ$ to $\theta=110^\circ$ ($D_{2h}$ $\to$ $D_{4h}$ $\to$ $D_{2h}$) for fixed $R$ values of 0.8, 1.2 and 1.7 {\AA} are presented in Figure \ref{fig:H4theta}.

Like the PECs for the other \ce{H} clusters, the $\dno$ and MRSCF PECs lie above those of the methods designed to capture all of the correlation energy.  As $R$ increases, and static correlation increases while dynamic correlation decreases, the $\dno$ and MRSCF PECs approach the others (note the scale of the $y$-axis in Figure \ref{fig:H4theta}).

In the case of CR-CCL, coupled-cluster methods are known to have a concave (downward pointing) derivative discontinuity,\cite{Robinson2012c, RamosCordoba2015, Marie2021} or cusp, at $\theta=90$, which becomes more prominent as $R$ increases. This is in contrast to the convex (upward pointing) cusp seen for RHF and spin-restricted density functional methods. The cusp in the CR-CCL PEC is barely noticable at $R=0.8$ \AA, but is evident at the larger $R$ values.  The MRMP2 PEC closely resembles the FCI PEC at all $R$ values, and the absolute error decreases with increasing $R$.  The height of the PEC is significantly underestimated by B3LYP at all $R$ values.

The $\dnoof$ and $\dnocs$ methods accurately model the PEC near $\theta=70^\circ$ and 90$^\circ$, however there is an obvious cusp at $\theta \sim 85^\circ$ and $\theta \sim 95^\circ$.  This is due to an inadequate description of the 2-RDM upon transitioning from a $D_{4h}$ to $D_{2h}$ arrangement, which is analysed, and elaborated upon, later.

The barrier height, defined as the difference in energy at $\theta= 90^\circ$ and 70$^\circ$, predicted by each method at each $R$ value is presented in Table \ref{tab:BH}.
\begin{table}
	\caption{\label{tab:BH} Barrier height\footnote{Barrier height is defined as $E[90^\circ] - E[70^\circ]$} (in kJ mol$^{-1}$) for $D_{2h}\to D_{4h} \to D_{2h}$ transition of \ce{H4} cluster at varying $R$.}
	\begin{ruledtabular}
	\begin{tabular}{lccc}
			& \multicolumn{3}{c}{$R$ (\AA)}	\\
			\cline{2-4}
	method		& 0.8		& 1.2		& 1.7		\\
	\hline
	$\dno$		&	305	&	212	&	76	\\
	$\dnoof$	&	287	&	200	&	76	\\
	$\dnocs$	&	284	&	200	&	78	\\
	B3LYP		&	238	&	157	&	50	\\
	CR-CCL		&	292	&	186	&	51	\\
	MRSCF		&	281	&	190	&	72	\\
	MRMP2		&	286	&	204	&	80	\\
	FCI		&	293	&	208	&	81			
	\end{tabular}
	\end{ruledtabular}
\end{table}
At $R=0.8$ \AA, $\dnoof$ and $\dnocs$ underestimate the barrier by 6 and 9 kJ mol$^{-1}$, respectively.  Compare that to $\dno$, which overestimates the barrier by 12 kJ mol$^{-1}$.  MRMP2 provides accuracy similar to $\dnoof$ and $\dnocs$, while CR-CCL only differs from the FCI barrier by 1 kJ mol$^{-1}$.  As $R$ is increased, the agreement of $\dno$ with FCI improves, which highlights the importance of static correlation to these PECs.  At $R=1.2$, $\dnoof$ and $\dnocs$ both differ from FCI by 8 kJ mol$^{-1}$, and differ by 5 and 3 kJ mol$^{-1}$, respectively, at $R=1.7$ \AA. At the same time, the agreement of MRMP2 with FCI also improves, to within 4 kJ mol$^{-1}$ at $R=1.2$ {\AA} and within 1 kJ mol$^{-1}$ at $R=1.7$ \AA.  In contrast, due to the cusp, the CR-CCL values worsen.

\begin{figure}
	\begin{center}
	\includegraphics[width=0.48\textwidth]{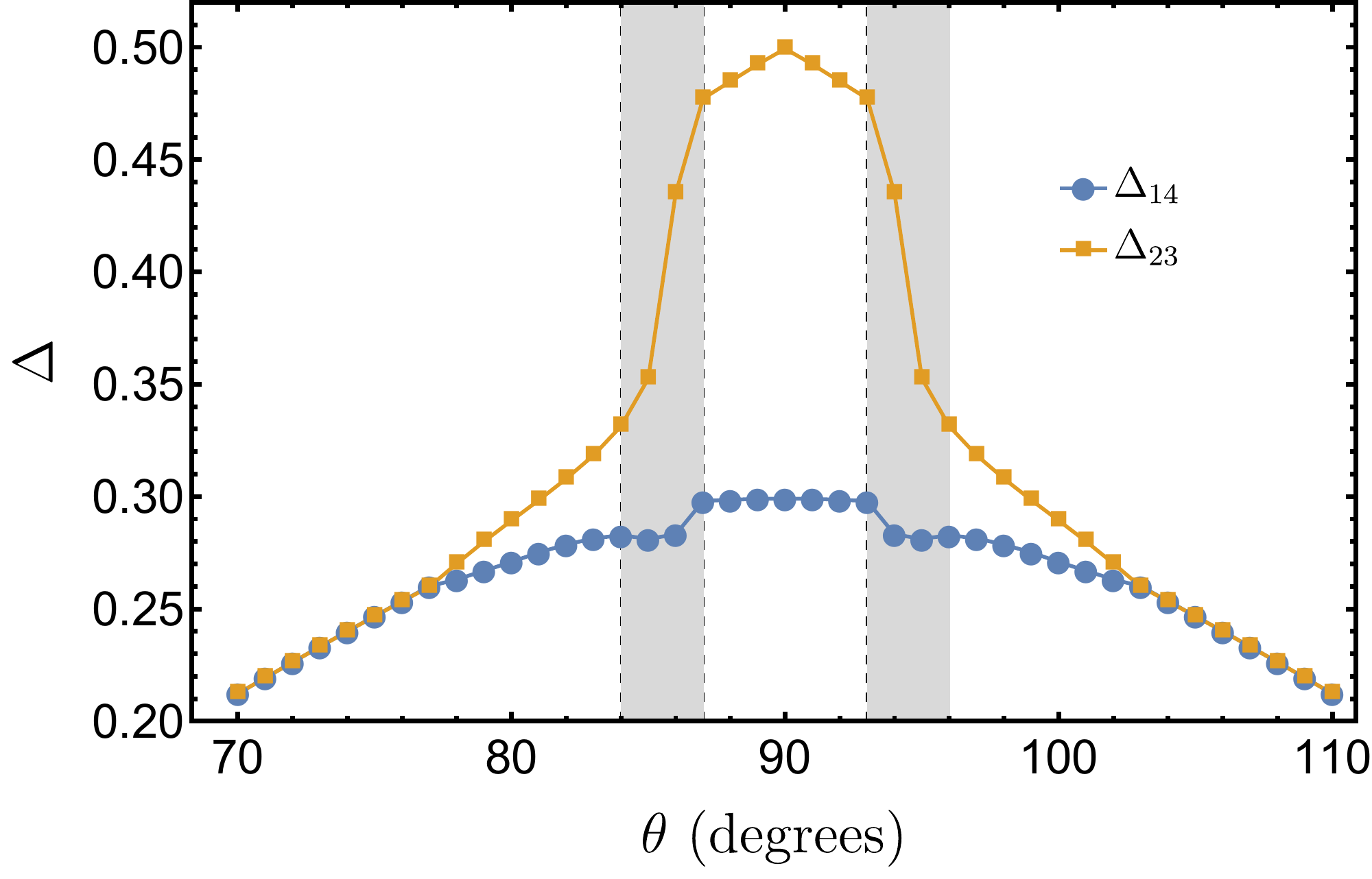}
	\caption{\label{fig:Deltas} $\Delta$ values for rectangle-to-square \ce{H4} cluster as a function of $\theta$ at $R= 1.7$ \AA.}
	\end{center}
\end{figure}

Although the barrier heights predicted by $\dnoof$ and $\dnocs$ are in good agreement with FCI values, the deviation of the PECs from FCI approximately 5$^\circ$ from $\theta = 90^\circ$ is concerning.  An initial analysis of the correlation captured by $\dno$ is provided by the electron transfer variables, $\{ \Delta_{ia} \}$.  The values of $\Delta_{14}$ and $\Delta_{23}$ from $\theta=70^\circ$ to $\theta=110^\circ$, at $R = 1.7$ \AA, are presented in Figure \ref{fig:Deltas}. 
From $\theta=70^\circ$ to $\theta=77^\circ$, $\Delta_{14}$ and $\Delta_{23}$ are equivalent, indicating that the motion of both electron pairs is correlated in the same manner.  This points to the rectangular cluster behaving as two \ce{H2} molecules, which is not surprising at particularly large separations like $R=1.7$ \AA.  Beyond $\theta=77^\circ$, the values of $\Delta_{14}$ and $\Delta_{23}$ diverge, and there is an obvious transition that occurs from $\theta = 84^\circ$ to $\theta = 87^\circ$, and from $\theta = 93^\circ$ to $\theta = 96^\circ$ (grey rectangles).  At the $\theta=90^\circ$ side of the transition, the electron pairs have significantly different amounts of static correlation.  The $\Delta_{23}$ electron pair is very near the strong correlation limit, $\Delta \to 1/2$, while $\Delta_{14} \approx 0.28$.  Essentially, there are two regimes of correlation.  For $\theta$ near 70$^\circ$ (and 110$^\circ$) there is the \ce{2H2} regime, and for $\theta$ near 90$^\circ$ there is the \ce{H4} regime.

\begin{figure}
	\begin{center}
	\begin{tabular}{c|c|c|c|c|}
			\cline{2-5}
			& \multicolumn{4}{c|}{ $\theta$ (degrees) }	\\
			\hline
	\multicolumn{1}{|c|}{orbital}	& 84	& 85	& 86	& 87	\\
			\hline \hline
	\multicolumn{1}{|c|}{\raisebox{0.035\textwidth}{1}}	
			& \includegraphics[width=0.08\textwidth]{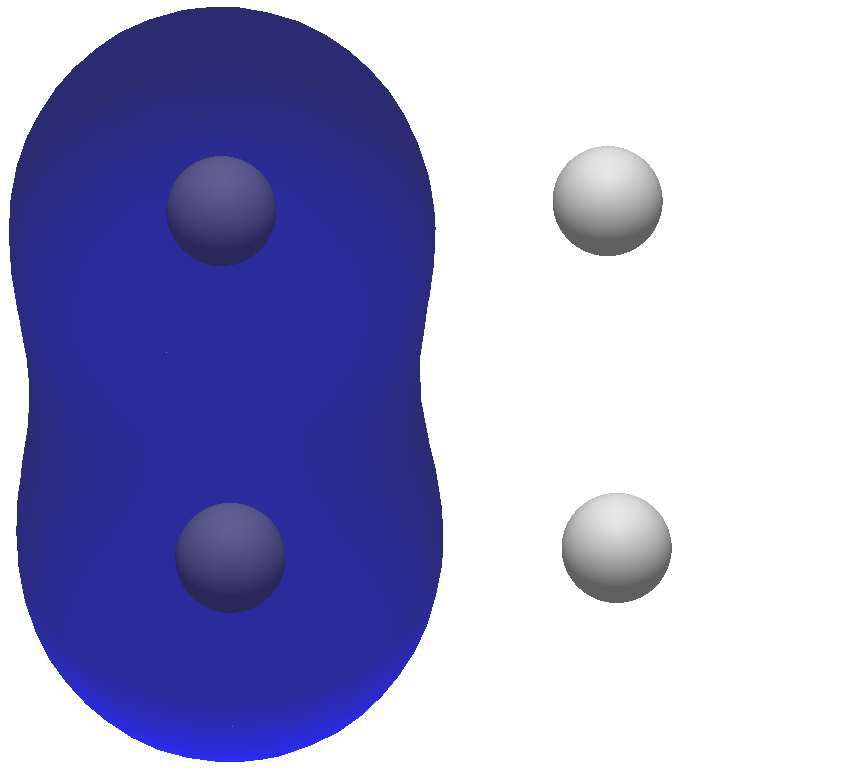}
			& \includegraphics[width=0.08\textwidth]{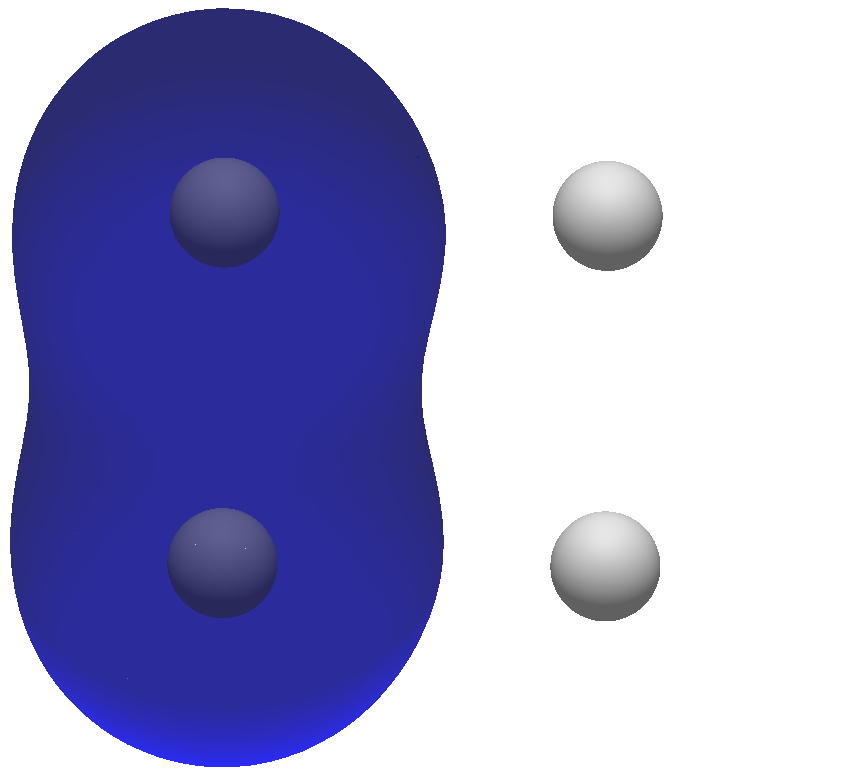}
			& \includegraphics[width=0.08\textwidth]{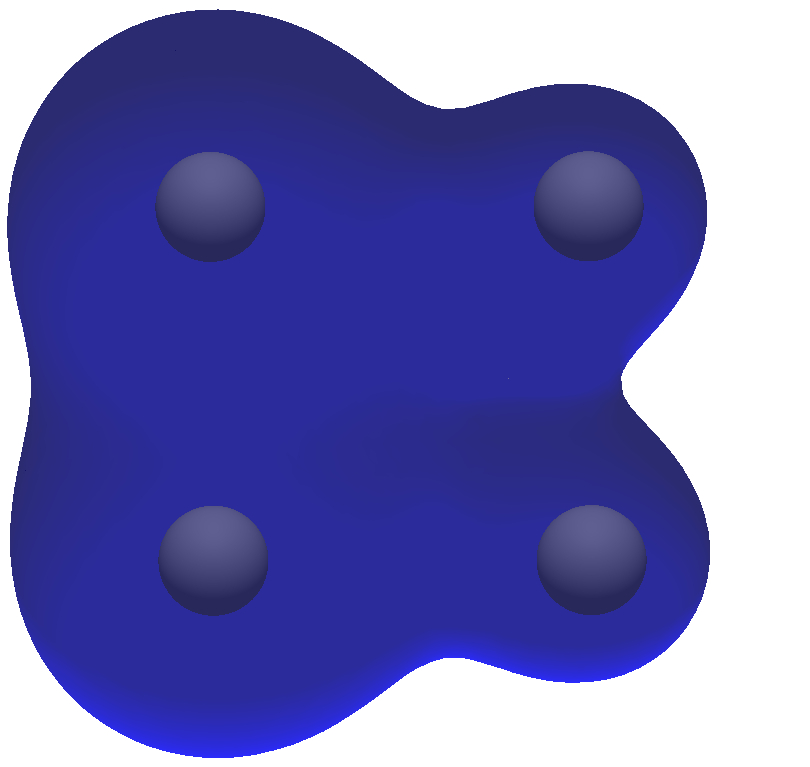}
			& \includegraphics[width=0.08\textwidth]{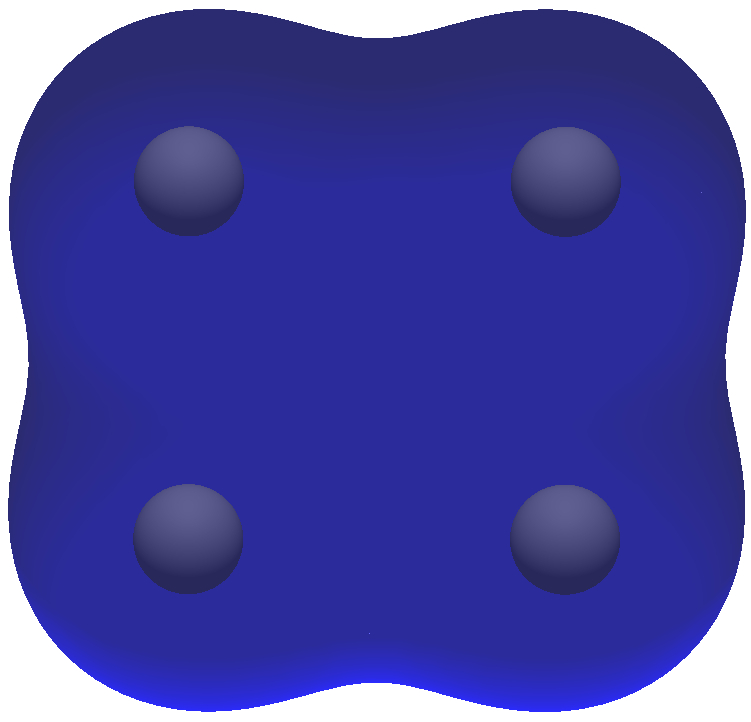} \\ \hline\hline
	\multicolumn{1}{|c|}{\raisebox{0.035\textwidth}{2}}		
			& \includegraphics[width=0.08\textwidth]{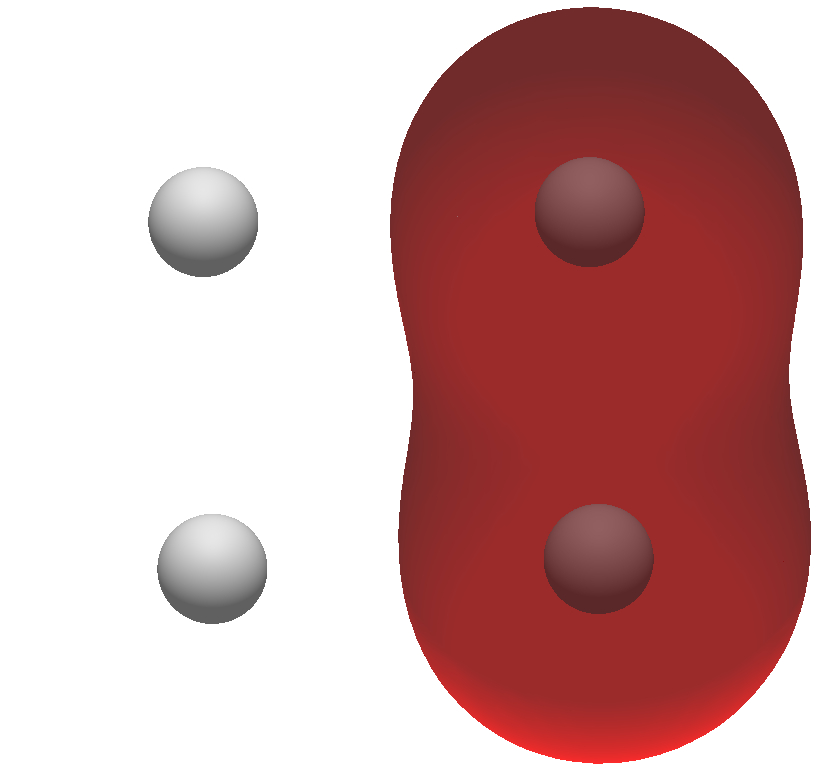}
			& \includegraphics[width=0.08\textwidth]{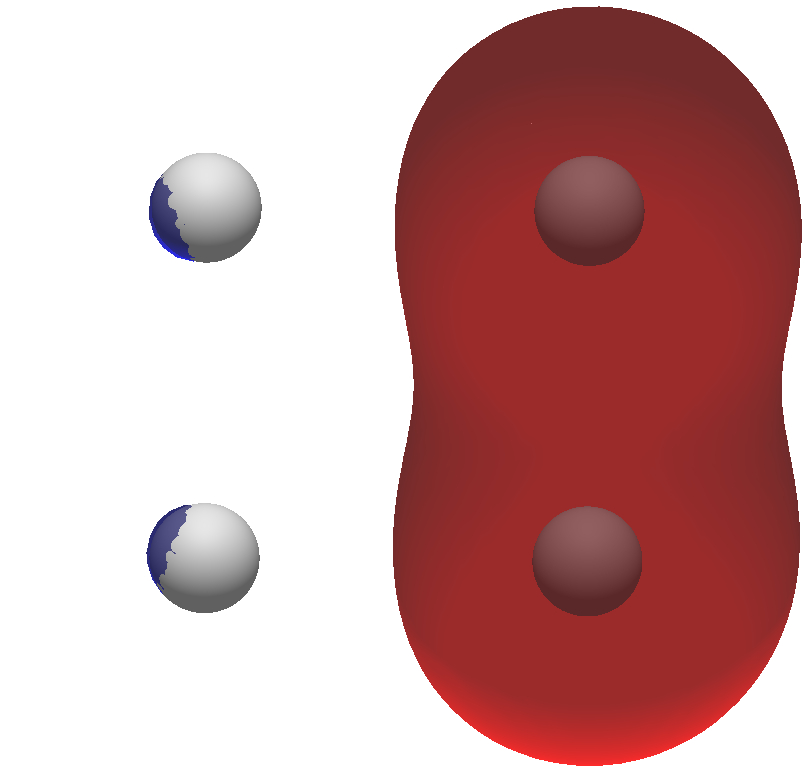}
			& \includegraphics[width=0.08\textwidth]{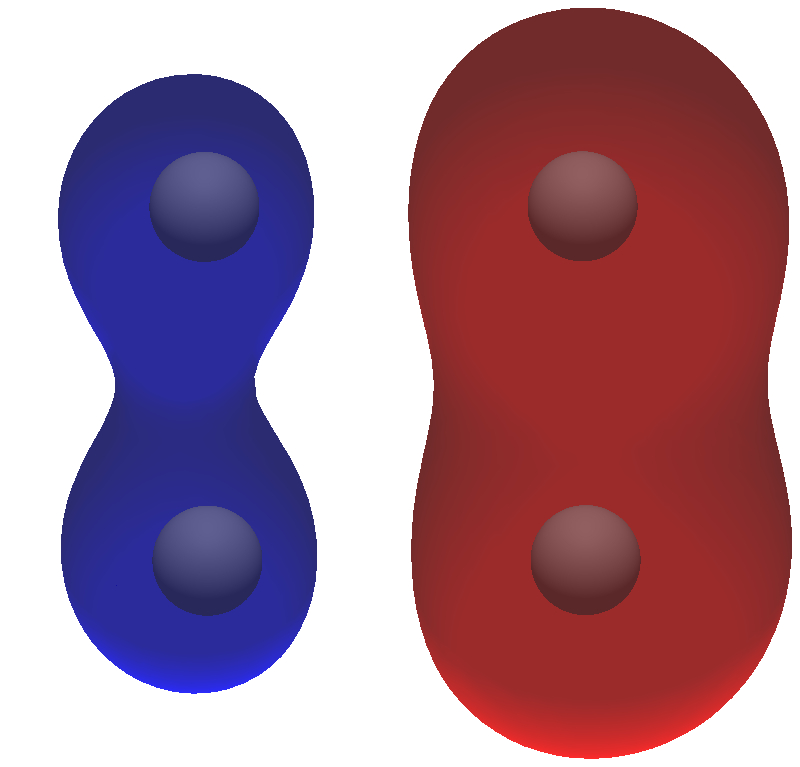}
			& \includegraphics[width=0.08\textwidth]{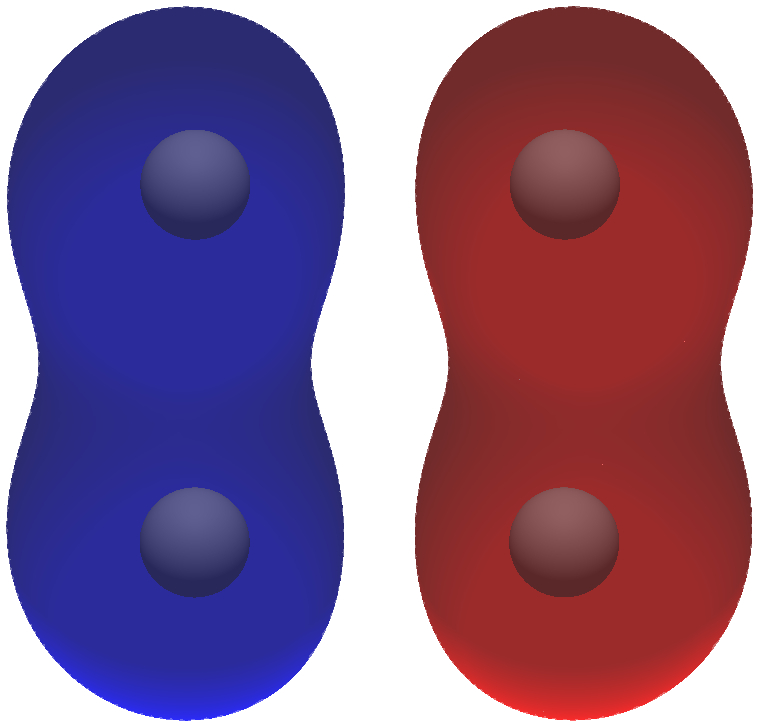} \\ \hline\hline
	\multicolumn{1}{|c|}{\raisebox{0.035\textwidth}{3}}		
			& \includegraphics[width=0.08\textwidth]{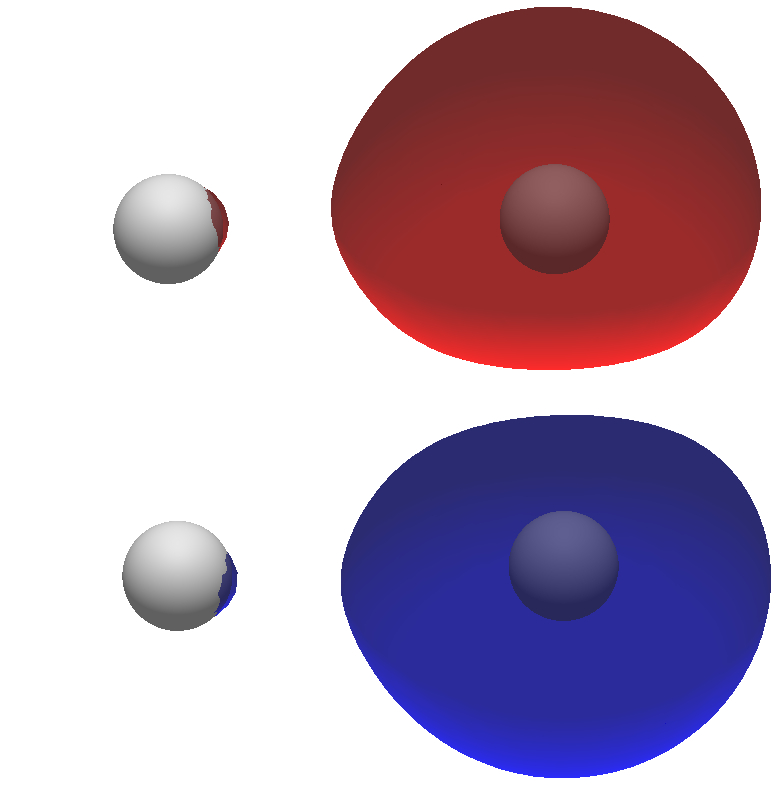}
			& \includegraphics[width=0.08\textwidth]{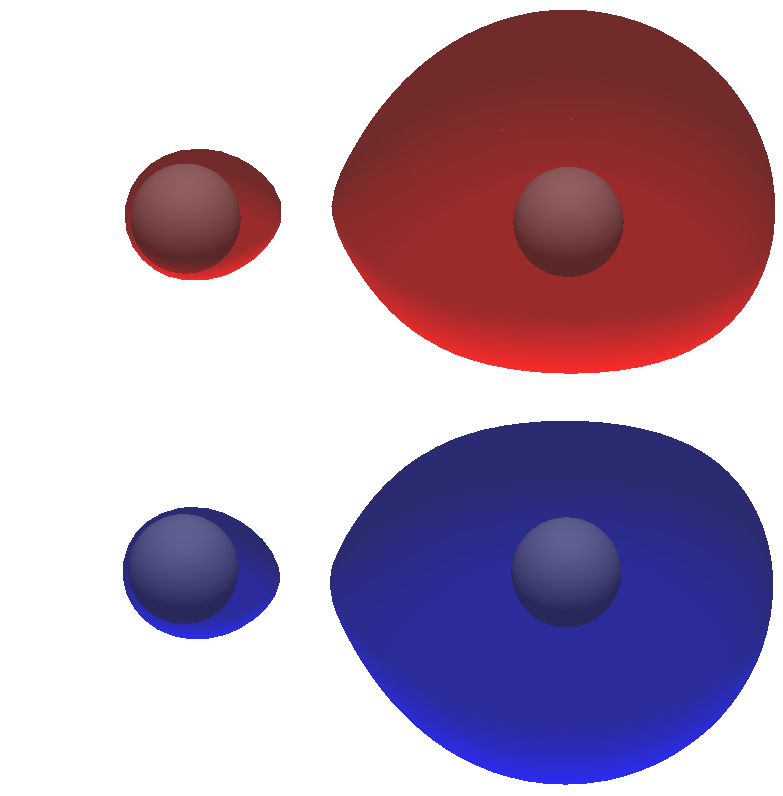}
			& \includegraphics[width=0.08\textwidth]{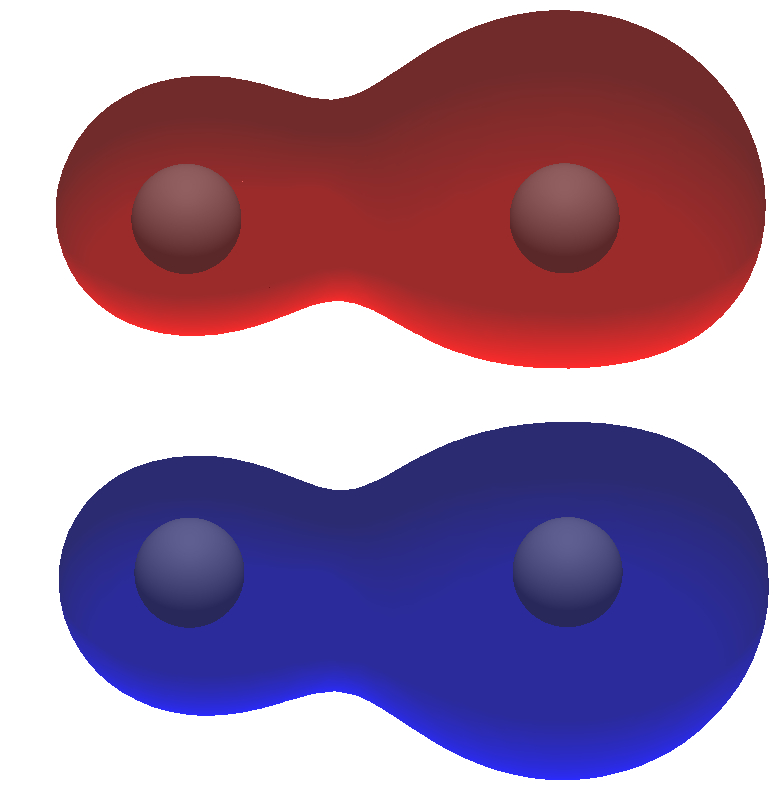}
			& \includegraphics[width=0.08\textwidth]{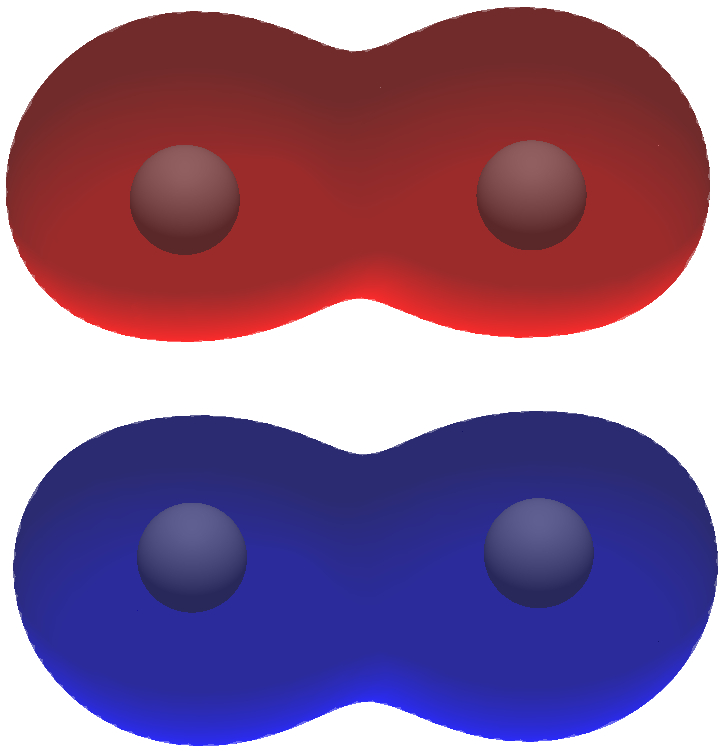} \\ \hline\hline
	\multicolumn{1}{|c|}{\raisebox{0.035\textwidth}{4}}		
			& \includegraphics[width=0.08\textwidth]{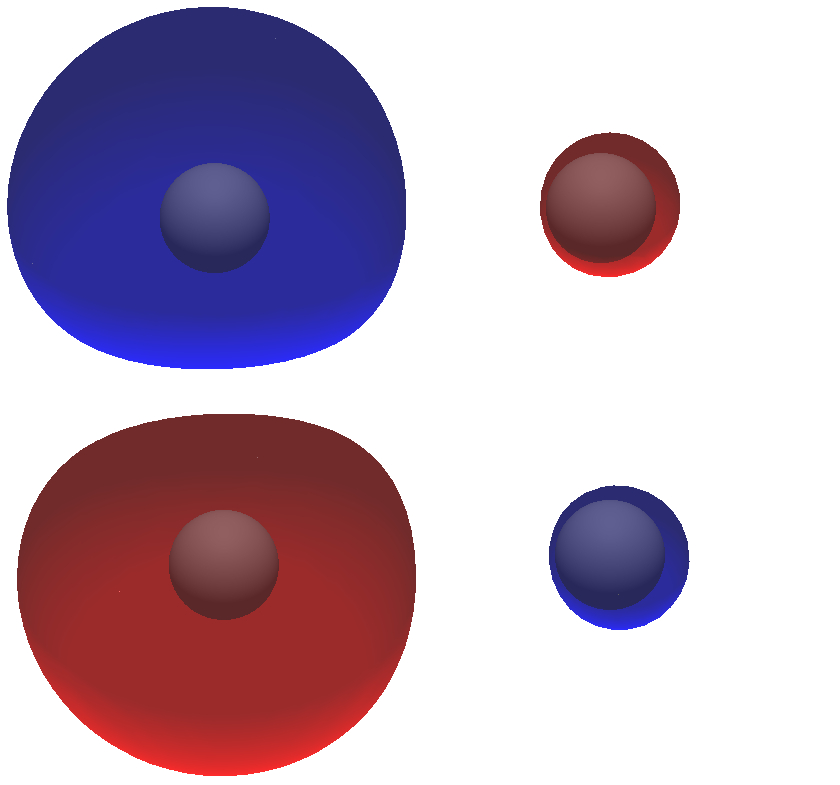}
			& \includegraphics[width=0.08\textwidth]{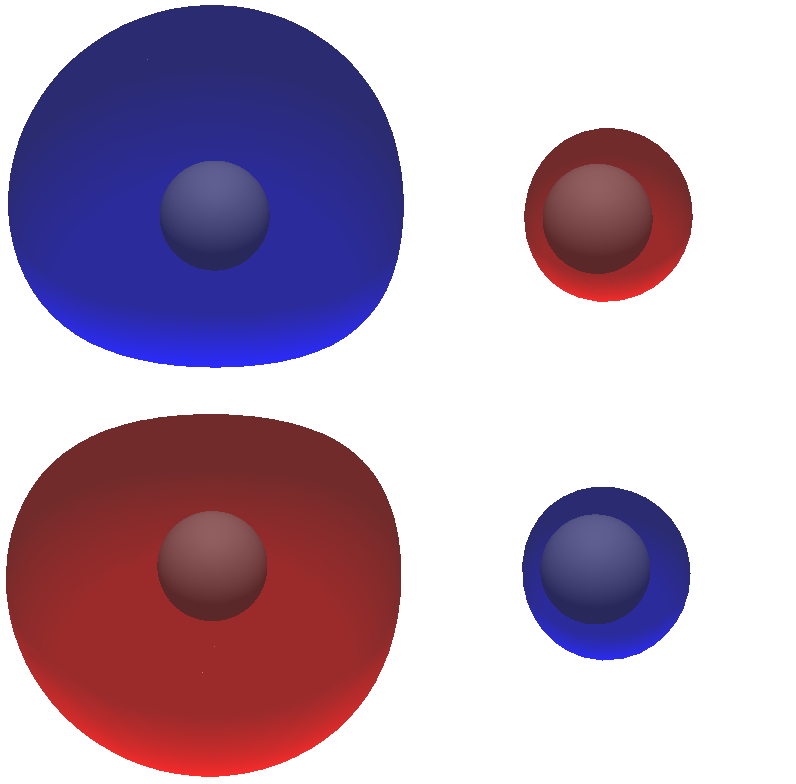}
			& \includegraphics[width=0.08\textwidth]{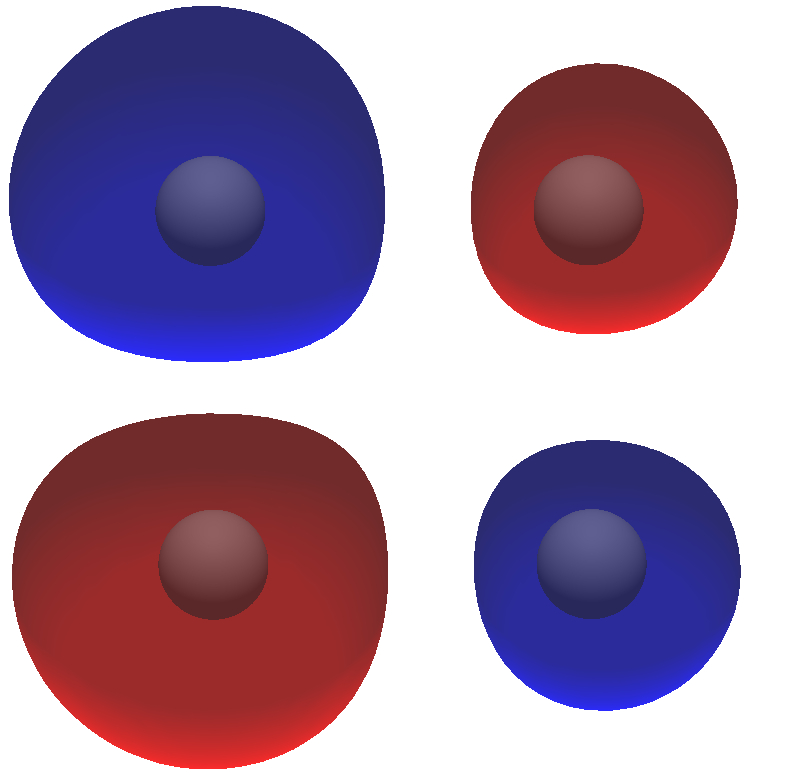}
			& \includegraphics[width=0.08\textwidth]{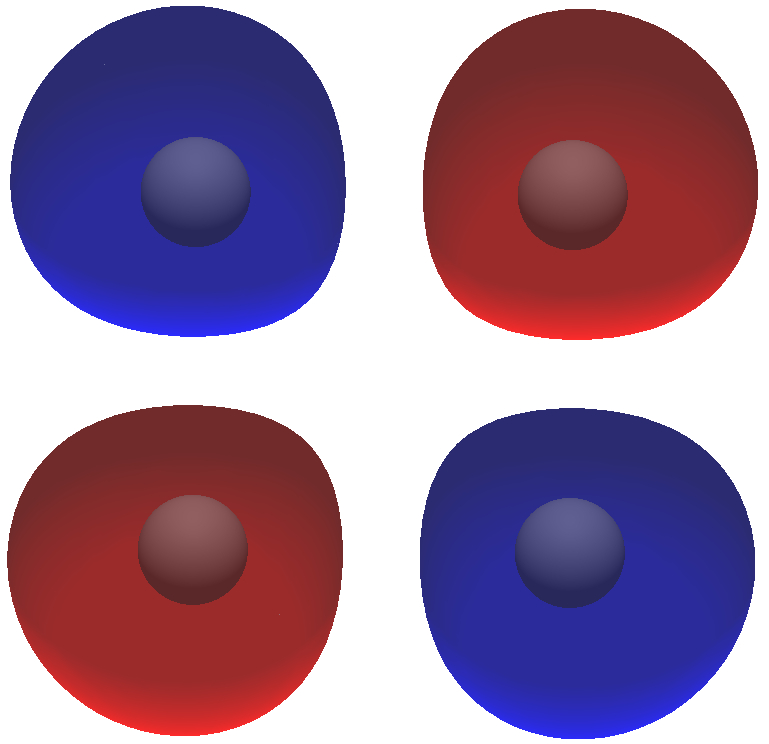} \\\hline
	\end{tabular}
	\end{center}
	\caption{\label{fig:H4orb} Change in $\dno$ active orbitals of \ce{H4} with $\theta$.}
\end{figure}

The presence of two different regimes, or models, for the correlation in \ce{H4} by $\dno$, and their transition, is supported further by an analysis of the $\dno$ orbitals (Figure \ref{fig:H4orb}).  
From Figure \ref{fig:H4orb}, it is seen that from $\theta=84^\circ$ to $85^\circ$ the $\dno$ orbitals resemble those of two separate \ce{H2} molecules, albeit with small tails on the opposite pair of protons.  At $\theta=86^\circ$, the orbitals begin to extend significantly to the neighbouring pair of protons, and by $\theta=87^\circ$ they are almost evenly distributed.  This transition occurs quickly with respect to the PEC, in just 3$^\circ$.

\begin{figure*}
	\begin{center}
	$$\Gamma^{\alpha\beta}(\br,\br_0)$$
	\includegraphics[width=0.325\textwidth]{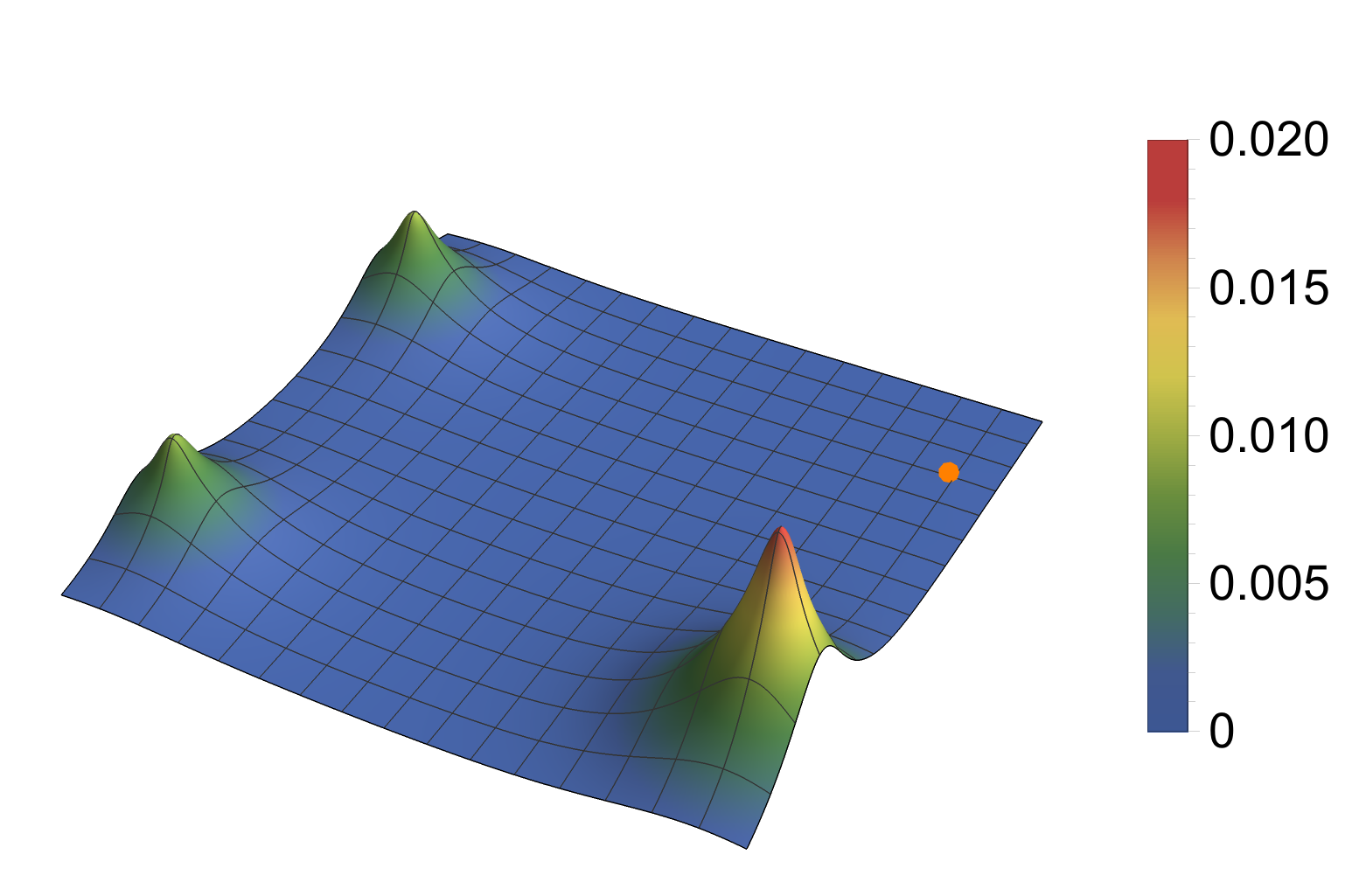}
	\includegraphics[width=0.325\textwidth]{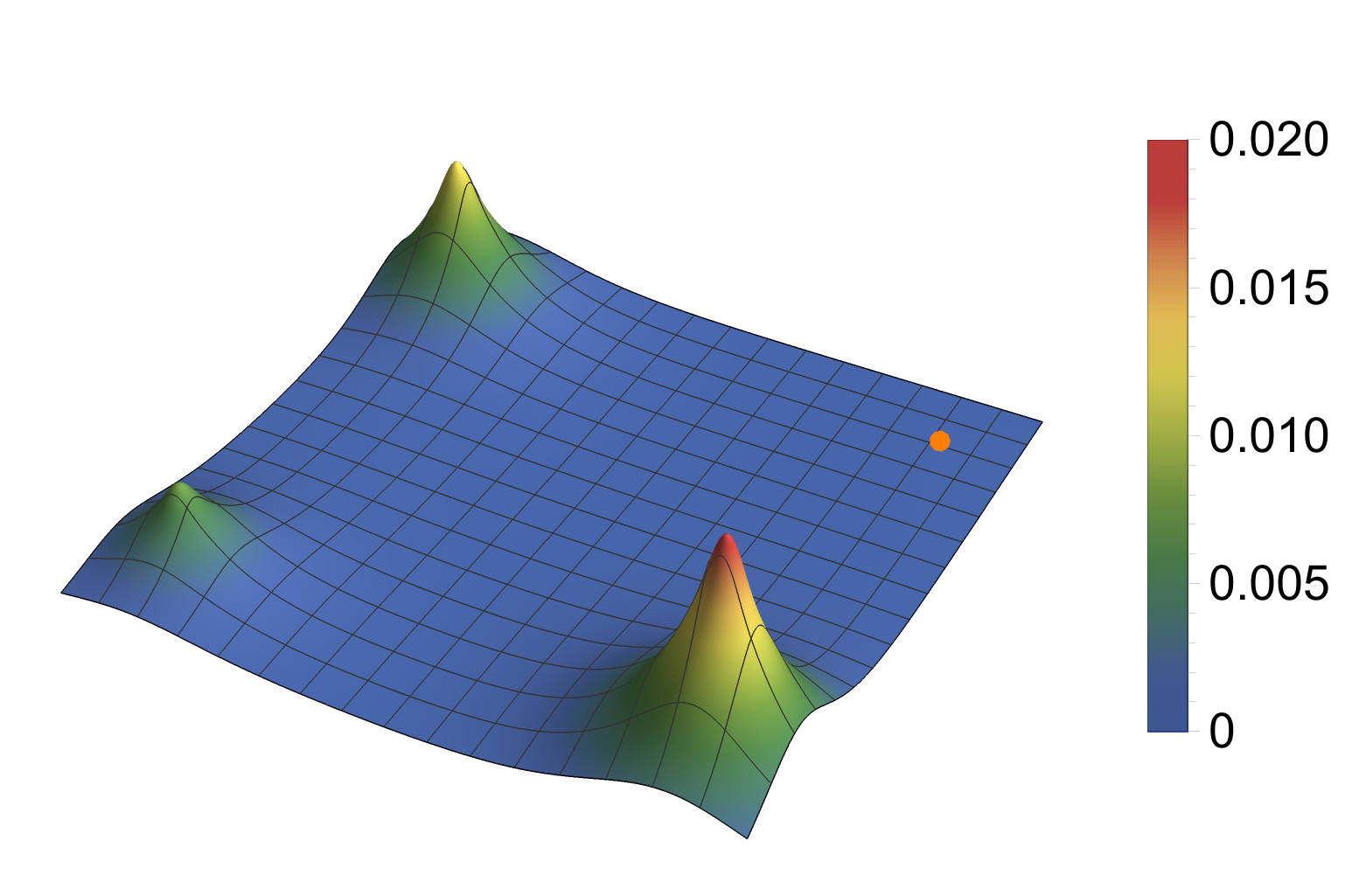}
	\includegraphics[width=0.325\textwidth]{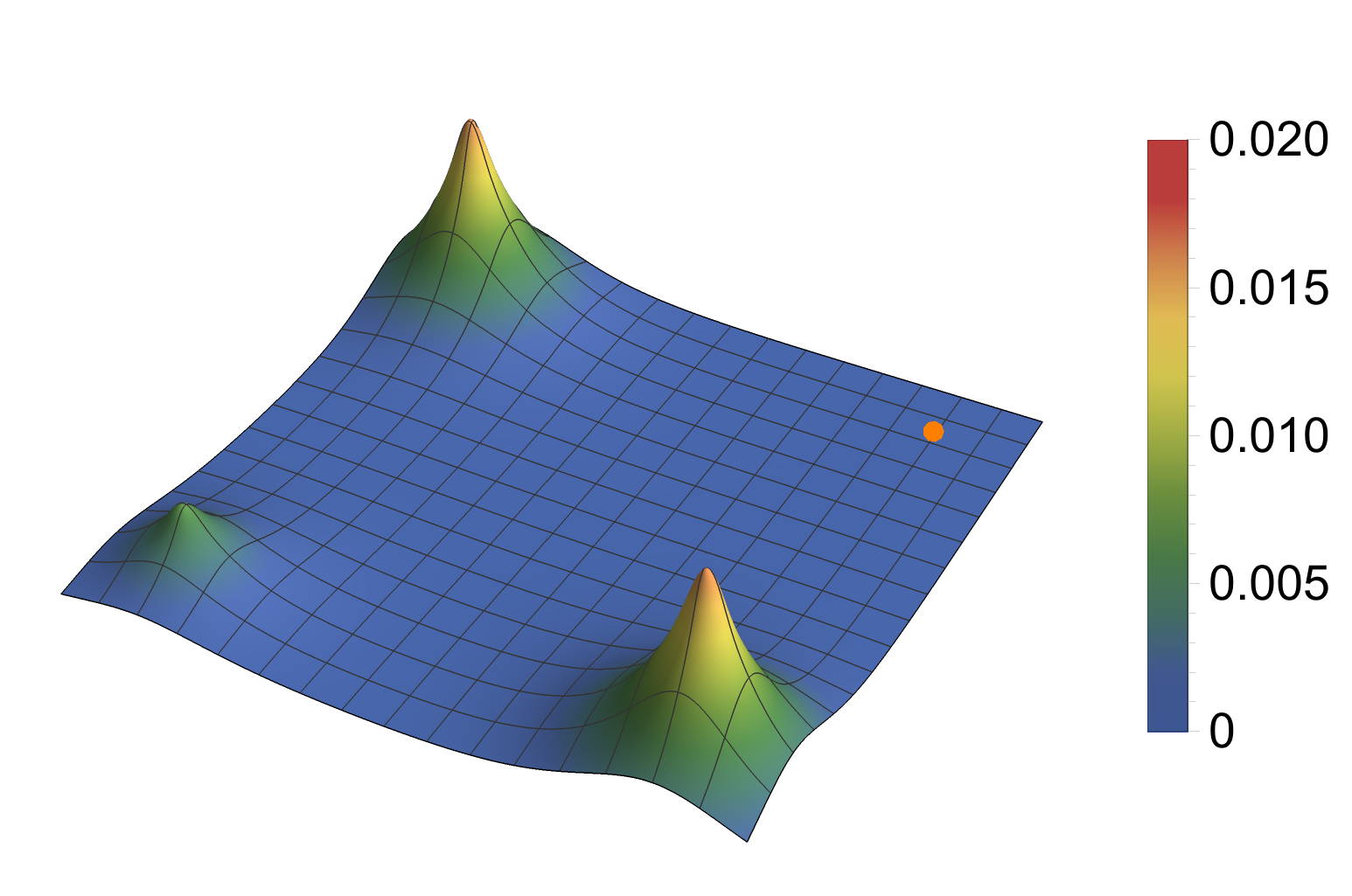}
	$$\Gamma^{\dno,\alpha\beta}(\br,\br_0)$$
	\includegraphics[width=0.32\textwidth]{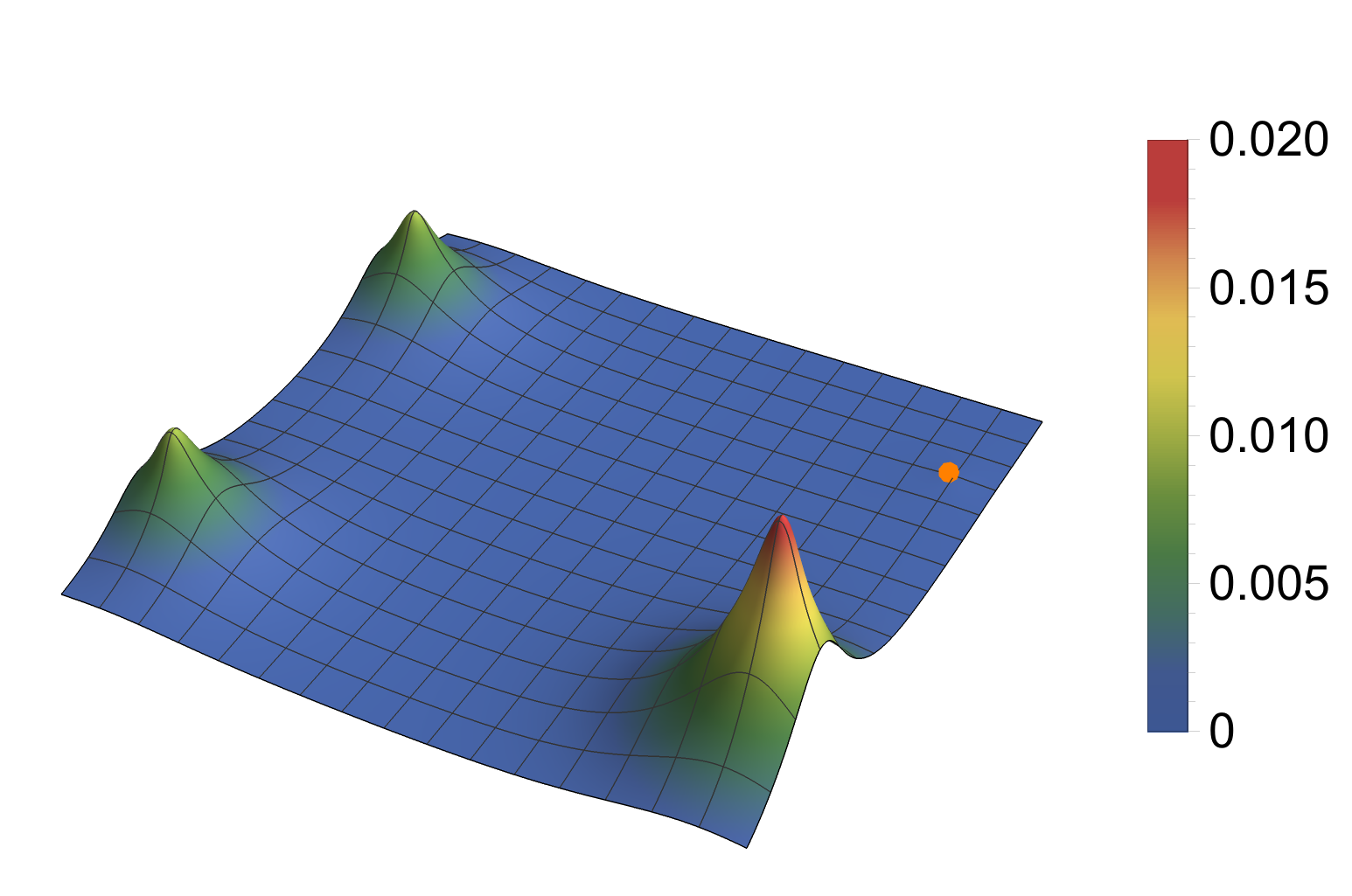}
	\includegraphics[width=0.325\textwidth]{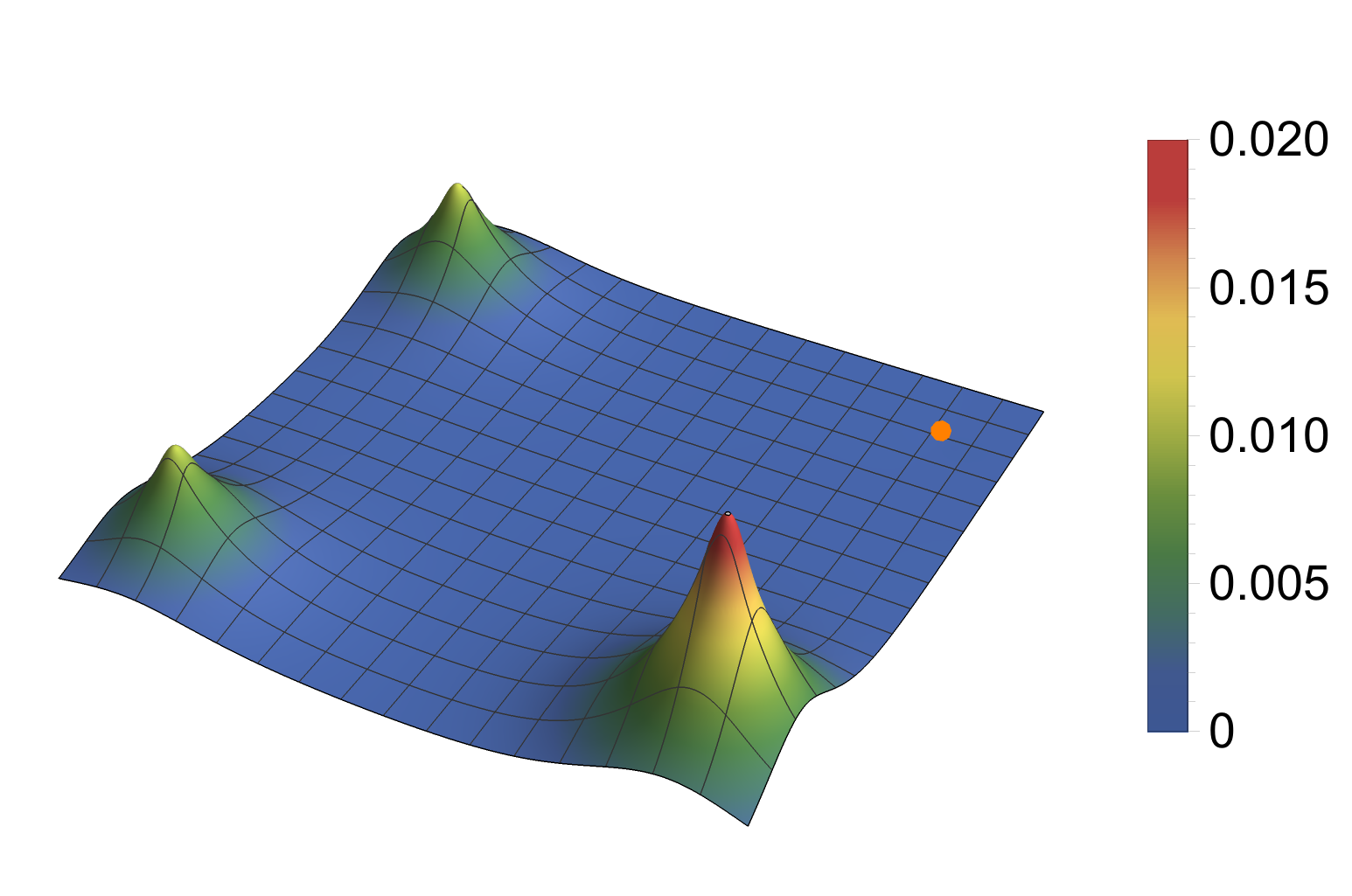}
	\includegraphics[width=0.325\textwidth]{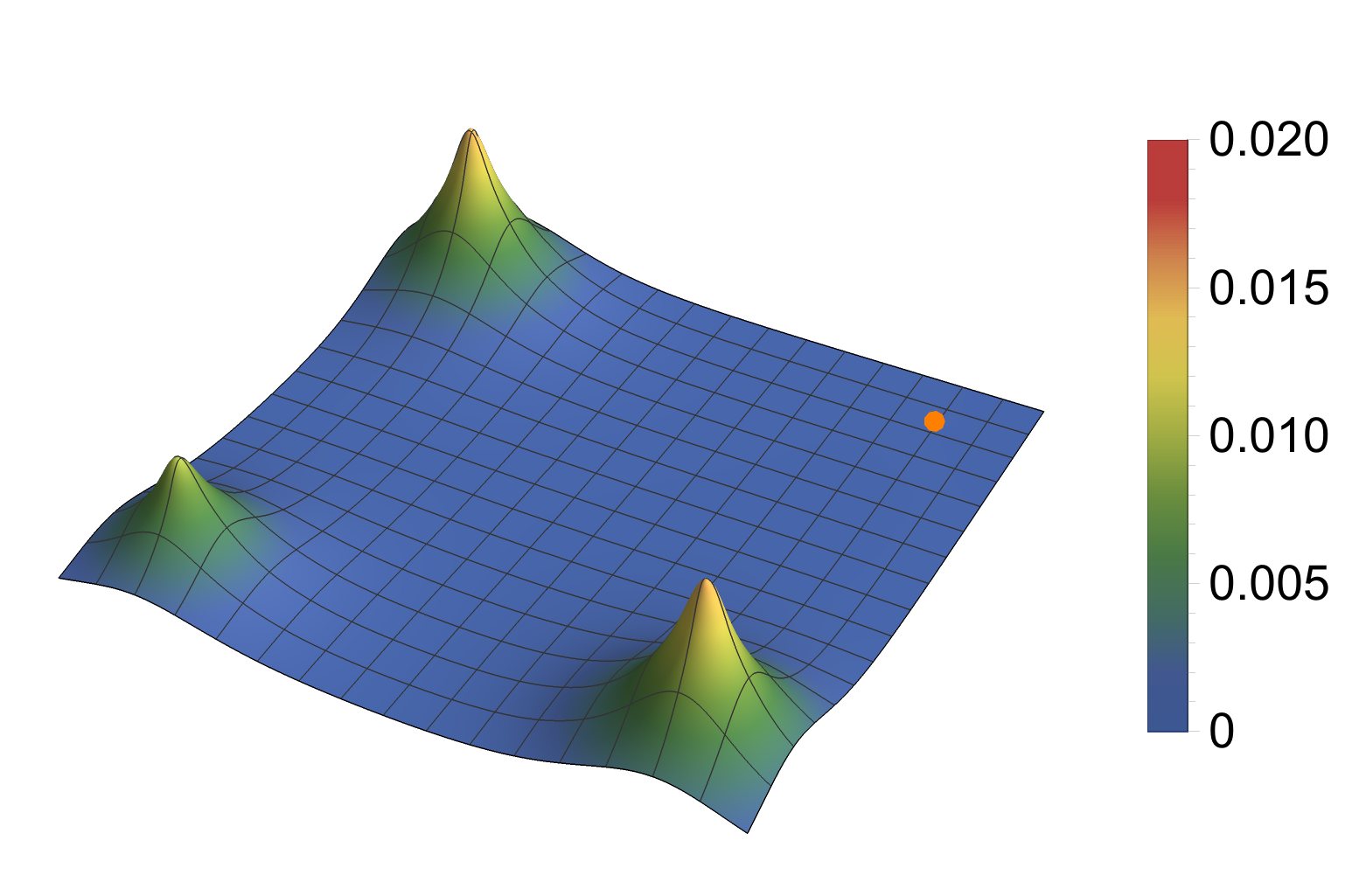}\\
	$$\Gamma^{\alpha\alpha}(\br,\br_0)$$
	\includegraphics[width=0.325\textwidth]{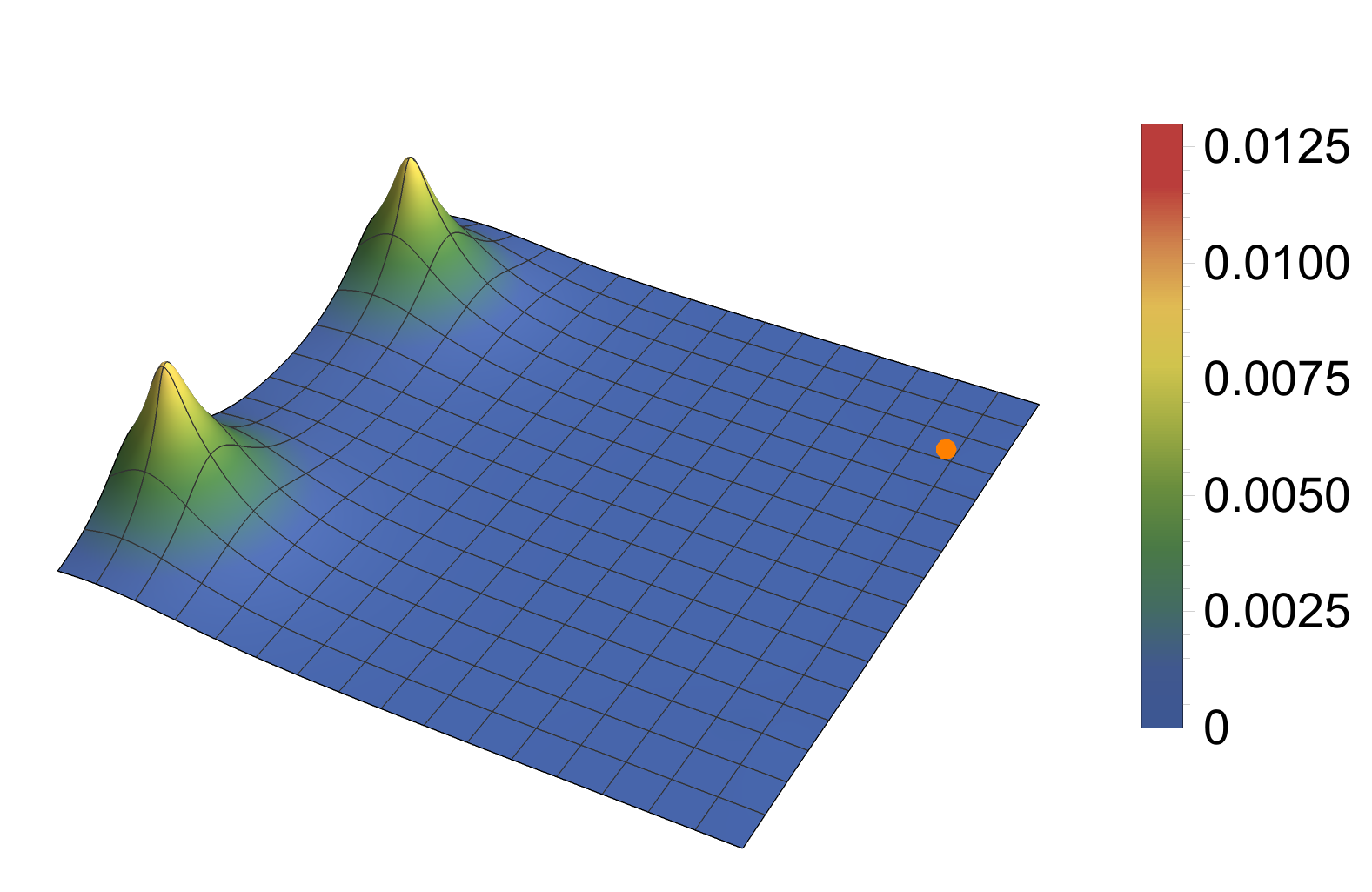}
	\includegraphics[width=0.325\textwidth]{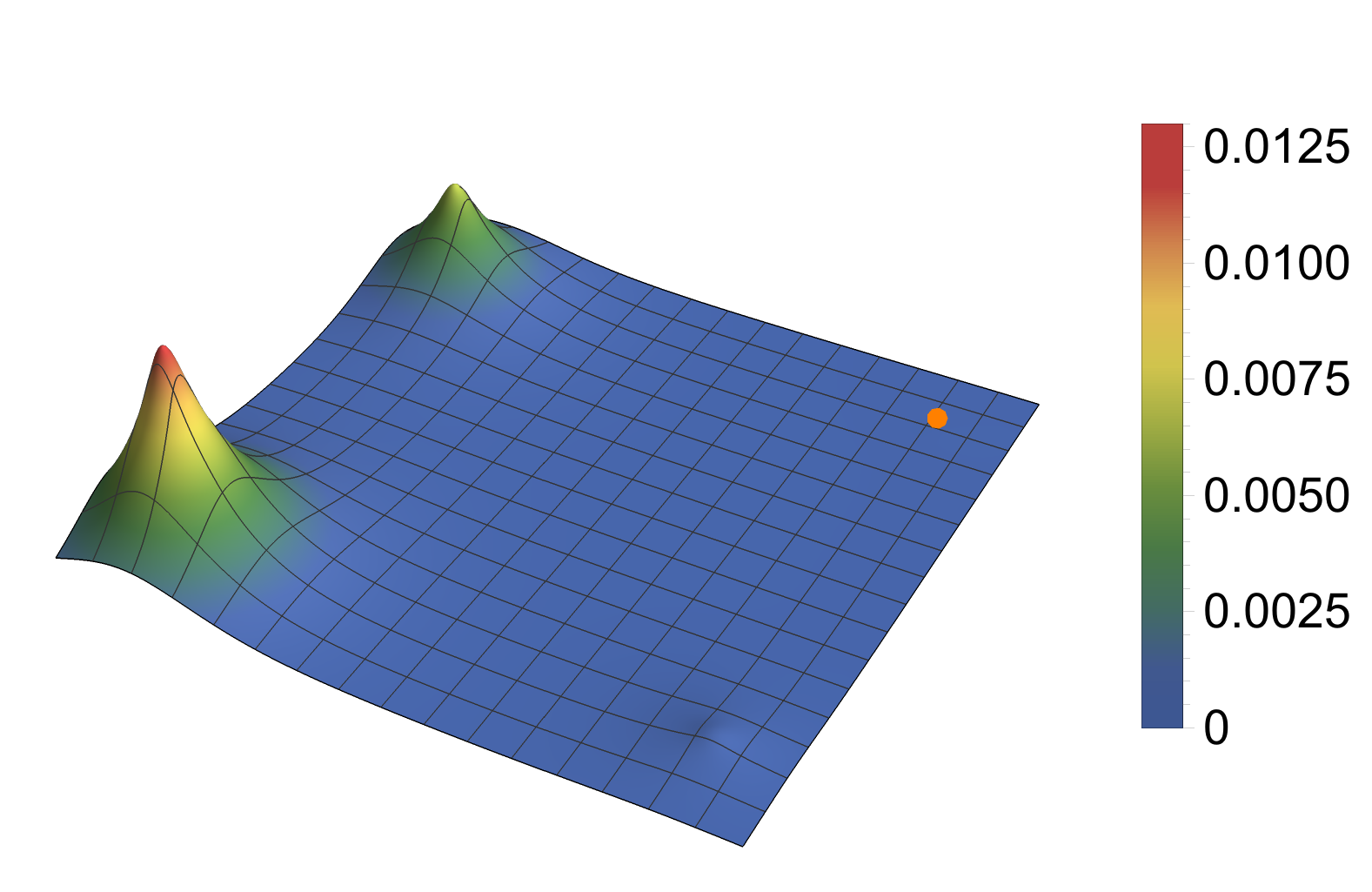}
	\includegraphics[width=0.325\textwidth]{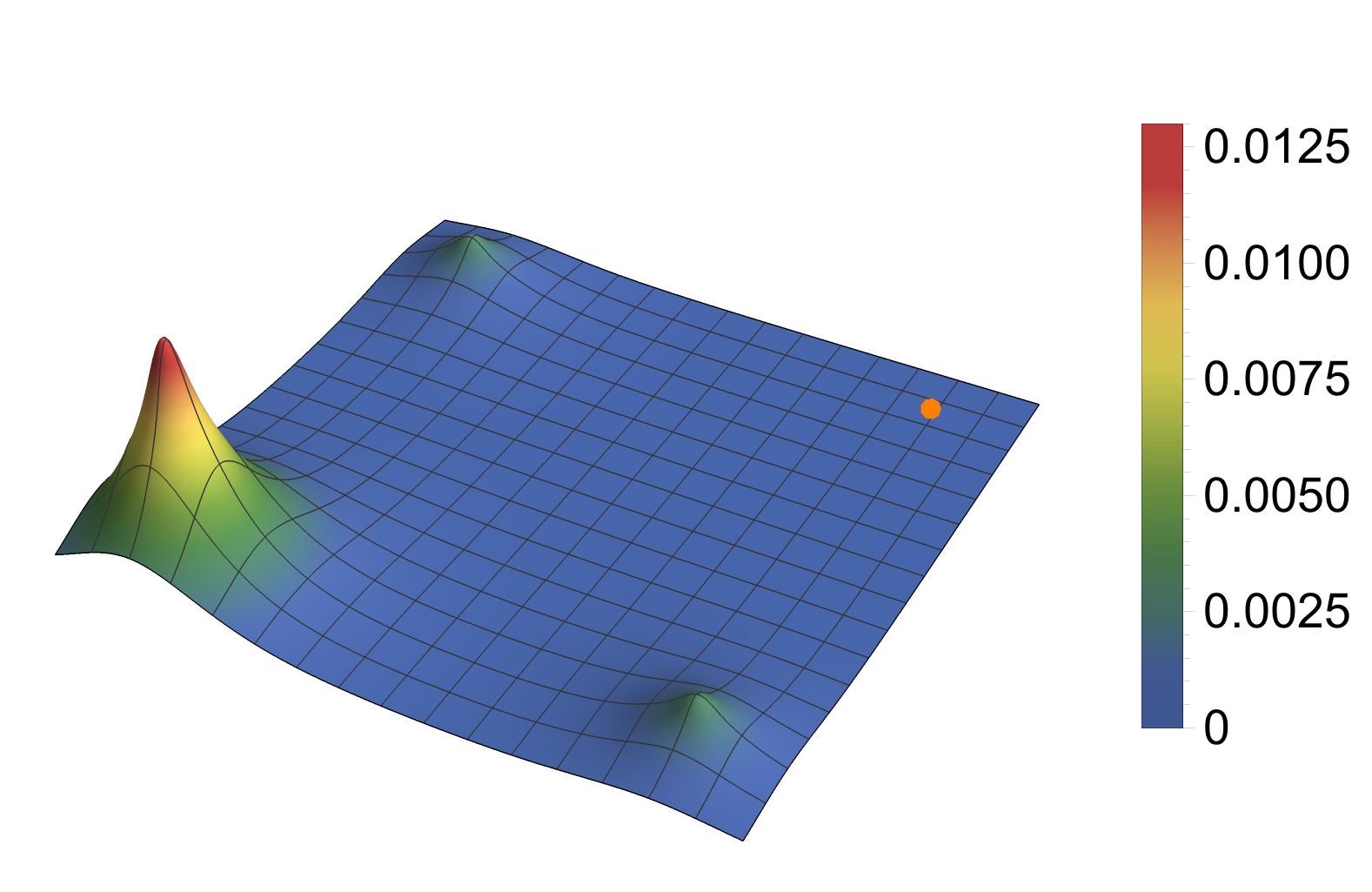}\\
	$$\Gamma^{\dno,\alpha\alpha}(\br,\br_0)$$
	\includegraphics[width=0.325\textwidth]{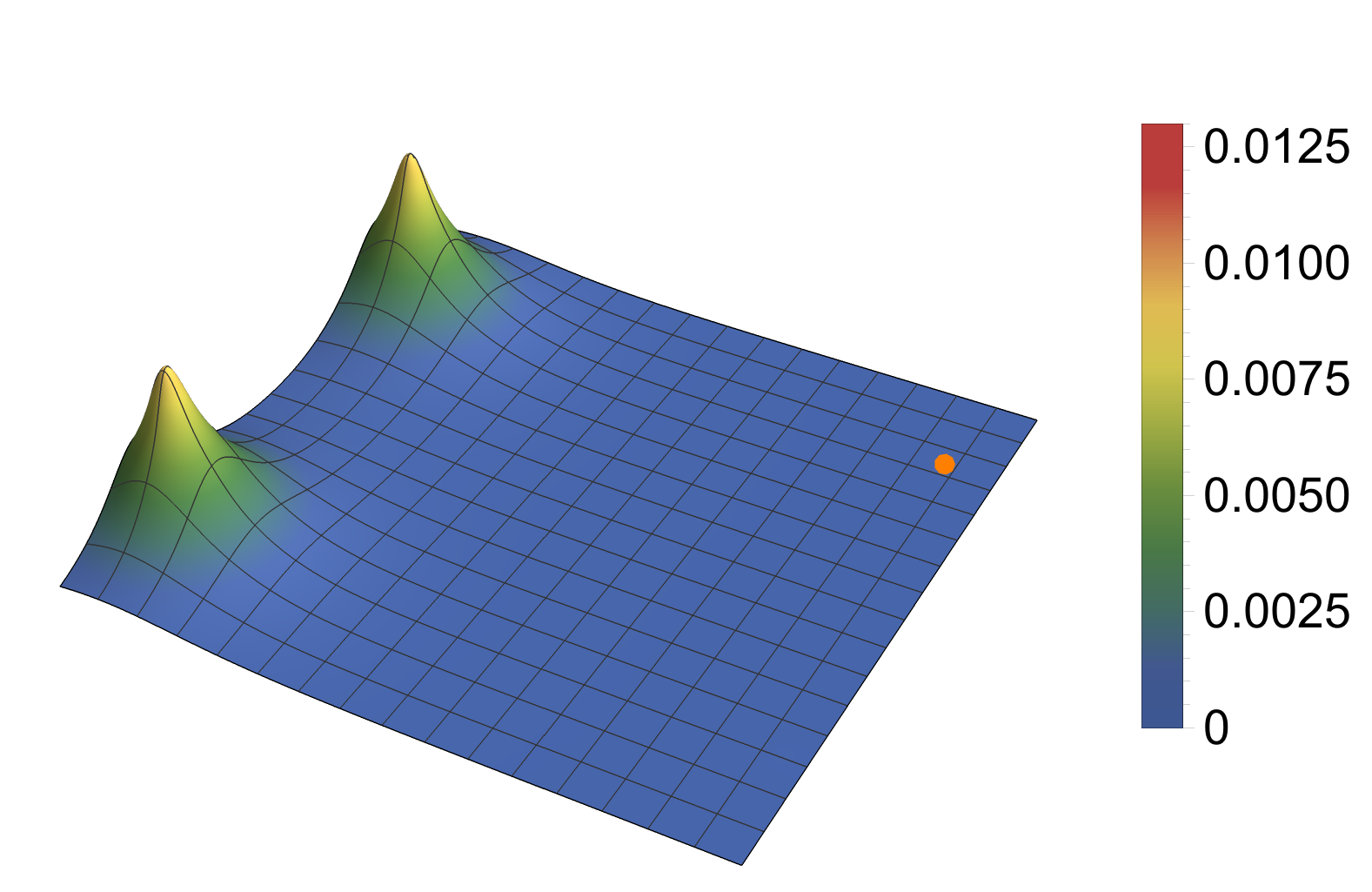}
	\includegraphics[width=0.325\textwidth]{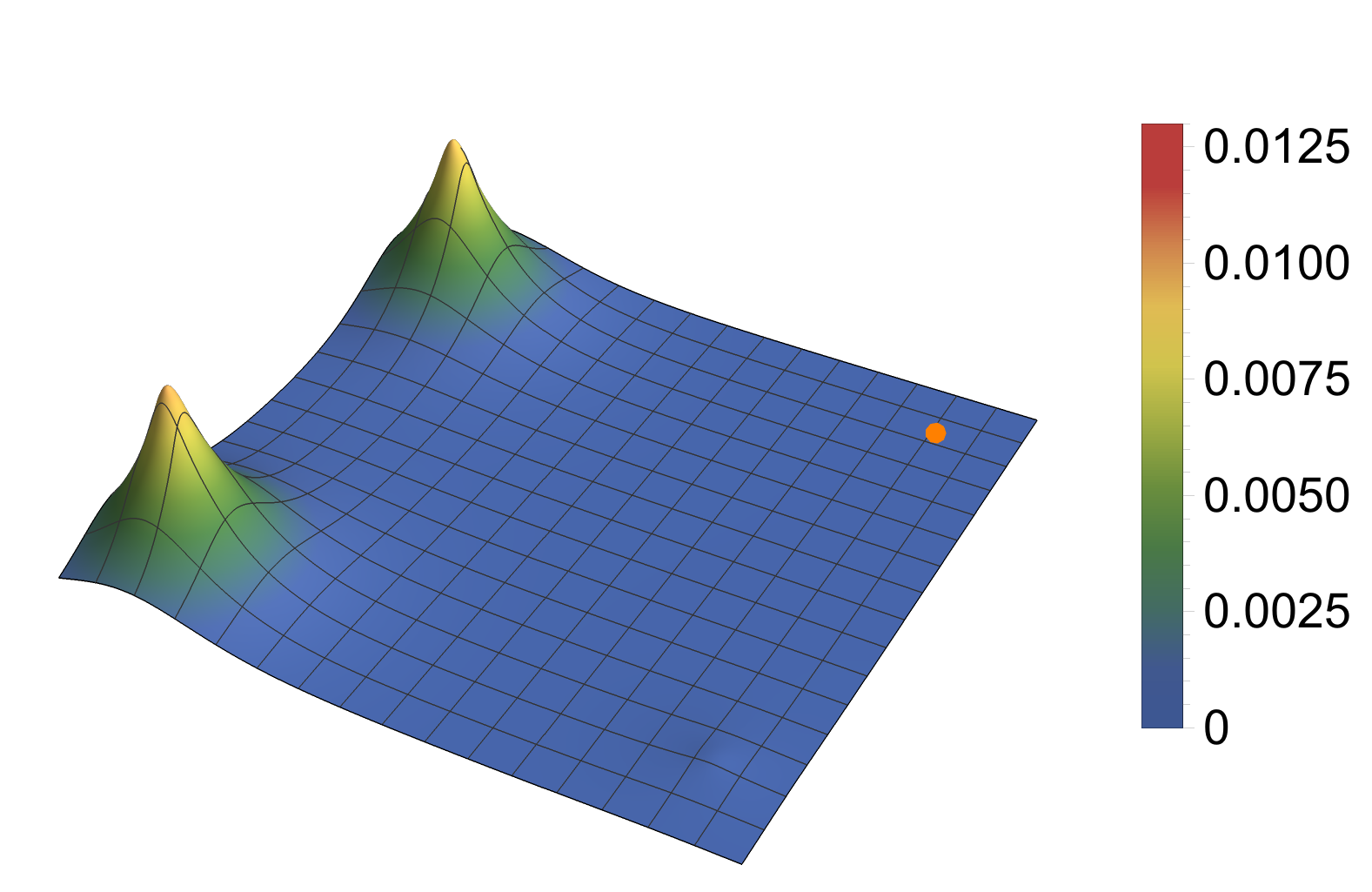}
	\includegraphics[width=0.325\textwidth]{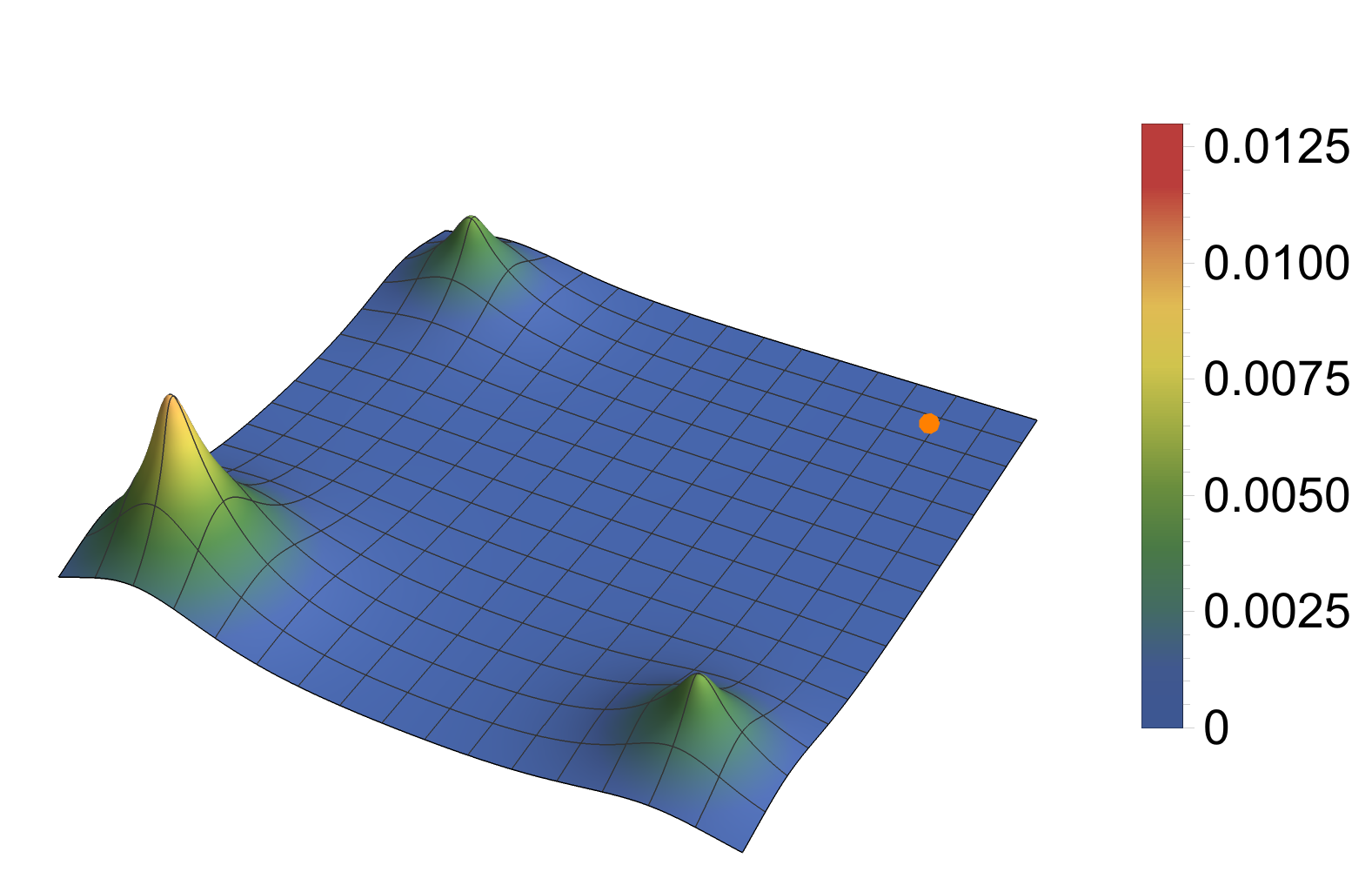}
	\caption{\label{fig:2eden} FCI and $\dno$ two-electron densities of \ce{H4} cluster when a test electron is placed at a \ce{H} nucleus, $\br_0$ (orange dot). Densities are shown for \ce{H4} at $R = 1.7$ {\AA} and $\theta =$  $70^\circ$ (left), $85^\circ$ (middle), and $90^\circ$ (right).}
	\end{center}
\end{figure*}

Analysis of the $\Delta$ values and the orbitals of $\dno$ partially explains the behaviour of the model for the this challenging \ce{H4} PEC, but it is does not explain why it differs from the FCI result.  A comparison of the $\dno$ and FCI two-electron densities, $\Gamma^{\dno}(\br_1,\br_2)$ and   $\Gamma(\br_1,\br_2)$, can show conclusively how the $\dno$ model of electron correlation differs from how the electrons actually behave.  The $\dno$ and FCI opposite-spin ($\alpha\beta$) and parallel-spin ($\alpha\alpha$) two-electron densities for \ce{H4} at $\theta= 70^\circ, 85^\circ$ and 90$^\circ$, with $R=1.7$ {\AA}, are shown in Figure \ref{fig:2eden}.
The two-electron densities are plotted for a test-electron placed at the upper right \ce{H} atom (indicated by an orange dot).  Therefore, the plots show the probability of finding the other electron(s), opposite or parallel-spin, within the \ce{H4} cluster.

At $\theta = 70^\circ$ (left column of Figure \ref{fig:2eden}), the $\dno$ and FCI two-electron densities closely resemble each other.  For opposite-spin electrons (top half of Figure \ref{fig:2eden}), if an $\alpha$ electron is found at the upper-right \ce{H} atom, then it is most likely to find a $\beta$ electron on the closest \ce{H} atom, to which it is paired.  It would then be equally likely to find the other $\beta$ electron on either of the two more distant \ce{H} atoms.  In the case of parallel-spin electrons (bottom half of Figure \ref{fig:2eden}), if an $\alpha$ electron is found at the upper-right \ce{H} atom then the other $\alpha$ electron must be found in the ``other" \ce{H2} molecule, with an equal probability of being found at either proton.

At $\theta = 90^\circ$ (right column of Figure \ref{fig:2eden}), the behaviour is much different. For electrons of opposite-spin, when an $\alpha$ electron is found at the upper-right \ce{H} atom, it is equally likely to find the $\beta$ electron at either of the nearest protons, with a reduced probability of finding a $\beta$ electron at the furthest proton. Comparing the $\dno$ and FCI two-electron densities, the same qualitative behaviour is seen but the $\dno$ 2-RDM results in too high a probability of finding the $\beta$ electron on the furthest proton compared to the probability at the nearest protons.  In the case of parallel-spin electrons, if an $\alpha$ electron is found at the upper-right \ce{H} atom then it is most likely that another $\alpha$ electron is on the furthest proton, with only a slight chance of an $\alpha$ electron on the nearest protons.  Again, $\dno$ models this behaviour qualitatively, but compared to FCI, $\dno$ predicts the probability of finding an $\alpha$ electron on the nearest protons to be too high relative to the furthest proton.  This suggests that the amount of exchange, or Fermi correlation, which occurs between parallel-spin electrons, in the $\dno$ 2-RDM is slightly inadequate at $\theta=90^\circ$.

Finally, at $\theta=85^\circ$ (middle column of Figure \ref{fig:2eden}), the difference between $\dno$ and FCI two-electron densities is the most pronounced.  It is evident that the electrons of the $\dno$ 2-RDM at $\theta=85^\circ$ behave more like the electrons at $\theta=70^\circ$ (\ce{2H2} regime) than the actual behaviour given by the FCI two-electron density.  In the case of the $\dno$ two-electron densities, both opposite and parallel-spin, the probabilities of finding electrons on the left-side protons (other \ce{H2}) is near equivalent.  Whereas, the FCI two-electron density shows uneven probabilities for the left-side protons,  which more resembles the behaviour at $\theta=90^\circ$ (\ce{H4} regime).

The inability of $\dno$ to accurately model the PEC from $\theta = 70^\circ$ to $90^\circ$, is due to the energetic preference of $\dno$ to model \ce{H4} as \ce{2H2}, at $\theta$ values close to $90^\circ$. The error observed in the parallel-spin two-electron density suggests that $\dno$ is incorrectly describing the correlation between different electron pairs (interpair correlation), which is similar to the conclusion reached in a study with PNOF6.\cite{RamosCordoba2015}  For $\dno$, this could mean an error in the exchange component (Fermi correlation), or possibly something else, but it implies that the high-spin correction (HSC) is not adequate. The HSC provides the correct description in the strong correlation limit (both pairs), but it has deficiencies for moderate amounts of static correlation and must be improved upon.

\begin{figure}
	\begin{center}
	\includegraphics[width=0.48\textwidth]{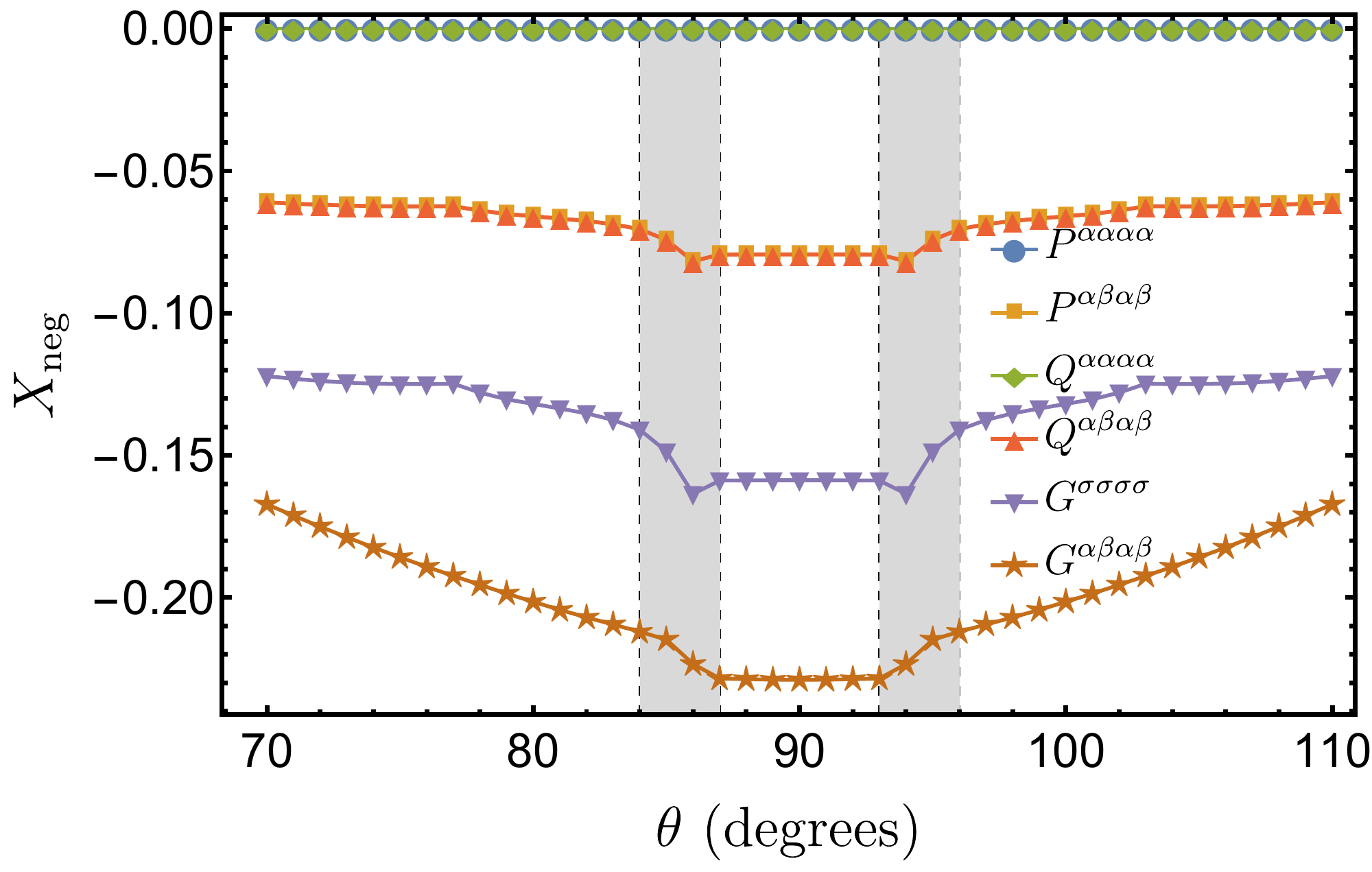}
	\caption{\label{fig:PQGneg} Sum of the negative eigenvalues of $\bm{P}$, $\bm{Q}$ and $\bm{G}$ for the rectangle-to-square \ce{H4} cluster as a function of $\theta$ at $R= 1.7$ \AA.}
	\end{center}
\end{figure}

The inadequacy of the HSC is highlighted further by examining the $N$-representabiltity of the $\dno$ 2-RDM; in particular, the violation of the $PQG$-conditions (Subsection \ref{ssec:Nrep}).  The sums of the negative eigenvalues of the $\bm{P}$, $\bm{Q}$, and $\bm{G}$ matrices corresponding to the $\dno$ 2-RDM for rectangle-to-square \ce{H4} are shown in Figure \ref{fig:PQGneg}.
The sums are separated into contributions from different spin-blocks of the $\bm{P}$, $\bm{Q}$, and $\bm{G}$ matrices. The $\bm{P}$ and $\bm{Q}$ matrices are separated into spin-blocks for each specific spin-pair, whereas the $\bm{G}$ matrix is separated into only three blocks, $\bm{G}^{\alpha\beta\alpha\beta}$, $\bm{G}^{\beta\alpha\beta\alpha}$, and $\bm{G}^{\sigma\sigma\sigma\sigma}$, due to the coupling of the parallel-spin blocks [Equation \eqref{eq:Gmat}]. For all values of $\theta$, the parallel-spin blocks of the $\bm{P}$ and $\bm{Q}$ matrices have no negative eigenvalues.  This is completely expected considering that all parallel-spin terms of the $\dno$ 2-RDM are derived from the $\dno$ wave function.  The only terms not originating from the $\dno$ wave function are those corresponding to the HSC, which are applied to the opposite-spin block of the 2-RDM only [Equation \eqref{eq:2RDMos}].  The off-diagonal, $\bm{G}^{\alpha\alpha\beta\beta}$ and $\bm{G}^{\beta\beta\alpha\alpha}$, spin-blocks of $\bm{G}^{\sigma\sigma\sigma\sigma}$ contain 2-RDM elements corresponding to the HSC and they are responsible for the contribution to $G^{\sigma\sigma\sigma\sigma}_\text{neg}$. The most severe violation of the $PQG$-condition is seen for $\bm{G}^{\alpha\beta\alpha\beta}$, also known as the particle-hole density matrix.  The sum of the negative eigenvalues of $\bm{G}^{\alpha\beta\alpha\beta}$ reach a minimum at $\theta=90^\circ$, where $G^{\alpha\beta\alpha\beta}_\text{neg} = -0.229$.  Due to the fact that the HSC is applied to the exchange-like terms ({\it i.e.}~$\Gamma^{\alpha\beta}_{pqqp}$) of the $\dno$ 2-RDM, the sum of the negative eigenvalues of the $\bm{P}^{\alpha\beta\alpha\beta}$ and $\bm{Q}^{\alpha\beta\alpha\beta}$ matrices are equivalent for all values of $\theta$.  In the case of $P_\text{neg}$ and $Q_\text{neg}$, the minimum value observed is -0.0818, which occurs at $\theta = 86^\circ$ and $\theta = 94^\circ$.  As seen for the energy and orbital occupancies, the violation of the $PQG$-conditions exhibits different behaviour, with respect to $\theta$, in the neighbourhood of $\theta = 90 ^\circ$ (between the grey rectangles) compared to $\theta$ closer to $70^\circ$ or $110^\circ$ (outside the grey rectangles).

\FloatBarrier
%---------------------------
\section{Conclusions}
%---------------------------
\label{sec:conc}

The $\dno$ method is one of many that combines different models for different aspects of electron correlation in an effort to efficiently capture the total.  The $\dno$-ODF approach is unique in the manner that it combines an approximate 2-RDM for static correlation with an ODF for dynamic correlation.  With the exception of the high-spin correction, the $\dno$ 2-RDM and energy can be derived from a multideterminantal wave function with restricted $2n$-tuple excitations.  Combining the $\dno$ 2-RDM for static correlation with an ODF for dynamic correlation requires the inclusion of a double-counting correction, which can be done by replacing the short-range correlation counted by the $\dno$ 2-RDM.

Previously, $\dno$, and other NOF methods, optimized the 2-RDM through alternating optimization of the occupation numbers and the orbitals until convergence was reached.  The orbitals themselves were optimized by iterative diagonalization of a pseudo-Fock matrix.  The trust-region Newton method introduced here, which allows for simultaneous optimization of the orbitals and occupancies, drastically reduces the number of optimization steps required and has the ability to identify and avoid saddle points.

Even simple hydrogen clusters can prove challenging for electronic structure methods, particularly single-reference methods.  Overall, the $\dno$-ODF approach performs well when describing the potential energy surfaces of \ce{H2}, \ce{H3} and \ce{H4}, relative to the more accurate, but costly, multireference wave function method, MRMP2.

Analysis of the two-electron density of \ce{H4} reveals that $\dno$ incorrectly describes interorbital correlation.  Despite being accurate in the strong-correlation limit, the {\it ad hoc} HSC does have deficiencies, particularly near the $D_{2h} \to D_{4h}$ transition.  A more complete description of interorbital correlation within $\dno$ is a target of future development.

%-----------------------------------
\begin{acknowledgments}
%-----------------------------------
JWH thanks the Natural Sciences and Engineering Research Council of Canada (NSERC) for a Discovery Grant, Compute/Calcul Canada for computing resources and the Discovery Institute for Computation and Synthesis for useful consultations.
\end{acknowledgments}

%-----------------------------------
\section*{Data Availability}
%-----------------------------------
The data that support the findings of this study are available from the corresponding author upon request.

%--------------------------
\appendix
%--------------------------

%--------------------------
\section{$\dno$ energy derivatives}
%--------------------------
\label{app:derivs}

The optimization of the $\dno$ 2-RDM is acheived by simultaneously finding the optimal $\dno$ orbitals, $\{\phi_p\}$, and electron transfer variables, $\{\Delta_{pq}\}$.  The optimization is performed using a trust-region Newton algorithm, which requires the analytical calculation of the energy gradient, $\bg$, and hessian, $\bH$, with respect to the skew-matrix parameters, $\by$, and the variables used to parameterize the electron-transfer variables, $\mathbf{\theta}$.

The gradient, with respect to $\by$, is evaluated at $\by = 0$.  The first and second derivatives of the individual orbitals,
\begin{equation}
	\left. \frac{\partial \phi_a}{\partial y_{pq}} \right|_{\by=0} = \delta_{aq} \phi_p - \delta_{ap} \phi_q
\end{equation}
and
\begin{align}
	\left. \frac{\partial^2 \phi_a}{\partial y_{pq}\partial y_{rs}} \right|_{\by=0} & = 	 \left( \delta_{qr}\delta_{as} - \delta_{qs}\delta_{ar} \right) \phi_p \nonumber \\
	& - \left( \delta_{pr}\delta_{as} - \delta_{ps}\delta_{ar} \right) \phi_q \nonumber \\
	& + \left( \delta_{ps}\delta_{aq} - \delta_{qs}\delta_{ap} \right) \phi_r \nonumber \\
	& - \left( \delta_{pr}\delta_{aq} - \delta_{qr}\delta_{ap} \right) \phi_s,
\end{align}
can be used to derive expressions for derivatives of the total energy, $E^{\dno}$.  The total gradient with respect to the orbital-mixing parameters, is given by
\begin{equation}
	g^\phi_{pq} = 2\left( \lambda_{qp} - \lambda_{pq} \right),
\end{equation}
where
\begin{align} \label{eq:lambda}
\lambda_{pq} & = (2 - W_p) n_p \left( h_{pq} + G_{pq}^{\cl} \right) \nonumber \\
		& + 2(1-W_p) n_p G_{pq}^{\oc} + W_p n_p G_{pq}^{\open} \nonumber \\ 
		& + 2 \sum_r \eta_{pr} \left( 2\langle pr|qr \rangle - \langle pr | rq \rangle \right) \nonumber \\
		& + 2 \sum_r \left( \zeta_{pr} - \xi_{pr} \right) \langle pq|rr \rangle \nonumber \\
		& - 2 \sum_r \kappa_{pr} \langle pr | rq \rangle
\end{align}
and the two-electron integrals are defined as follows
\begin{equation}
	\langle pq|rs \rangle = \int \frac{\phi^*_p(\br_1)\phi^*_q(\br_2)\phi_r(\br_1)\phi_s(\br_2)}{r_{12}} d\br_1 d\br_2,
\end{equation}
and
\begin{equation}
	G_{pq}^{\cl} = \sum_{\mu \nu} P_{\nu \mu}^{\cl} \left( \langle p \mu | q \nu \rangle - \frac{1}{2} \langle p \mu | \nu q \rangle \right),
\end{equation}
\begin{equation}
	G_{pq}^{\open} = \sum_{\mu \nu} P_{\nu \mu}^{\open} \left( \langle p \mu | q \nu \rangle - \langle p \mu | \nu q \rangle \right),
\end{equation}
\begin{equation}
	G_{pq}^{\oc} = \sum_{\mu \nu} P_{\nu \mu}^{\open} \left( \langle p \mu | q \nu \rangle - \frac{1}{2} \langle p \mu | \nu q \rangle \right),
\end{equation}
where $\mu$ and $\nu$ are indices of atomic basis functions.  The density matrices are defined as
\begin{equation}
	P_{\mu\nu}^{\cl} = \sum^{\cl}_p 2 n_p C_{\mu p} C_{\nu p}^*
\end{equation}
and
\begin{equation}
	P_{\mu\nu}^{\open} = \sum^{\open}_p n_p C_{\mu p} C_{\nu p}^*,
\end{equation}
where $C_{\mu p}$ is a $\dno$ orbital coefficient.

For the second derivatives with respect to the orbital-mixng parameters, it is useful to decompose $E^{\dno}$. The one-electron energy is given by
\begin{equation}
	E^{\dno}_{\text{1e}} = \sum^\cl_p 2n_p h_{pp} + \sum^\open_p n_p h_{pp},
\end{equation}
and the contribution of $E^{\dno}_{\text{1e}}$ to the orbital hessian is given by
\begin{align}
	& H^{\phi\phi,\text{1e}}_{pq,rs} = \nonumber \\
	& \delta_{pr} \left( \na_p + \nb_p + \na_r + \nb_r - \na_q - \nb_q - \na_s -\nb_s \right) h_{qs} \nonumber \\
	+ & \delta_{qs} \left( \na_q + \nb_q + \na_s + \nb_s - \na_p - \nb_p - \na_r -\nb_r \right) h_{pr} \nonumber \\
	+ & \delta_{ps} \left( \na_q + \nb_q + \na_r + \nb_r - \na_p - \nb_p - \na_s -\nb_s \right) h_{qr} \nonumber \\
	+ & \delta_{qr} \left( \na_p + \nb_p + \na_s + \nb_s - \na_q - \nb_q - \na_r -\nb_r \right) h_{ps}.
\end{align}
The two-electron energy associated with the 0-1RDM component of the energy is given by
\begin{equation}
	E^{\dno}_{\text{0-1RDM-2e}} = E^{\dno}_{\text{0-1RDM}} - E^{\dno}_{\text{1e}},
\end{equation}
and the corresponding contribution to the orbital hessian is
\begin{align}
	& H^{\phi\phi,\text{0-1RDM-2e}}_{pq,rs} = \nonumber \\
	& \delta_{pr} \Big( \left[ \na_p + \nb_p + \na_r + \nb_r - \na_q - \nb_q - \na_s -\nb_s \right]   G^{\cl}_{qs} \nonumber \\
	 & + \left[ \na_p + \na_r - \na_q - \na_s \right] G^{\open}_{qs} \nonumber \\
	 & + \left[ \nb_p + \nb_r - \nb_q - \nb_s \right] \left[ 2 G^{\oc}_{qs} - G^{\open}_{qs} \right] \Big) \nonumber \\
	& + \delta_{qs} \Big( \left[ \na_q + \nb_q + \na_s + \nb_s - \na_p - \nb_p - \na_r -\nb_r \right]   G^{\cl}_{pr} \nonumber \\
	& +  \left[ \na_q + \na_s - \na_p - \na_r \right] G^{\open}_{pr} \nonumber \\
	& +  \left[ \nb_q + \nb_s - \nb_p - \nb_r \right] \left[ 2 G^{\oc}_{pr} - G^{\open}_{pr} \right] \Big) \nonumber \\
	& + \delta_{ps} \Big( \left[ \na_q + \nb_q + \na_r + \nb_r - \na_p - \nb_p - \na_s -\nb_s \right]   G^{\cl}_{qr} \nonumber \\
	& +  \left[ \na_q + \na_r - \na_p - \na_s \right] G^{\open}_{qr} \nonumber \\
	& +  \left[ \nb_q + \nb_r - \nb_p - \nb_s \right] \left[ 2 G^{\oc}_{qr} - G^{\open}_{qr} \right] \Big) \nonumber \\
	& + \delta_{qr} \Big( \left[ \na_p + \nb_p + \na_s + \nb_s - \na_q - \nb_q - \na_r -\nb_r \right]   G^{\cl}_{ps} \nonumber \\
	& +  \left[ \na_p + \na_s - \na_q - \na_r \right] G^{\open}_{qr} \nonumber \\
	& +  \left[ \nb_p + \nb_s - \nb_q - \nb_r \right] \left[ 2 G^{\oc}_{qr} - G^{\open}_{qr} \right] \Big) \nonumber \\
	& + 4 \left[ \left(\na_p - \na_q\right)\left(\na_r - \na_s\right) + \left(\nb_p - \nb_q\right)\left(\nb_r - \nb_s\right) \right] \langle pr||qs \rangle \nonumber \\
	& + 4 \left[ \left(\na_p - \na_q\right)\left(\nb_r - \nb_s\right) + \left(\nb_p - \nb_q\right)\left(\na_r - \na_s\right) \right] \langle pr|qs \rangle,
\end{align}
where the antisymmetrized two-electron integrals are defined as
\begin{equation}
	\langle pr||qs \rangle = \langle pr|qs \rangle - \langle pr|sq \rangle.
\end{equation}

For the contributions to the orbital hessian from the remaining components of $E^{\dno}$, it is useful to consider a general expression for component $X$,
\begin{equation}
	E^{\dno}_X = \sum_{abcd} \Gamma_{abcd}^X \langle ab|cd \rangle
\end{equation}
where $\Gamma_{abcd}^X$ is a 2-RDM element, and in this case $X=\text{pair}, \text{stat}$ or HSC.  The second derivative with respect to the skew-matrix parameters for this general expression is given by
\begin{align}
	H^{\phi\phi,X}_{pq,rs} = 2 \sum_{bcd} \Big[ \big( & \delta_{ps} \Gamma^X_{rbcd} - \delta_{pr} \Gamma^X_{sbcd} \big) \langle qb|cd \rangle \nonumber \\
	+ \big( & \delta_{qr} \Gamma^X_{pbcd} - \delta_{pr} \Gamma^X_{qbcd} \big) \langle sb|cd \rangle \nonumber \\
	+ \big( & \delta_{ps} \Gamma^X_{qbcd} - \delta_{qs} \Gamma^X_{pbcd} \big) \langle rb|cd \rangle \nonumber \\
	+ \big( & \delta_{qr} \Gamma^X_{sbcd} - \delta_{qs} \Gamma^X_{rbcd} \big) \langle pb|cd \rangle \Big] \nonumber \\
	+ 4 \sum_{ab} \Big[ & \Gamma^X_{prab} \langle qs|ab \rangle - \Gamma^X_{qrab} \langle ps|ab \rangle \nonumber \\
	- & \Gamma^X_{psab} \langle qr|ab \rangle + \Gamma^X_{qsab} \langle pr|ab \rangle \nonumber \\
	+ & \Gamma^X_{parb} \langle qa|sb \rangle - \Gamma^X_{pasb} \langle qa|rb \rangle \nonumber \\
	- & \Gamma^X_{qarb} \langle pa|sb \rangle + \Gamma^X_{qasb} \langle pa|rb \rangle \nonumber \\
	+ & \Gamma^X_{pabr} \langle qa|bs \rangle - \Gamma^X_{qabr} \langle pa|bs \rangle \nonumber \\
	- & \Gamma^X_{pabs} \langle qa|br \rangle + \Gamma^X_{qabs} \langle pa|br \rangle \Big]
\end{align}
and can be found elsewhere.\cite{Bozkaya2011}  In the case of $\dno$, for which the 2-RDM is a JKL-functional, the above expression can be simplified due to the fact that each term of the 2-RDM is described by only two indices, rather than four.  The simplification can be applied specifically to terms involving $J$, $K$ or $L$ integrals.  In the case of Coulomb integrals, $J$, the 2-RDM simplifies as $\Gamma_{abcd} = \delta_{ac}\delta_{bd} \gamma_{ab}$, giving
\begin{equation}
	E^{\dno}_{X_J} = \sum_{ab} \gamma_{ab} J_{ab},
\end{equation}
and leading to the following orbital hessian contribution,
\begin{align}
	H^{\phi\phi,{X_J}}_{pq,rs} = 4 \big( \gamma_{pr} & + \gamma_{qs} - \gamma_{ps} - \gamma_{qr} \big) 
	\left( \langle pr | qs \rangle + \langle ps | qr \rangle \right) \nonumber \\
	+ 2\sum_a \Big[ & \delta_{ps} \left( \gamma_{qa} + \gamma_{ra} - \gamma_{pa} - \gamma_{sa} \right) \langle qa | ra \rangle \nonumber \\
	+ & \delta_{pr} \left( \gamma_{pa} + \gamma_{ra} - \gamma_{qa} - \gamma_{sa} \right) \langle qa | sa \rangle \nonumber \\
	+ & \delta_{qr} \left( \gamma_{pa} + \gamma_{sa} - \gamma_{qa} - \gamma_{ra} \right) \langle pa | sa \rangle \nonumber \\
	+ & \delta_{qs} \left( \gamma_{qa} + \gamma_{sa} - \gamma_{pa} - \gamma_{ra} \right) \langle pa | ra \rangle \Big]
\end{align}
In the case of exchange integrals, $K$, the 2-RDM simplifies as $\Gamma_{abcd} = \delta_{ad}\delta_{bc} \gamma_{ab}$, giving
\begin{equation}
	E^{\dno}_{X_K} = \sum_{ab} \gamma_{ab} K_{ab} 
\end{equation}
and leading to the following orbital hessian contribution,
\begin{align}
	H^{\phi\phi,{X_K}}_{pq,rs} = 4 \big( \gamma_{pr} & + \gamma_{qs} - \gamma_{ps} - \gamma_{qr} \big) 
	\left( \langle pr | sq \rangle + \langle ps | rq \rangle \right) \nonumber \\
	+ 2\sum_a \Big[ & \delta_{ps} \left( \gamma_{qa} + \gamma_{ra} - \gamma_{pa} - \gamma_{sa} \right) \langle qa | ar \rangle \nonumber \\
	+ & \delta_{pr} \left( \gamma_{pa} + \gamma_{ra} - \gamma_{qa} - \gamma_{sa} \right) \langle qa | as \rangle \nonumber \\
	+ & \delta_{qr} \left( \gamma_{pa} + \gamma_{sa} - \gamma_{qa} - \gamma_{ra} \right) \langle pa | as \rangle \nonumber \\
	+ & \delta_{qs} \left( \gamma_{qa} + \gamma_{sa} - \gamma_{pa} - \gamma_{ra} \right) \langle pa | ar \rangle \Big]
\end{align}
Similarly, for time-inversion eschange integrals, $L$, the 2-RDM simplifies as $\Gamma_{abcd} = \delta_{ab}\delta_{cd} \gamma_{ac}$ and the orbital hessian contribution simplifies in the same manner.

Simultaneous optimization of the electron-transfer variables, $\{ \Delta_{pq} \}$, and consequently the orbital occupancies, requires including the gradient and hessian contributions from $\mathbf{\theta}$, where $\theta_{pq}$ is related to $\Delta_{pq}$ via Equation \eqref{eq:theta}.  The gradient contributions are given by,
\begin{equation}
	g^\theta_{pq} = -\frac{\sin(2\theta_{pq})}{2} \frac{\partial E^{\dno}}{\partial \Delta_{pq}}
\end{equation}
The derivative of $E^{\dno}_\text{0-1RDM}$ with respect to $\Delta_{pq}$ is derived using the following identity, 
\begin{equation}
	\frac{\partial n_t}{\partial \Delta_{pq}} = \delta_{qt} - \delta_{pt},
\end{equation}
which leads to
\begin{equation}
	\frac{\partial E^{\dno}_\text{0-1RDM}}{\partial \Delta_{pq}} =
	2 \left( h_{qq} - h_{pp} + G_{qq}^{\cl} - G_{pp}^{\cl} + G_{qq}^{\oc} - G_{pp}^{\oc} \right).
\end{equation}
The contribution to the gradient from the remaining components of the energy can be calculated by replacing the coefficients, $\eta_{tu}$, $\zeta_{tu}$, $\xi_{tu}$ and $\kappa_{tu}$ [Equations \eqref{eq:epair}, \eqref{eq:estat} and \eqref{eq:ehsc}] with their corresponding derivatives.  The derivatives of the coefficients are given by
\begin{align}
	\frac{\partial \eta_{tu}}{\partial \Delta_{pq}} = \delta_{tu} \Bigg[ & \delta_{qt} \left( 1- 2 n_q \right) -\delta_{pt} \left( 1 - 2 n_p \right) \Bigg] \nonumber \\
	+ \left(1 - \delta_{tu}\right) \Bigg[ & O_t V_u \Big\{ \delta_{pt} \delta_{qu} \left( n_u - n_t - \Delta_{tu} \right) \nonumber \\
	& \hphantom{O_tV_t} + \Delta_{tu} \left( \delta_{qu} + \delta_{pt} - \delta_{pt}\delta_{qu} \right) \Big\} \nonumber \\
	+ & V_t O_u \Big\{ \delta_{pu} \delta_{qt} \left( n_t - n_u - \Delta_{ut} \right) \nonumber \\
	& \hphantom{O_tV_t} + \Delta_{ut} \left( \delta_{qt} + \delta_{pu} - \delta_{pu}\delta_{qt} \right) \Big\} \nonumber \\
	+ & V_t V_u \left( \delta_{qt} \Delta_{pu} + \delta_{qu} \Delta_{pt} \right) \Bigg],
\end{align}
\begin{equation}
	\frac{\partial \zeta_{tu}}{\partial \Delta_{pq}} = \frac{V_t V_u}{2} \left( \delta_{qt}\sqrt{\frac{\Delta_{pu}}{\Delta_{pt}}} + \delta_{qu}\sqrt{\frac{\Delta_{pt}}{\Delta_{pu}}} \right),
\end{equation}
\begin{align}
\frac{\partial \xi_{tu}}{\partial \Delta_{pq}} & = \frac{O_t V_u}{2} \left( \delta_{pt}\delta_{qu} \sqrt{\frac{n_t}{\Delta_{tu}}} - \delta_{pt} \sqrt{\frac{\Delta_{tu}}{n_t}} \right) \nonumber \\ 
	& + \frac{O_u V_t}{2} \left( \delta_{pu}\delta_{qt} \sqrt{\frac{n_u}{\Delta_{ut}}} - \delta_{pu} \sqrt{\frac{\Delta_{ut}}{n_u}} \right),
\end{align}
and
\begin{align}
	\frac{\partial \kappa_{tu}}{\partial \Delta_{pq}} & = \left( 1 - \delta_{tu} \right)
	\Bigg[ \sum_{v\ne w} \left( \xi_{uw} \frac{\partial \xi_{tv}}{\partial \Delta_{pq}} + \xi_{tv} \frac{\partial \xi_{uw}}{\partial \Delta_{pq}} \right) \nonumber \\
	& + W_t \frac{n_t}{2\sqrt{2}} \sum_v \frac{\partial \xi_{uv}}{\partial \Delta_{pq}}
	+ W_u \frac{n_u}{2\sqrt{2}} \sum_v \frac{\partial \xi_{tv}}{\partial \Delta_{pq}} \Bigg].
\end{align}

Using the chain rule, the $\mathbf{\theta}$ contribution to the hessian can be expressed in terms of derivatives with respect to the $\{ \Delta_{pq} \}$,
\begin{align}
	H_{pq,rs}^{\theta\theta} & = \frac{\sin (2 \theta_{pq}) \sin(2 \theta_{rs})}{4} \frac{\partial^2 E^{\dno}}{\partial \Delta_{pq} \partial \Delta_{rs}} \nonumber \\
	& - \delta_{pr} \delta_{qs} \cos(2\theta_{pq}) \frac{\partial E^{\dno}}{\partial \Delta_{pq}}
\end{align}
In the case of the second derivative, it is useful to express the 0-1RDM and pair contributions together, where
\begin{equation}
	E^{\dno}_{\text{0-1RDM-pair}} = E^{\dno}_{\text{0-1RDM}} + E^{\dno}_{\text{pair}},
\end{equation}
and 
\begin{align}
	\frac{\partial^2 E^{\dno}_{\text{0-1RDM-pair}}}{ \partial \Delta_{pq} \partial \Delta_{rs} } & = 4\left( J_{pr} + J_{qs} - J_{ps} - J_{qr} \right) \nonumber \\
	& - 2 \left(  K_{pr} + K_{qs} - K_{ps} - K_{qr} \right)
\end{align}

For the remainder of the contributions to the hessian, as with the gradient, the second derivatives can be calculated by replacing the coefficients, $\zeta_{tu}$, $\xi_{tu}$ and $\kappa_{tu}$, with their corresponding second derivative,
\begin{align}
	\frac{\partial^2 \zeta_{tu}}{\partial \Delta_{pq} \partial \Delta_{rs}} & =
	\delta_{pr} \frac{V_t V_u}{4} \Bigg( \frac{\delta_{pt} \delta_{su} + \delta_{qu} \delta_{st}}{\sqrt{ \Delta_{pt} \Delta_{pu}}} \nonumber \\
	& - \delta_{qt}\delta_{st} \sqrt{\frac{\Delta_{pu}}{\Delta_{pt}^3}} - \delta_{qu}\delta_{su} \sqrt{\frac{\Delta_{pt}}{\Delta_{pu}^3}} \Bigg),
\end{align}
\begin{align}
	\frac{\partial^2 \xi_{tu}}{\partial \Delta_{pq} \partial \Delta_{rs}} & =
	\delta_{pr} \frac{O_t V_u}{4} \Bigg( \frac{-\delta_{pt} \delta_{qu} - \delta_{pt} \delta_{su}}{\sqrt{n_t \Delta_{tu}}} \nonumber \\ & - \delta_{pt} \delta_{qu} \delta_{su} \sqrt{\frac{n_t}{\Delta_{tu}^3}} - \delta_{pt} \sqrt{\frac{\Delta_{tu}}{n_t^3}} \Bigg) \nonumber \\
	& + \delta_{pr} \frac{O_u V_t}{4} \Bigg( \frac{-\delta_{pu} \delta_{qt} - \delta_{pu} \delta_{st}}{\sqrt{n_u \Delta_{ut}}} \nonumber \\
	& - \delta_{pu} \delta_{qt} \delta_{st} \sqrt{\frac{n_u}{\Delta_{ut}^3}} - \delta_{pu} \sqrt{\frac{\Delta_{ut}}{n_u^3}} \Bigg)
\end{align}
and
\begin{align}
	\frac{\partial^2 \kappa_{tu}}{\partial \Delta_{pq} \partial \Delta_{rs}} = \left( 1 - \delta_{tu} \right) \Bigg[ \sum_{v\ne w} \Big( & \xi_{uw} \frac{\partial^2 \xi_{tv}}{\partial \Delta_{pq} \partial \Delta_{rs}} + \xi_{tv} \frac{\partial^2 \xi_{uw}}{\partial \Delta_{pq}\partial \Delta_{rs}} \nonumber \\
	+ & \frac{\partial \xi_{tv}}{\partial \Delta_{pq}} \frac{\partial \xi_{uw}}{\partial \Delta_{rs}} + \frac{\partial \xi_{uw}}{\partial \Delta_{pq}} \frac{\partial \xi_{tv}}{\partial \Delta_{rs}} \Big) \nonumber \\
	+ W_t \frac{n_t}{2\sqrt{2}} \sum_v \frac{\partial^2 \xi_{uv}}{\partial \Delta_{pq}\partial \Delta_{rs}} & + W_u \frac{n_u}{2\sqrt{2}} \sum_v \frac{\partial^2 \xi_{tv}}{\partial \Delta_{pq}\partial \Delta_{rs}} \Bigg].
\end{align}

The hessian terms involving derivatives with respect to both $\by$ and $\mathbf{\theta}$, can be expressed in terms of derivatives of $\mathbf{\lambda}$,
\begin{equation}
	H^{\phi\theta}_{pq,rs} = -\sin(2\theta_{rs}) \left( \frac{\partial \lambda_{qp}}{\partial \Delta_{rs}} - \frac{\partial \lambda_{pq}}{\partial \Delta_{rs}} \right).
\end{equation}
The derivative of the $\lambda_{pq}$ component corresponding to $E^{\dno}_\text{0-1RDM}$ [first three terms of Equation \eqref{eq:lambda}] is given by 
\begin{align}
	\frac{\partial \lambda^{\text{0-1RDM}}_{pq}}{\partial \Delta_{rs}} = \big( & 2 - W_p\big) \nonumber \\
	\times \Big[ & \left( \delta_{ps} - \delta_{pr} \right) \left( h_{pq} + G_{pq}^{\cl}
	+ (1-W_p)G_{pq}^{\oc} + W_p G_{pq}^{\open} \right) \nonumber \\
	& + 2\left( \langle ps|qs \rangle - \langle pr|qr \rangle \right) -\left( \langle ps|sq \rangle - \langle pr|rq \rangle \right) \Big].
\end{align}
The remaining components of $\frac{\partial \lambda_{pq}}{\partial \Delta_{rs}}$ are derived by substituting the coefficients, $\zeta_{tu}$, $\xi_{tu}$ and $\kappa_{tu}$, with their corresponding derivative.

\bibliography{DNO_Hn}

\end{document}